\documentclass[aps,pra,superscriptaddress,showpacs,floatfix]{revtex4-2}

\usepackage{amssymb,amsmath,amsfonts,mathrsfs,bm,bbm,graphicx,epsfig,epstopdf}
\usepackage[version=3]{mhchem}
\usepackage{braket}
\usepackage{hyperref}
\usepackage{color}
\usepackage{xcolor}
\usepackage{adjustbox}
\usepackage{csquotes}
\usepackage{float}
\usepackage{colortbl}
\usepackage{siunitx}
\DeclareSIUnit{\nothing}{\relax}
\usepackage{makecell}
\usepackage[T1]{fontenc}
\usepackage{soul}
\usepackage{booktabs}
\usepackage{bibunits}
\usepackage{multirow}

\usepackage{bibunits}
\defaultbibliographystyle{apsrev4-2}
\defaultbibliography{bibliography}

\begin{document}

\begin{bibunit}

\date{\today}

\author{Yisheng Lei}
\thanks{Corresponding author}
\email{YishengLei2025@u.northwestern.edu}
\affiliation{Institut de Recherche de Chimie Paris, Chimie ParisTech, CNRS, PSL University, 75005 Paris, France}

\author{Senthil Kumar Kuppusamy}
\affiliation{Institute for Quantum Materials and Technologies (IQMT), Karlsruhe Institute of Technology, Karlsruhe, Germany
}

\author{Idris Tlemsani}
\affiliation{Institut de Recherche de Chimie Paris, Chimie ParisTech, CNRS, PSL University, 75005 Paris, France}
\affiliation{Institut de Chimie Moléculaire et des Matériaux d’Orsay CNRS, Université Paris-Saclay UMR 8182, 17, avenue des Sciences, 91400, Orsay (France)}

\author{Suma Al-Hunaishi}
\affiliation{Institut de Recherche de Chimie Paris, Chimie ParisTech, CNRS, PSL University, 75005 Paris, France}

\author{Pengrui Jiao}
\affiliation{Institut de Recherche de Chimie Paris, Chimie ParisTech, CNRS, PSL University, 75005 Paris, France}

\author{Olaf Fuhr}
\affiliation{Institute of Nanotechnology (INT), Karlsruhe Institute of Technology, Karlsruhe, Germany}
\affiliation{Karlsruhe Nano Micro Facility (KNMFi), Karlruhe Institute of Technology (KIT), Kaiserstraße 12, 76131 Karlsruhe}

\author{Mario Ruben}
\affiliation{Institute for Quantum Materials and Technologies (IQMT), Karlsruhe Institute of Technology, Karlsruhe, Germany
}
\affiliation{Institute of Nanotechnology (INT), Karlsruhe Institute of Technology, Karlsruhe, Germany}
\affiliation{Centre Européen de Sciences Quantiques (CESQ), Institut de Science et d’Ingénierie Supramoléculaires (ISIS), Strasbourg, France}

\author{Philippe Goldner}
\affiliation{Institut de Recherche de Chimie Paris, Chimie ParisTech, CNRS, PSL University, 75005 Paris, France}

\author{Diana Serrano}
\thanks{Corresponding author}
\email{diana.serrano@chimieparistech.psl.eu}
\affiliation{Institut de Recherche de Chimie Paris, Chimie ParisTech, CNRS, PSL University, 75005 Paris, France}

\begin{abstract}
Broadband quantum memory devices are essential elements for future quantum networks \cite{kimble2008quantum, wehner2018quantum, lei2023quantum}. Here we propose a broadband quantum memory scheme called Hole Anti-hole Grating Echo Memory (HAGEM) for rare-earth ions in solids, which can be implemented with most of the present rare-earth ions hosted in various types of solids. Using HAGEM, a quantum memory device with one billion temporal modes per second can be developed with current available technologies \cite{businger2022non}. To validate the memory protocol, we conducted some proof-of-concept experiments. We provide a Eu$^\text{{3+}}$ molecular complex with special hyperfine level structures of which the hyperfine level separations are in a specific mathematical correlation that can be obtained by harnessing chemical engineering \cite{serrano2022ultra}. Using the proposed memory protocol and material, we experimentally demonstrate a single-mode quantum optical storage efficiency of 14.9\% and a memory bandwidth of 200~MHz, which can easily be extended to a few GHz. In addition, a multi-mode storage of four temporal modes is achieved with a storage efficiency of 1.3\%. Broadband \& large-multimode quantum memory device with a long storage time enabled by HAGEM integrating with the current available entangled photon pair source and tunable delay lines, a quantum repeater network with an entanglement distribution rate on the order of 1~kHz - 10~kHz can be created with a communication distance around 100~km \cite{simon2007quantum, collins2007multiplexed, lago2021telecom, liu2021heralded}. This will greatly facilitate the development of quantum networks in near future.
\end{abstract}


\title{Broadband and Large-Multimode Quantum Optical Storage with Rare-Earth Ions in Solids}
\maketitle{}

\section{Introduction}
Quantum memory devices are essential elements for quantum networks and many other quantum applications \cite{kimble2008quantum, wehner2018quantum}. For quantum repeaters based on absorptive quantum memories and photon pair sources \cite{simon2007quantum}, storage time will determine the maximum communication distance, and storage bandwidth will determine the information processing density, which is multi-mode capacity. Multi-mode capacity will greatly increase the entanglement distribution rate, especially with multiplexing \cite{collins2007multiplexed}. Rare-earth (RE) ions in crystals undergo optical transitions within 4f levels, which are electromagnetically shielded by outer shells of 5s and 5p, resulting in quantum emitters with long optical and spin coherence times. In recent two decades, rarely doped RE ions in solids have attracted enormous attention for quantum memory applications, and great progress has been made so far. RE ions doped in crystals have been used to demonstrate quantum storage with long storage time, large memory bandwidth, high storage efficiency, and multi-mode capacity separately \cite{lvovsky2009optical, lei2023quantum,tittel2025quantum}. Eu$^\text{{3+}}$ \& Pr$^\text{{3+}}$ ions doped in various types of crystals have been demonstrated with high storage efficiency, long storage time, multi-mode capacity, nanophotonic integration, and on-demand characteristics \cite{jobez2015coherent, meng2026efficient, duranti2024efficient, teller2025solid, feldmann2025cavity}. 
Restricted by narrow hyperfine level separations, memory bandwidth with Eu$^\text{{3+}}$ \& Pr$^\text{{3+}}$ ions doped in solids has been limited to a few MHz to tens of MHz \cite{cruzeiro2018characterization}. Efficient broadband quantum storage has been achieved only with one type of non-Kramer ion but with short storage time, which is Tm$^\text{{3+}}$ \cite{saglamyurek2011broadband, davidson2020improved,  askarani2020entanglement, askarani2021long}. Broadband quantum storage was demonstrated with Pr$^\text{{3+}}$ doped in crystal, but large background absorption seriously limits the storage efficiency, and the storage time is short \cite{nicolle2021gigahertz}. \textit{Lei et al.} recently proposed a commensurate scheme to achieve high memory bandwidth for non-Kramer RE ions with two hyperfine levels, which is to match all holes and anti-holes with atomic frequency comb (AFC) peaks and valleys by applying a specific amplitude of magnetic field \cite{lei2025efficient}. Quantum memory based on Er$^\text{{3+}}$ ions doped in solids has been shown to have multi-mode capacity and large memory bandwidth, but limited storage time and efficiency \cite{zhang2023telecom, jiang2023quantum, wei2024quantum}, high storage efficiency but narrow memory bandwidth \cite{stuart2021initialization}. Yb$^\text{{3+}}$ ions doped in solids have been shown to have multi-mode capacity, high storage efficiency, long storage time and on-demand capability \cite{businger2020optical, businger2022non, rodriguez2026entanglement}. However, a single quantum memory device with all these features has not been shown so far; furthermore, the memory bandwidth and multi-mode capacity that have been shown so far are not large enough to develop a quantum repeater network. With the capacity of the available quantum memory devices based on RE ions in solids, only one elementary link in a quantum repeater network has been demonstrated with a limited entanglement distribution rate \cite{liu2021heralded, lago2021telecom}. In this article, we proposed a quantum memory protocol HAGEM, which will enable this type of device with different types of RE ions and bring a quantum repeater network into reality by integrating with frequency matched entangled photon pair sources, owing to the enormous amount of temporal modes; in addition, we introduce a chemically tunable Eu$^\text{{3+}}$ molecular material and use it to experimentally demonstrate HAGEM. 

Chemical engineering of molecular qubits and quantum emitters has been extensively studied over the past decade, enabled by its flexibility of chemical designs \cite{toninelli2021single, yu2021molecular, kuppusamy2024spin, bayliss2022enhancing}. RE ion molecules (REIM) can offer a variety of exceptional functionalities, which go significantly beyond materials currently studied for quantum technologies: (1) chemical synthesis enables precise control over the local coordination environment and atomic-scale positioning of the RE ion, providing a route to long-lived quantum coherence and tailored spin \& optical properties that are difficult to achieve in conventional crystalline hosts; (2) the molecular structure and ligand environment can be systematically engineered to tune key spectroscopic parameters, including the optical transition wavelength and hyperfine level splittings. Together, these two levels of control provide a unique platform for developing new materials for quantum applications. 

In this article, we report the first proposal of broadband and large-multimode quantum optical memory protocol, and some proof-of-concept experiments with a tailored Eu$^\text{3+}$ molecular material that implement it. In addition, we show that a quantum repeater network based on quantum memories with RE ions in solids, variable delay elements, and entangled photon pair sources can be readily developed with current available technologies for a communication distance around 100~km and an entanglement distribution rate on the order of 1~kHz to 10~kHz. 

\section{Memory Protocol}

\subsection{Protocol description}
\begin{figure}[!h]
\centerline{\includegraphics[width=1\columnwidth]{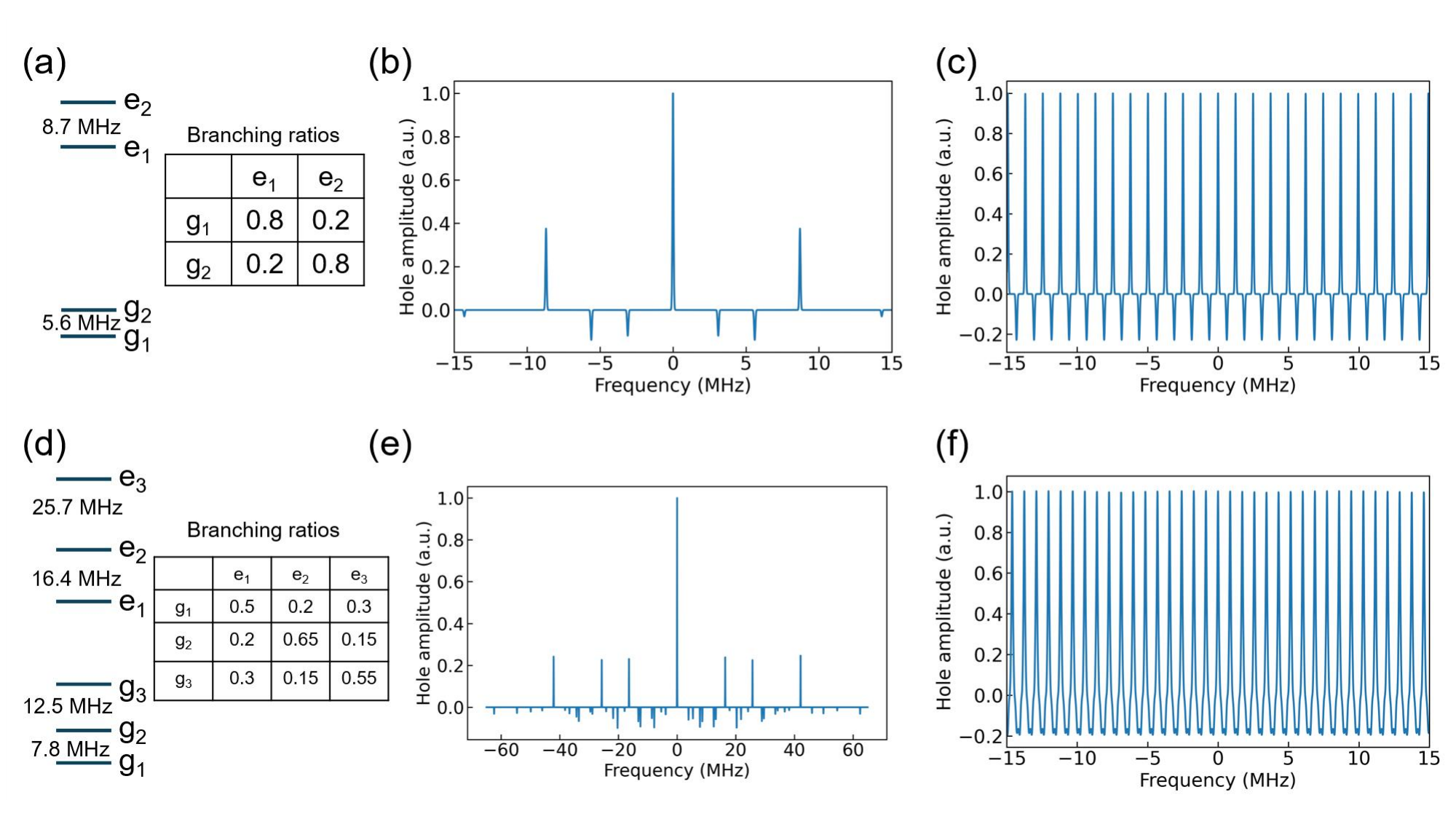}}
	\caption{HAGEM memory protocol principle: (a) One example of a two-level hyperfine structure and its branching ratio values for hyperfine levels between the optical transition states used in (b-c) spectral hole burning simulations. (b) Simulation of a single spectral hole burning for (a). (c) Simulation of HAGEM spectrum for (a) with $\delta = 0.6217$~MHz. (d) One example of a three-level hyperfine structure and its branching ratio values for hyperfine levels between the optical transition states used in (e-f) spectral hole burning simulations. (e) Simulation of a single spectral hole burning for (d). (f) Simulation of HAGEM spectrum for (d) with $\delta = 0.4296$~MHz.}
	\label{Fig_HAGEM}
\end{figure}

This memory protocol is inspired by AFC memory protocol, which is to create atomic grating structure equally spaced by $\Delta$ in frequency domain with an inhomogeneous broadened atomic ensemble, and after photon absorption the atomic ensemble will be rephased after a time period of 1/$\Delta$ then a photon echo will emit \cite{afzelius2009multimode}. Let us look at a simple case first, for RE ions with two hyperfine levels in both the ground and excited states, if the ground state splitting is $\delta$ and the excited state splitting is $2\delta$, broadband storage can be achieved with the HAGEM memory protocol (SI S1.2). There is a special case for Tm$^\text{{3+}}$ ions doped in some solids, of which the excited state energy gap is much smaller than the ground state energy gap. During a single spectral hole burning, three spectral holes overlap to form a hole cluster and six anti-holes overlap to form two anti-hole clusters and they are equally located to the hole cluster on two sides, and this can be used for broadband storage.

Due to the inhomogeneous broadening, there are four classes of ions resonant with a single-frequency laser pulse, if both the ground and excited states have two hyperfine levels with optical transitions between different hyperfine levels. After spectral hole burning (SHB), each class of ions will create two holes (reduced absorption) and two anti-holes (increased absorption). In total, there are one center hole located at f$_\text{0}$, 2 side holes located around f$_\text{0}$ de-tuned by $\pm\Delta e$, and 6 anti-holes located around f$_\text{0}$ de-tuned by values equal to $\pm \Delta_g$, $\pm (\Delta_g - \Delta_e)$ and $\pm (\Delta_g + \Delta_e)$. Here $\Delta_e$ = 8.7~MHz and $\Delta_g$ = 5.6~MHz, and the branching ratio values are listed in Fig. \ref{Fig_HAGEM}(a). The single spectral hole burning spectrum is simulated with spectral linewidth of 0.1~MHz (Full-Width-Half-Maximum value) and a characteristic pumping duration corresponding to approximately 20 excitation–relaxation cycles, as shown in Fig. \ref{Fig_HAGEM}(b). If $\Delta_e = k_1* \delta$ and $\Delta_g = k_2* \delta$, where $k_1$ is an even integer and $k_2$ is an odd integer, all anti-holes are located at odd multiples of $\delta$ while all holes are located at even multiples of $\delta$ after a single SHB. For this case, $\delta = 0.6217$~MHz, $k_1$ = 14 and $k_2$ = 9. If all $2n \delta$ frequency points are spectrally pumped, holes are located at all $2n \delta$ frequency points, and anti-holes are located at all $(2n+1) \delta$ frequency points, so a grating spectrum is created with a grating distance of $2 \delta$, as shown in Fig. \ref{Fig_HAGEM}(c). This type of spectrum can be created with the entire inhomogeneous broadening line. The simulation method for $\delta$ and the degree of mismatch of the simulated spectrum compared to a perfect grating spectrum $C_{norm}$ are explained in SI S1.1, and the simulation method of the SHB spectrum is explained in SI S3.3.

There are nine classes of ions resonant with a single-frequency laser pulse, if both the ground and excited states have three hyperfine levels with optical transitions between different hyperfine levels. After spectral pumping, each class of ions will create three holes and six anti-holes. In total, there are one center hole located at f$_\text{0}$, 6 side holes located around f$_\text{0}$ de-tuned by values equal to terms in [$\pm\Delta e_{1}$, $\pm\Delta e_{2}$, $\pm\left( \Delta e_{1} + \Delta e_{2} \right)$], and 42 anti-holes located around f$_\text{0}$ de-tuned by values equal to one term in [$\pm\Delta g_{1}$, $\pm\Delta g_{2}$, $\pm\left( \Delta g_{1} + \Delta g_{2} \right)$] + one term in [$0$, $\pm\Delta e_{1}$, $\pm\Delta e_{2}$, $\pm\left( \Delta e_{1} + \Delta e_{2} \right)$]. In the example of Fig. \ref{Fig_HAGEM}(d), $\Delta g_{2}$ is assumed to be 12.5~MHz, $\Delta g_{1}$ = 7.8~MHz, $\Delta e_{2}$ = 25.7~MHz and $\Delta e_{1}$ = 16.4~MHz. One single spectral hole burning spectrum is simulated with the same linewidth and pumping time as above, as shown in Fig. \ref{Fig_HAGEM}(e). If $\Delta_{e_1} = k_1* \delta$, $\Delta_{e_2} = k_2* \delta$, $\Delta_{g_1} = k_3* \delta$ and $\Delta_{g_2} = k_4* \delta$, where $k_1$ \& $k_2$ are even integers, at least one of $k_3$ \& $k_4$ is an odd integer, and the other one can be even or odd. For this case, $\delta = 0.4296$~MHz, $k_1$ = 38, $k_2$ = 60, $k_3$ = 18 and $k_4$ = 29. Next, We analyze whether or not the ions at one frequency location can be completely emptied and shifted to the anti-hole positions located at frequencies of odd multiples of $\delta$. Categorical \text{ 1:} atoms are transferred from $g_{3}$ to $g_{1}$ or $g_{2}$; the anti-holes are at the frequencies of odd multiples of $\delta$, since the frequency changes at the ground hyperfine levels are $\Delta g_{2}$ and $\Delta g_{2}$ + $\Delta g_{1}$, which are odd multiples of $\delta$. \text{ Categorical 2:} atoms are transferred from $g_{1}$ or $g_{2}$ to $g_{3}$; the anti-holes are at the frequencies of odd multiples of $\delta$.\text{ Categorical 3:} atoms are transferred from $g_{1}$ to $g_{2}$ or vice versa; the anti-holes are at incorrect positions (frequencies of even multiples of $\delta$), since the frequency changes at the ground hyperfine levels is $\Delta g_{1}$. For laser frequencies at $f_{0} + k*2\delta$ (k is an integer), atoms at \text{ Categorical 3} will eventually be transferred to $g_{3}$, so the anti-holes are at the correct positions. In summary, with this particular $\delta$, a nearly perfect spectral grating structure can be created and no background absorption will be left. The HAGEM spectrum grating distance $\Delta_{HAGEM}$ is $2\delta$, which corresponds to a storage time of $1/\Delta_{HAGEM}$. The HAGEM spectrum is shown in Fig. \ref{Fig_HAGEM}(f). 


Based on the analysis above, we give the general rules: $\delta$ is a specially selected value (many possible choices); one ground state energy gap is an odd multiple of $\delta$ and another one can be even or odd multiples of $\delta$; both excited state energy gaps have to be even multiples of $\delta$. This rule can be extended to RE ions with four or more hyperfine levels (SI S1.4), that is, at least one ground state energy gap is an odd multiple of $\delta$ and the rest can be even or odd multiples of $\delta$; all excited state energy gaps must be even multiples of $\delta$. If the atoms have $n_g$ hyperfine levels in the ground states and $n_e$ hyperfine levels in the excited states, there are in total $n_e \times (n_e-1)$ spectral side holes and $\left(n_g \times (n_g-1)\right)\times\left(n_e \times (n_e-1)+1\right)$ spectral anti-holes. Eu$^\text{{3+}}$ ions doped in some solids under zero magnetic field may not meet the requirements for certain storage times. Applying an external magnetic field, the three degenerate hyperfine levels will split into six hyperfine levels, and the HAGEM protocol may be implemented with some particular amplitudes of the magnetic field, depending on the Zeeman splittings \cite{zhong2015optically}. For some types of RE ions, coupling their electron or nuclear spins with nuclear spins from the host materials through superhyperfine interactions may also be used to implement the HAGEM protocol under some particular amplitudes of magnetic field, and superhypefine interactions can also greatly increase their ground state lifetime and optical coherence time \cite{ahlefeldt2015optical}. 

For Kramer ions (SI S1.6), it is still possible to implement the HAGEM protocol even with the large hyperfine level splittings. There are three scenarios: for Kramer ions without nuclear spins and which have no superhyperfine interactions, the ground electron spin splitting is $\Delta_{g} = \delta + k_1 * 2\delta$ (odd multiples of $\delta$) and the excited electron spin splitting is $\Delta_{e} = k_2 * 2\delta$ (even multiples of $\delta$), where $k_1$ and $k_2$ are integers. Typical electron spins in solids have large splittings, so $k_1$ and $k_2$ are large values. For this case, HAGEM can be implemented with $\Delta_{HAGEM} = 2\delta$, which is the same as HAGEM for two hyperfine levels; for Kramer ions without nuclear spins \cite{chai2026hybrid}, their electronic spins that interact with nuclear spins from host materials may give a suitable superhyperfine level structure for the HAGEM protocol. The broadband storage scheme for this case is the HAGEM mixed with AFC, if another ground electronic spin state has a long lifetime. A $\delta$ value chosen for this superhyperfine level structure needs to match HAGEM memory protocol, and it is combined with AFC since there is a branching ratio into another electron spin level, so some of the atoms are pumped away from the storage frequency window; for Kramer ions with nuclear spins and have no superhyperfine interactions, if a $\delta$ value chosen for the hyperfine level splittings matches HAGEM memory protocol, they can be used for broadband storage \cite{rancic2018er}. Other electron spin hyperfine levels can be emptied during spectral preparation, and after HAGEM storage, these levels can be used for spin-wave storage with control pulses for the last two scenarios, which will enable long storage time and on-demand capability. This can only be achieved under low temperature and a moderate magnetic field (0.1 - 0.3~T), since a high magnetic field will greatly enhance spin-phonon relaxation, which greatly reduces the electron spin lifetime. Other electron spin hyperfine levels can also be used as shelving states for spectral preparation, which can help to prepare a storage spectrum (HAGEM + AFC) with a higher finesse, which leads to a higher storage efficiency. For the last scenario, such as $^\text{167}$Er$^\text{{3+}}$ ions, the storage bandwidth of AFC is limited by atomic state transition frequency separation of optical transitions $\Delta m_I = 0$, around 100~MHz, if the same electronic spin hyperfine levels are used for shelving \cite{stuart2021initialization}. It also requires long and complicated spectral preparation sequences, such as spin initialization and back-burn. For Er$^\text{{3+}}$ ions, this is particularly critical since their optical lifetimes are over 10~ms, which makes it challenging to perform efficient spectral tailoring. In this context, HAGEM offers a distinct advantage, by relying on a substantially simpler optical-pumping sequence, thereby reducing the spectral-preparation overhead and the number of excitation--relaxation cycles required to prepare the memory spectrum.  

The finesse of the grating spectrum is intrinsically 2. The storage efficiency can be estimated based on the formula for squarish combs \cite{bonarota2010efficiency, jobez2016towards},
\begin{equation}\label{StorageEfficiency}
\eta = \Tilde{d}^2\text{exp}(-\Tilde{d})\text{sinc}^2(\frac{\pi}{F})\text{exp}(-d_0), 
\end{equation}
where $\Tilde{d}$ is the optical depth (OD) of the atomic medium after optical pumping, $\Tilde{d}$ = $d$ for HAGEM where $d$ is the original OD of the atomic ensemble (this is different from AFC, where atoms are pumped away from the storage frequency window), $F$ = 2 is the finesse of the combs, and $d_0$ is the OD corresponding to the background absorption after spectral tailoring. It can be easily calculated that the maximum storage efficiency is 21.9\% when $\Tilde{d}$ = 2 and $d_0$ = 0, so to achieve the maximum storage efficiency, the OD must be 2 and the spectral preparation must be efficient to eliminate background absorption. 

As explained above, for RE ions with some shelving states, HAGEM combined with AFC can be used to increase the finesse of the storage spectrum. For RE ions without any shelving state, the finesse of 2 may not be a hard limit. Frequency-selective spectral compression of inhomogeneous broadening at the anti-hole regions can be explored to increase the finesse of the storage spectrum \cite{yavuz2013suppression}.

\subsection{Protocol implementation}

\begin{figure}[!h]
\centering
\includegraphics[width=1\textwidth]{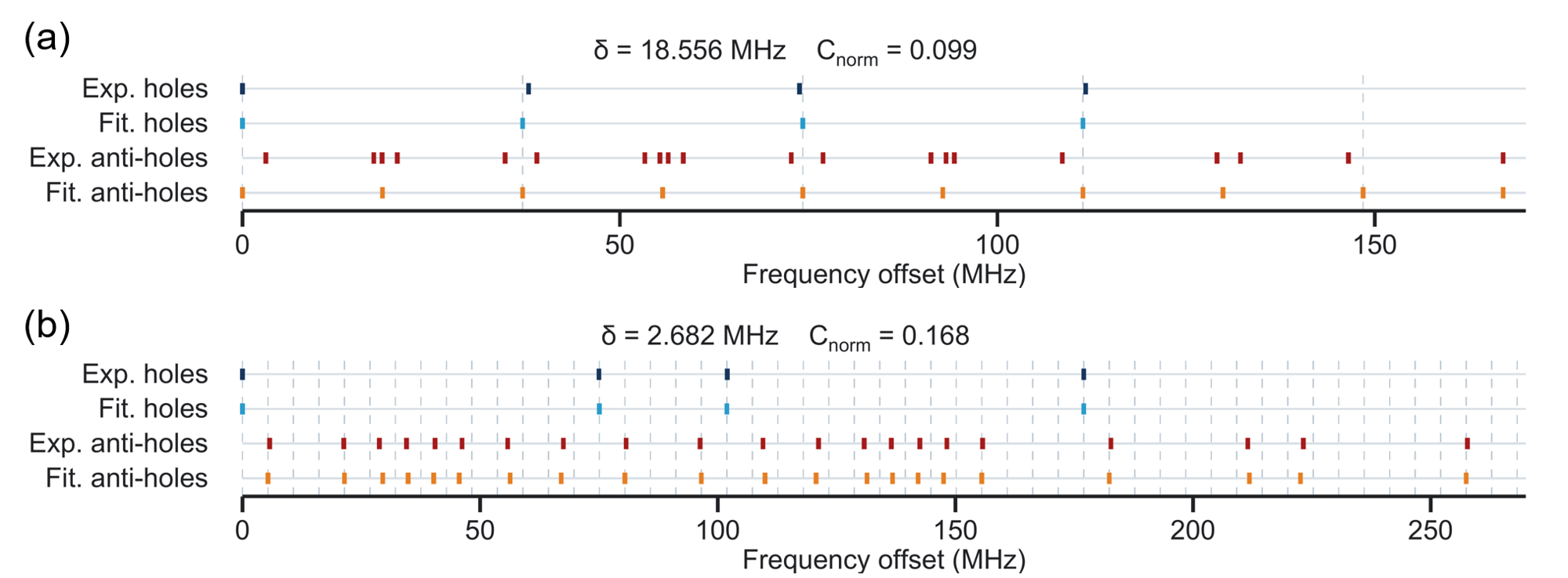}
\caption{Positive-half-spectrum comparisons for $[{}^{151}\mathrm{Eu}^{3+}(\mathrm{TMHD})_3(\mathrm{phen})]$ (a) and $^{151}\mathrm{Eu}^{3+}:\mathrm{Y}_2\mathrm{SiO}_5$ site 1 (b). ``Exp.'' and ``Fit.'' denote experimental and fitted line positions. Dashed lines mark the $2n\delta$ grid. Holes are expected on the grid, 14 anti-holes midway between the lines at odd multiples of $\delta$, and 7 anti-holes on the grid, corresponding to atoms in \text{Categorical 3} as discussed above. For some hyperfine level structures with correlated values, some spectral features may overlap, such as the case in (a). C$_{\text{norm}}$ represents the mismatch degree between a perfect HAGEM spectrum with a grating distance of $2 \delta$ and the spectrum created by a single SHB with the simulated material. A lower value of C$_{\text{norm}}$ leads to a better HAGEM implementation.}
\label{fig_EuHAGEMfit}
\end{figure}

Based on the analysis above, hyperfine level separations do not need to be correlated and can be any random unrelated values, especially for smaller values of $\delta$. To implement the HAGEM memory protocol, we simulated most existing Eu materials of interest for quantum applications, 17 in total. The results are listed in the Extended Data Table. 1. Two examples of $[{}^{151}\mathrm{Eu}^{3+}(\mathrm{TMHD})_3(\mathrm{phen})]$ and $^{151}\mathrm{Eu}^{3+}:\mathrm{Y}_2\mathrm{SiO}_5$ site 1 are shown in Fig. \ref{fig_EuHAGEMfit}. We also simulated Pr$^{3+}$, Tm$^{3+}$, Er$^{3+}$ and Yb$^{3+}$ materials (SI S2). Some materials are difficult to implement HAGEM with large $\delta$, but with smaller $\delta$, nearly perfect matching conditions can always be found for any RE ion material, and an example is shown in the Extended Data Table. 2. This is reasonable, since if $\delta$ is much smaller compared to the hyperfine level splitting values, there is more mathematical freedom to match with a perfect HAGEM spectrum. From the simulations, we found that $^\text{151}$Eu$^\text{{3+}}$(TMHD)$_\text{3}$(phen) is the only material with direct-matched hyperfine level structures compatible with HAGEM. If the grating distance of the HAGEM spectrum reaches the value of spin inhomogeneous broadening, the HAGEM spectrum will be messed up at some frequency regions. A rough grating structure can still exist, so storage can still work but with a lower efficiency. RE ions in solids can, in principle, be engineered to exhibit a broad optical inhomogeneous absorption profile while maintaining a narrow inhomogeneous distribution of spin transition frequencies by exploiting the different susceptibilities of the optical and spin transitions to static local-field perturbations. Structural strain, crystal-field variations, and static electric-field disorder can produce large site-to-site shifts of the optical transition frequency, whereas an appropriately chosen host and a ZEFOZ spin transition can strongly suppress the corresponding distribution of hyperfine transition frequencies.

\section{Experiments \& analysis}

To validate the above memory protocol, we tend to find a material with perfectly direct-matched hyperfine level structures. Screening all existing materials of rare-earth ions with three hyperfine levels in solids, there is no single one which meets the requirements. We turn to the chemical engineering of molecular rare-earth ions. For host materials with highly symmetric point group, the energy level gaps at the ground states and excited states will have the features of which one gap is double of the other one governed by crystal field theory. The only task to meet HAGEM rules is then to match the ratio between the ground state and the excited state. Low site symmetry is however preferred both for a sizeable $^{5}D_{0} \leftrightarrow {}^{7}F_{0}$ oscillator strength and for the hyperfine state mixing that enables optical pumping. This removes the fixed 1:2 gap ratio of the axial sites, but the gap ratios become then a tunable parameter of the molecular environment. We therefore provide a material, Eu$^\text{{3+}}$(TMHD)$_\text{3}$(phen), with a low point group symmetry of C$_\text{{1}}$ \cite{al2026robust}, while still fulfilling the required mathematical correlation for the hyperfine level separations to implement HAGEM directly.
  
\subsection{HAGEM molecular Eu$^\text{3+}$ material}

The micro-crystalline powders have typical sizes of a few to tens of $\mu$m as shown in Extended Data Fig. \ref{Fig_Material}(b) and the single crystal has a dimension of 1.8 X 1.6 X 0.7~mm$^3$ as shown in Extended Data Fig. \ref{Fig_Material}(c). The optical measurements were performed at a temperature of 1.4~K. The inhomogeneous broadening of the single crystal and powder is measured to be 9.06~GHz and 20.03~GHz, respectively. The ground state lifetime of the single crystal has a fast decay of 12~ms and a slow decay of 24.6~s. The ground state lifetime of the powder has a fast decay of 128~ms and a slow decay of 13.1~s. The optical coherence time of the single crystal and powder are 5.5 $\pm$ 0.4~$\mu$s and 3.3 $\pm$ 0.2~$\mu$s, respectively, measured by two-pulse photon echo. The optical lifetimes of the single crystal and powder are 0.798~ms and 0.808~ms, respectively, measured by photoluminescence. The spectral hole pumping efficiencies of the single crystal and powder are determined to be 86\% and 65\% respectively. All of the results are shown in Extended Data Fig. 1 \& 2.

After a single SHB with laser frequency f$_\text{0}$, there are 6 side holes and 42 anti-holes, as described above. To determine the hyperfine state structures, we used the SHB technique and the spectrum is shown in the Extended Data Fig. \ref{Fig_SHB}(a). After obtaining the SHB spectrum, we first determined the hyperfine level gaps (SI S3.2), then simulated the SHB spectrum to determine the branching ratio values between $^{7}F_{0}$ and $^{5}D_{0}$ (SI S3.3). The results are shown in the Extended Data Fig. \ref{Fig_SHB}(b)\&(c). The branching ratio values will determine the pumping times for spectral tailoring. Larger off-diagonal values (spin-crossed transitions) need less spectral preparation time. In an extreme case, all off-diagonal values are zero, then the atomic distribution cannot be altered by optical spectral hole burning.

\subsection{Broadband quantum storage}

\begin{figure}[!h]
\centerline{\includegraphics[width=1\columnwidth]{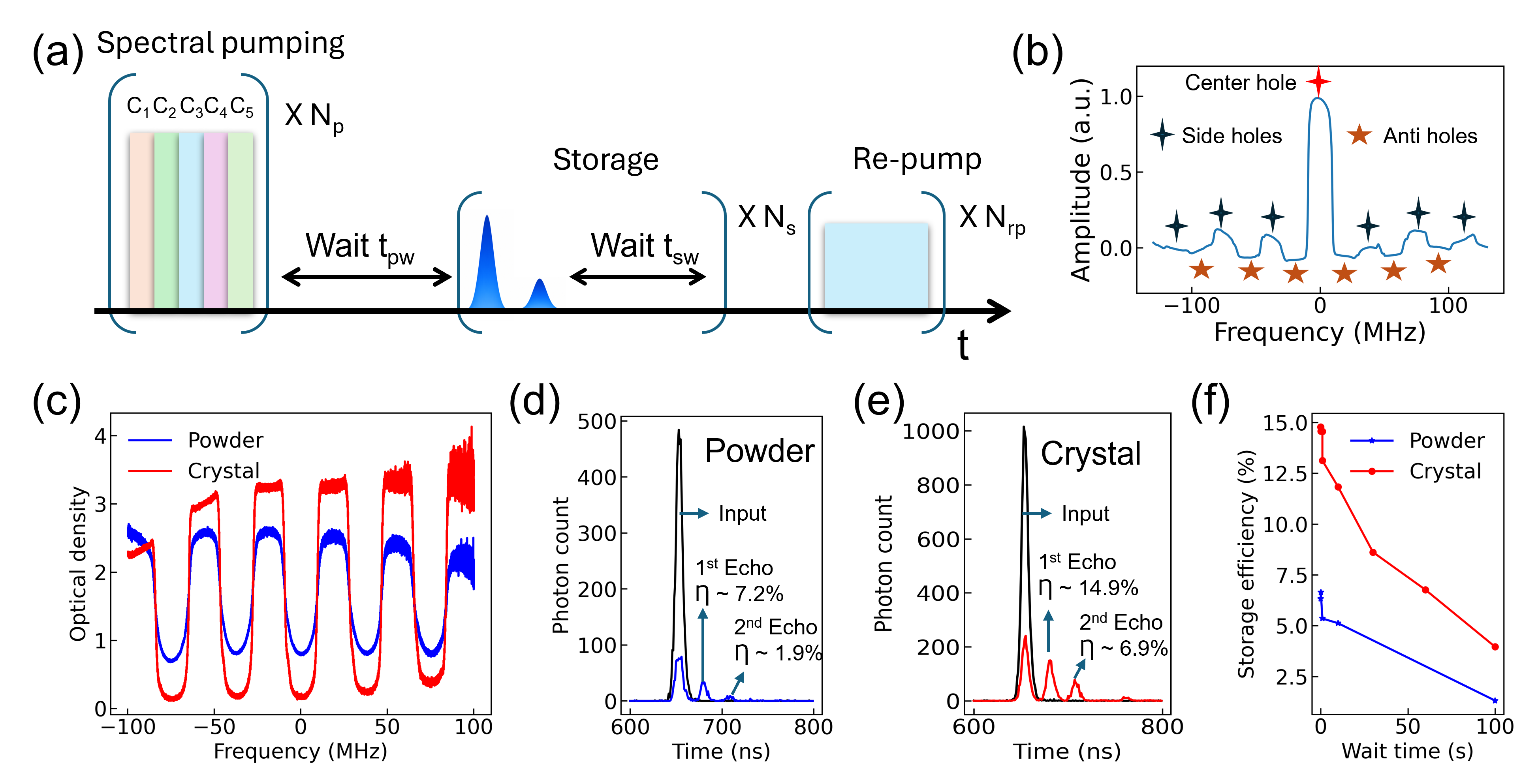}}
	\caption{Single-mode quantum storage: (a) Experimental sequence of quantum storage experiments. (b) Spectral pumping of a single comb. (c) 200~MHz optical density spectrum of powder and crystal after spectral pumping sequence. (d) SPD (single photon detection) measurements of quantum storage for powder. (e) SPD measurements of quantum storage for crystal. (f) Storage efficiencies with different wait times after spectral preparation for powder and crystal.}
	\label{Fig_Storage}
\end{figure}

\textbf{Single-mode storage}: the hyperfine level structure of [$^\text{151}$Eu$^\text{{3+}}$(TMHD)$_\text{3}$(phen)] can be used for broadband storage. The grating distance is set to 37.2~MHz. Five combs are created, and each comb is pumped by laser pulses with a bandwidth of 18.6~MHz. Persistent spectral hole burning is used to prepare the atomic grating spectrum. The storage experiments were carried out at a temperature of 1.4~K. Each comb is pumped by 0.1~ms and repeated for 400 cycles (N$_\text{p}$). After spectral preparation, 10~ms wait (t$_\text{pw}$) is added for the atoms to decay back to the ground states, then bright coherent laser pulses or single-photon level pulses are sent for storage. The input storage pulse is 10~ns, so the corresponding spectral linewidth is 100~MHz, which is to match the spectral grating frequency window of 200~MHz. The single photon detector has a dead time of 45~ns and the input pulse is 10~ns, so the average photon numbers of the input pulses have to be well below one to avoid the detector saturation.  The average photon number of the input pulses is 0.1. Each storage event is 20~$\mu$s (one input storage pulse added with a wait time t$_\text{sw}$ for the storage process.) and is repeated 1000 times (N$_\text{s}$), as shown in Fig. \ref{Fig_Storage}(a) `Storage' part. After storage events are done, spectral re-pump is performed to restore the initial atomic distribution. The experimental sequence was repeated for 100 times for single photon measurements. The experiments were carried out with both powder and crystal. The storage experiments were performed at the frequency window of optical density around 2. Due to insufficient pumping efficiency for powder, the frequency window was selected at optical density $\sim$ 1.6. After spectral preparation, the background absorption is 0.15 for crystal and 0.7 for powder. For $\Delta$ = 37.2~MHz, the finesse is estimated to be $\sim$2 and $d_0$ is estimated to be $\sim$0.15. For powder, the first \& second echo efficiencies are 7.2\% and 1.9\%, respectively. For the crystal, the first \& second echo efficiencies are 14.9\% and 6.9\% respectively. The theoretical storage efficiencies for crystal \& powder are calculated to be 18.9\% and 10.4\% respectively. These agree with our experimental results, considering that any imperfection in the comb shape leads to stronger self-dephasing after photon absorption, and powder scattering will reduce rephasing efficiency, lowering storage efficiency. Storage efficiencies with different waiting times are shown in Fig. \ref{Fig_Storage}(f). Due to the long ground state lifetime, after spectral preparation, the atomic ensemble can be used for efficient storage for a long period, which ensures a high duty cycle of the quantum memory device.

\textbf{Multi-mode storage}: next we try to search for a smaller $\delta$ between 0.4~MHz and 1.5~MHz for HAGEM implementation, and $\delta$ = 0.499~MHz is determined to be a good fit, as shown in Extended Data Fig. \ref{fig_HAGEMEuMMultimode}. The storage spectrum cannot be prepared properly due to the instability of the laser. Next, we used $\delta$ = 2.325~MHz, optimizing the spectral preparation sequence to create a 100~MHz spectrum to store photons of 10~ns. We perform quantum storage of 4 temporal modes, with a storage time of 215~ns and an average storage efficiency of 1.3\%, as shown in Extended Data Fig. \ref{Fig_StorageMultimode}.


\section{Quantum repeater performance}

\begin{table}[htbp]
\centering
\caption{Numerical results for the degenerate $^{167}$Er$^{3+}$:Y$_2$SiO$_5$ model and the non-degenerate Eu$^{3+}$:Y$_2$SiO$_5$ model. Here, $D$ denotes the total communication distance in kilometers, $\eta_m$ is the quantum-memory efficiency, and $P_{\rm elem}$ denotes the elementary-link success probability. $P_{\rm end}$ is the corresponding end-to-end success probability. $R_{\rm rep}$ is the entanglement-distribution rate achieved with the repeater protocol, whereas $R_{\rm dir}$ is the corresponding direct-transmission rate without the repeater, both expressed in hertz.}

\label{tab:combined_results}
\begin{tabular}{c c c c c c c}
\hline
Material & $D$ (km) & $\eta_m$ & $P_{\mathrm{elem}}$ & $P_{\mathrm{end}}$ & $R_{\mathrm{rep}}$ (Hz) & $R_{\mathrm{dir}}$ (Hz) \\
\hline

\multirow{7}{*}{Er$^{3+}$:YSO}
& 80  & 0.204 & $1.532\times10^{-2}$ & $1.609\times10^{-4}$ & $1.609\times10^{4}$ & $8.238\times10^{4}$ \\
& 100 & 0.195 & $1.217\times10^{-2}$ & $9.186\times10^{-5}$ & $9.186\times10^{3}$ & $3.281\times10^{4}$ \\
& 120 & 0.185 & $9.664\times10^{-3}$ & $5.244\times10^{-5}$ & $5.244\times10^{3}$ & $1.306\times10^{4}$ \\
& 140 & 0.176 & $7.677\times10^{-3}$ & $2.994\times10^{-5}$ & $2.994\times10^{3}$ & $5.199\times10^{3}$ \\
& 160 & 0.167 & $6.098\times10^{-3}$ & $1.709\times10^{-5}$ & $1.709\times10^{3}$ & $2.070\times10^{3}$ \\
& 180 & 0.159 & $4.844\times10^{-3}$ & $9.759\times10^{-6}$ & $9.759\times10^{2}$ & $8.238\times10^{2}$ \\
& 200 & 0.151 & $3.848\times10^{-3}$ & $5.572\times10^{-6}$ & $5.572\times10^{2}$ & $3.281\times10^{2}$ \\
\hline

\multirow{7}{*}{Eu$^{3+}$:YSO}
& 80  & 0.134 & $1.532\times10^{-2}$ & $1.037\times10^{-4}$ & $1.037\times10^{5}$ & $8.238\times10^{5}$ \\
& 100 & 0.110 & $1.217\times10^{-2}$ & $4.847\times10^{-5}$ & $4.847\times10^{4}$ & $3.281\times10^{5}$ \\
& 120 & 0.090 & $9.664\times10^{-3}$ & $2.266\times10^{-5}$ & $2.266\times10^{4}$ & $1.306\times10^{5}$ \\
& 140 & 0.074 & $7.677\times10^{-3}$ & $1.059\times10^{-5}$ & $1.059\times10^{4}$ & $5.199\times10^{4}$ \\
& 160 & 0.060 & $6.098\times10^{-3}$ & $4.950\times10^{-6}$ & $4.950\times10^{3}$ & $2.070\times10^{4}$ \\
& 180 & 0.049 & $4.844\times10^{-3}$ & $2.314\times10^{-6}$ & $2.314\times10^{3}$ & $8.238\times10^{3}$ \\
& 200 & 0.040 & $3.848\times10^{-3}$ & $1.081\times10^{-6}$ & $1.081\times10^{3}$ & $3.281\times10^{3}$ \\
\hline
\end{tabular}
\end{table}

We compare two quantum-repeater architectures based on quantum memories using rare-earth (RE) ions doped in YSO crystals (SI S4), and the simulation parameters are readily achievable with current technologies, listed in the Extended Data Table. 3.  In the degenerate configuration, where the two photons of the entangled pair have the same frequency, the repeater is modeled using \(^{167}\mathrm{Er}^{3+}:\mathrm{YSO}\) memories in a two-elementary-link geometry with Bell sate measurement at the midpoint, giving a source probability \(2p(1-p)\). 
After a successful single-photon heralding measurement, entanglement swapping is performed and the two outer memories are projected onto the single-excitation entangled state \((|01\rangle \pm |10\rangle)/\sqrt{2}\). In the non-degenerate configuration, where the photons of the entangled pair have different frequencies, the repeater is modeled using \(\mathrm{Eu}^{3+}:\mathrm{YSO}\) memories with midpoint heralding. Both simulations include fiber attenuation, finite memory lifetime, source repetition rate, temporal storage-mode capacity, detector efficiency, and Bell-state measurement success probability. On-demand quantum memory devices are necessary for multiplexing. For RE ions, this can be achieved by integrating with multi-mode variable fiber delay lines \cite{kaneda2019high, pang2020hybrid, hong2025multimode, crespo2026compact}. The temporal modes are separated by sub-ns to a few ns and the success rates of retrieving one photon from the quantum memory at the elementary links are limited to be the order of $10^{-3}$, so efficient variable fiber delay lines with storage times varying from sub-ns to tens of ns with fixed gaps and a multi-mode capacity of $10^2$ - $10^3$ can greatly enhance the entanglement swapping success rates, which are easy to develop since the storage times are very short.
In both architectures, increasing the communication distance reduces fiber transmission and memory efficiency, resulting in a decrease in the end-to-end entanglement-distribution rate. 
The calculated repeater rates are compared with the corresponding direct-transmission rates under the same optical-loss assumptions.

For some applications such as quantum key distribution, active feed-forward for the last level of entanglement swapping can be replaced by post-selection with the cost of wasting a fraction of the generated entanglement, so the entanglement distribution distance can be doubled with the same storage time. For HAGEM, if $\delta_1$ works well, $\delta_2 = \delta_1 / k$ will also work well, where $k$ is a small odd integer. In addition, we can search for $\delta$ between $\delta_1$ and $\delta_1 / k$, which may also work well for HAGEM. This will give us enough freedom to design longer storage times for the quantum memory devices at later stages of entanglement swapping and the two end nodes, giving enough time for earlier stages of entanglement swapping to do active feed-forward. Multiple stages of entanglement swapping will reduce effective storage modes.

For \(^{167}\mathrm{Er}^{3+}:\mathrm{YSO}\) quantum memory, the entanglement distribution rate of its quantum repeater network can exceed direct transmission with a communication distance of around 200~km, achieving the supremacy of quantum repeaters. With a longer optical coherence time, a higher maximum storage efficiency, and a better variable delay device, the communication distance to achieve quantum repeater supremacy will be shorter.

\section{Discussions}
The most distinct feature between HAGEM and AFC is whether some of the atoms are pumped away from the storage frequency window. The main advantages of the HAGEM over AFC or any other quantum storage protocol are the large storage bandwidth leading to enormous amounts of temporal modes and simple spectral preparation, since the AFC storage bandwidth is typically limited by atomic state level separations and requires complicated spectral tailoring, such as class cleaning, spin polarization, back burn, etc., but suffering from a lower storage efficiency. In addition, for AFC, to prepare the same storage spectrum as HAGEM, the initial OD is multiple times compared to that of HAGEM, so the background absorption will be stronger when storage is performed after some periods of spectral preparation, due to limited atomic ground state lifetime. In result, HAGEM will give a higher duty cycle during a continuous quantum repeater operation together with a shorter spectral preparation time. It can be implemented by most RE ion materials with a reasonable optical coherence time (for example, $\ge 1 \mu s$), not restricted by its hyperfine level structures, which shows the robustness of the HAGEM. Its storage bandwidth is only limited by the inhomogeneous broadening of the material, which can be chemically engineered to be broad; therefore, in theory, the storage bandwidth can be infinite times of AFC for a continuous storage spectrum with many materials. In practice, a few GHz storage bandwidth can be achieved easily, which is 10 to 1000 times higher compared to the storage bandwidth achieved with AFC with a reasonably long storage time for any type of RE ions, which shows the powerfulness of HAGEM. For example, a 3~GHz bandwidth can be created with grating distance 1~kHz by using the spectral preparation techniques with high spectral resolution \cite{jobez2016towards, businger2022non}, so it can store sub-ns photons with a storage time of 1~ms, in total, one million temporal modes of photons can be stored. Molecular engineering can tune RE ions transition wavelengths and inhomogeneous broadening, so spectral multiplexing can be achieved, for example, 10 frequency windows \cite{seri2019quantum}. Without frequency multiplexing, spatial multiplexing of 10 can be easily achieved \cite{yang2018multiplexed}. In summary, more than one billion modes per second of photon storage can be achieved by combining temporal \& frequency multiplexing or temporal \& spatial multiplexing by using HAGEM with current available technologies. Furthermore, during quantum repeater operation, the photon arrival probability is very low, so the temporal modes can be quite close to each other. Placing the crystal inside an impedance-matched cavity can increase the storage efficiency to a maximum of 40.5\%, which is mainly limited by dephasing, due to a low finesse value of 2 \cite{afzelius2010impedance}. It is noteworthy that for a quantum repeater with a communication distance of a few hundred kilometers, the requirement for quantum memory efficiency can be lower. The memory bandwidth will not be reduced by spectral dispersion in a cavity since the overall atomic density does not change after spectral tailoring, so the theoretical maximum storage can be reached together with a large storage bandwidth; whereas, spectral dispersion will greatly reduce the storage bandwidth with the AFC memory protocol in a cavity, so achieving a few MHz storage bandwidth will limit the storage efficiency to around 80\%. The trade-off for spin-wave storage also exists.

 For Kramer RE ions, multiplexing can be achieved by spin-wave storage as well, which has not been demonstrated for large memory bandwidth so far. Multiplexing for quantum-memory-based quantum repeaters is essential to achieve a high communication rate and a long communication distance \cite{collins2007multiplexed}. By improving the performance of spin-wave storage, variable fiber delay lines, or variable free-space cavity delays, communication distance can be further extended by more than one repeater station. Further improvements with RE ion materials can increase the optical coherence time and reduce spin inhomogeneous broadening, which will enable longer storage time and higher storage efficiency. For quantum memory devices with fixed storage times, in addition to temporal multiplexing with tunable delay lines as shown above, frequency multiplexing together with adjustable frequency shifts \& spectral filters can also be used for quantum repeater operations \cite{sinclair2014spectral}.

The limited multi-mode storage times and efficiencies reported in the experiments above are mainly caused by the laser instability and the short coherence time of the molecular material, which can be further improved by better laser frequency locking and some careful material engineering. The molecular techniques presented in this article can be used to engineer proper host molecules for various types of RE ions. Er$^\text{{3+}}$ ions in solids exhibit large spectral diffusion, which reduces photon indistinguishability and spin-spin entanglement (generated by photon interference) fidelity \cite{ourari2023indistinguishable}. Er$^\text{{3+}}$ ions hosted by molecules can be explored to mitigate this issue. Host molecules for Tm$^\text{3+}$ and Er$^\text{3+}$ ions can be explored in the future to find one with transition wavelengths that match those of rubidium atoms, which can be used to build hybrid quantum networks \cite{gu2025hybrid, chai2026hybrid}.


\renewcommand{\refname}{References}
\putbib

\section{Methods}

\subsection{Preparation of the microcrystalline powder and bulk crystalline forms of the $^\text{151}$Eu$^\text{{3+}}$ molecular complex}

The compound studied in this work is the neutral mononuclear europium(III) complex [${}^{151}$Eu(TMHD)$_3$(phen)], where TMHD denotes 2,2,6,6-tetramethyl-3,5-heptanedionate and phen denotes 1,10-phenanthroline. The synthesis followed the established procedure for natural-abundance [Eu(TMHD)$_3$(phen)]~\cite{al2026robust}, with the natural europium source replaced by the isotopically purified ${}^{151}$Eu precursor. The ${}^{151}$EuCl$_3\cdot$6H$_2$O precursor was prepared from the commercially available (BuyIsotope) ${}^{151}$Eu$_2$O$_6$ (about 99.2\% enriched) as described before \cite{serrano2022ultra}. Yield of the enriched complex: 578 mg (66\%). 
Single crystals were grown from approximately 30 mg of the isotopically purified powder dissolved in a mixed methanol/dichloromethane solvent. The solution was evaporated at \(60\,^\circ\mathrm{C}\) until the first precipitate appeared, then transferred without filtration into a smaller sealed vial and left undisturbed at \(60\,^\circ\mathrm{C}\) for two days. Slow evaporation through the imperfectly sealed cap promoted crystal growth on the vial walls and within the solution. Elevated temperature simultaneously drove dissolution of the finer crystallites and growth of the larger ones, consistent with Ostwald ripening.

\subsection{Cryogenic temperature optical spectroscopic setup}
Optical absorption, photoluminescence (PL) decays, PE and SHB measurements were performed in a He bath cryostat (Janis SVT-200) at a temperature of $1.4\,\text{K}$ under resonant excitation of the $^{5}D_{0} \leftrightarrow {}^{7}F_{0}$ transition of Eu$^{3+}$. The excitation source was a tunable continuous-wave dye laser (Sirah Matisse DS) with $\approx 300\,\text{kHz}$ linewidth. The temperature of the sample holder was monitored with a Si diode (Lakeshore DT-670). Pulse sequences were created using an acousto-optic modulator (AA Optoelectronic MT200-B100A0, 5-VIS, 200\,MHz central frequency), in a double-pass configuration, driven by an arbitrary waveform generator (Agilent N8242A) with a sampling rate of  625\,megasamples per second.
The crystalline powders were placed in a home-built sample holder made of brass. This holder is composed of seven individual sample containers that have front and rear optical access through glass windows (SI S3.4). Each container was filled with $\approx 5\,\text{mg}$ of powder, forming slabs of thickness about $500\,\mu\text{m}$. The excitation beam was focused on the container's front window by a $75\,\text{mm}$ focal-length lens placed in front of the cryostat window. Light scattered by the powders and transmitted through the back window was re-focused by a lens with a large numerical aperture sitting outside the cryostat. Signals were detected with an avalanche photodiode detectors (Hamamatsu C5460 with a bandwidth of  $10\,\text{MHz}$ for SHB, and Thorlabs 110 A/M with a bandwidth of $50\,\text{MHz}$ for all other optical measurements).
PL measurements at $15\,\text{K}$ were carried out for the isotopically purified $^{151}$Eu$^{3+}$ complex using a closed-cycle cold-finger cryostat. The $^{5}D_{0}$ level was resonantly excited at $580.4\,\text{nm}$ with a tunable optical parametric oscillator pumped by a Nd$^{3+}$:YAG Q-switched laser (Ekspla NT342BSH, $6\,\text{ns}$ pulse length and $10\,\text{Hz}$ repetition rate). Spectra were recorded using an Acton SP2300 spectrometer equipped with a holographic grating with $1{,}200$ grooves per mm and an ICCD camera (Princeton Instruments). PL decays were recorded at a temperature of $1.4\,\text{K}$ for the $^{151}$Eu$^{3+}$ complex (Extended Data Fig.~1d) after a single pulse ($1\,\text{ms}$ long), under resonant excitation of the $^{5}D_{0}$ level, yielding comparable decay curves and population lifetimes ($T_{1,\mathrm{opt}} \approx 800\,\mu\text{s}$) for all samples.

\subsection{Optical measurements}
Echo amplitudes from two-pulse PE experiments were measured through the fast Fourier transform of the beating signal due to interference between the PE signal and a frequency-detuned laser pulse (heterodyne pulse, with a frequency detuning of $30\,\text{MHz}$). During measurements, the laser wavelength was scanned over $500\,\text{MHz}$ in $1\,\text{s}$ to prevent echo signal loss due to SHB, in which atoms were pumped away from the laser absorption window. The length and intensity of the $\pi/2$ and $\pi$ pulses were optimized to obtain maximum PE amplitudes, with typical pulse lengths between $1$ and $2\,\mu\text{s}$, and a laser power before the cryostat of $200\,\text{mW}$. A band-pass filter was placed in front of the avalanche photodiode detector to block the strong PL emission with decays into other J-levels in the Eu$^{3+}$ complexes, which have shorter wavelengths compared to the optical transitions between $^{7}F_{0}$ and $^{5}D_{0}$. The echo amplitude obtained by fast Fourier transformation was averaged over $50$ experimental sequences to improve signal-to-noise ratio. For single exponential echo decays, the optical coherence lifetime ($T_{2}$) was directly derived as $\exp(-2\tau/T_{2})$, where the factor of 2 accounts for the transformation from echo amplitude to intensity. SHB spectra from isotopically ($^{151}$Eu$^{3+}$) enriched complex in powder and crystal were obtained by applying a single pulse, referred as the burn pulse, of $10\,\text{ms}$ length and excitation power of $5\,\text{mW}$. After a waiting time of $10\,\text{ms}$, the spectrum was read out with a $1\,\text{ms}$-long scanning pulse with $5\,\text{mW}$ power and $200\,\text{MHz}$ scanning range around the burning frequency (Fig. \ref{Fig_SHB}(a)). 

The wait time after spectral pumping \& before readout was set to $10\,\text{ms}$ (that is, $\gg T_{1,\mathrm{opt}}$) to enable spontaneous emission from the optical excited state, followed by progressive trapping into non-pumped nuclear spin levels. The SHB spectra were corrected for the frequency-dependent response of the acousto-optic modulator by dividing by a readout signal obtained with the burn laser pulse turned off. A series of high-power pulses ranging over $200\,\text{MHz}$ were applied at the end of the sequence to reset the ground-state population back to their initial atomic distribution. The population lifetime of the ground-state spin levels ($T_{1,\mathrm{spin}}$) was determined by monitoring the hole depth as a function of wait time between the burn and readout pulses. The hole decay curve presents two distinct decay rates, estimated by fitting a double exponential decay to the experimental data.

\subsection{HAGEM storage}
The microwave signals to control AOMs are generated by an arbitrary waveform generator and the experimental sequence is synchronized by its marker signals. An electronically controlled shutter (Thorlabs, SH05R/M \& controller KSC101) is used to block laser leakage during storage periods. A single photon detector (Laser Components, COUNT-10C) detects single photons during quantum storage experiments and a Time Tagger device (Swabian Instruments, Time Tagger 20) is deployed to analyze the output voltages from the SPD and record photon counts and their arrival times. A tunable beam blocker is placed in the storage pulse path to control the intensity of the storage pulses. 

\section{Acknowledgments}
This work received government funding managed by the French National Research Agency under grants ANR-20-CE09-0022-01 (UltraNanoSpec) and ANR-23-CE47-0011 (MoleQuBe) as well as under France 2030, reference ANR-23-PETQ-0007 (MolQIF). S.K.K. and M.R. thank the Helmholtz Association for support through the programs Natural, Artificial, and Cognitive Information Processing (NACIP) and Materials Systems Engineering (MSE). S.K.K. and M.R. also thank Prof. Dieter Fenske for his continuing support with SC-XRD studies. S.K.K., M.R. and O.F. thank Karlsruhe Nano Micro Facility (KNMFi) for providing the analytical infrastructure. Discussions with Mikael Afzelius are very helpful. 

\section{Author Contributions}
Y.L. proposed the memory protocol \& analyzed its quantum repeater performances, initiated \& performed the experiments, analyzed the data and carried out the theoretical modeling of the memory protocol \& experiments. S.K.K. proposed and synthesized the molecular complex. I.T. and S.K.K. contributed to the production of millimeter-sized crystals and the recrystallization of the powders. D.S. and S.A.H. assisted with the experiments and data analysis. P.J. assisted with the HAGEM material search simulations. O.F. performed the SC-XRD studies. P.G., M.R. and D.S. advised of all efforts. Y.L. and D.S. wrote the manuscript. All authors contributed to the preparation of the manuscript.

\section{Disclosures}
The authors Y.L. and D.S. declare that a patent application is being prepared for the memory protocol HAGEM described in this manuscript. The authors have no other competing interests to declare.

\section{Data availability}
Data underlying the results presented in this paper are not publicly available at this time but may be obtained from the authors upon reasonable request.

\clearpage
\section*{Extended Data}

\setcounter{figure}{0}
\renewcommand{\thefigure}{\arabic{figure}}
\renewcommand{\figurename}{Extended Data Fig.}
\setcounter{table}{0}
\renewcommand{\thetable}{\arabic{table}}
\renewcommand{\tablename}{Extended Data Table.}

\begin{table}[htbp]
\centering
\label{Eu17materials}
\caption{Optimized Eu$^{3+}$ material fits. The molecular complexes are placed at the end of the table. The asterisk marks the overlap-aware $[{}^{151}\mathrm{Eu}^{3+}(\mathrm{TMHD})_3(\mathrm{phen})]$ fit. C$_{\text{norm}}$ represents the mismatch degree between a perfect HAGEM spectrum with a grating distance of $2 \delta$ and the spectrum created by a single SHB with the simulated material. A lower value of C$_{\text{norm}}$ leads to a better HAGEM implementation.}
\label{tab:eu-summary}
\scriptsize
\renewcommand{\arraystretch}{1.10}
\begin{tabular*}{0.98\textwidth}{@{\extracolsep{\fill}}cccccc}
\toprule
Material & $\delta$ (MHz) & $2\delta$ (MHz) & $T_{\rm echo}$ (ns) & $C_{\rm norm}$ & $C_{\rm abs}$ (MHz) \\
\midrule
$^{151}\mathrm{Eu}^{3+}\!:\mathrm{CaF}_2$ ($C_{4v}$ site) & 1.630625 & 3.261250 & 306.6 & 0.124952 & 0.203750 \\
$^{151}\mathrm{Eu}^{3+}\!:\mathrm{CaF}_2$ ($C_{3v}$ site) & 5.560000 & 11.120000 & 89.9 & 0.053657 & 0.298333 \\
$^{153}\mathrm{Eu}^{3+}\!:\mathrm{CaF}_2$ ($C_{4v}$ site) & 2.553846 & 5.107692 & 195.8 & 0.337349 & 0.861538 \\
$^{153}\mathrm{Eu}^{3+}\!:\mathrm{CaF}_2$ ($C_{3v}$ site) & 5.683333 & 11.366667 & 88.0 & 0.033431 & 0.190000 \\
$^{151}\mathrm{Eu}^{3+}\!:\mathrm{CaWO}_4$ & 2.163333 & 4.326667 & 231.1 & 0.379277 & 0.820503 \\
$^{153}\mathrm{Eu}^{3+}\!:\mathrm{CaWO}_4$ & 5.525000 & 11.050000 & 90.5 & 0.343784 & 1.899405 \\
$^{151}\mathrm{Eu}^{3+}\!:\mathrm{EuCl}_3\!\cdot\!6\mathrm{H}_2\mathrm{O}$ & 2.417692 & 4.835385 & 206.8 & 0.425158 & 1.027901 \\
$^{153}\mathrm{Eu}^{3+}\!:\mathrm{EuCl}_3\!\cdot\!6\mathrm{H}_2\mathrm{O}$ & 2.626596 & 5.253191 & 190.4 & 0.358827 & 0.942492 \\
$^{151}\mathrm{Eu}^{3+}\!:\mathrm{Y}_2\mathrm{O}_3$ ($C_2$ site) & 2.625000 & 5.250000 & 190.5 & 0.356792 & 0.936579 \\
$^{153}\mathrm{Eu}^{3+}\!:\mathrm{Y}_2\mathrm{O}_3$ ($C_2$ site) & 4.892637 & 9.785275 & 102.2 & 0.271155 & 1.326662 \\
$^{151}\mathrm{Eu}^{3+}\!:\mathrm{Y}_2\mathrm{SiO}_5$ (site 1) & 2.681818 & 5.363636 & 186.4 & 0.168362 & 0.451515 \\
$^{151}\mathrm{Eu}^{3+}\!:\mathrm{Y}_2\mathrm{SiO}_5$ (site 2) & 2.227368 & 4.454737 & 224.5 & 0.202426 & 0.450877 \\
$^{153}\mathrm{Eu}^{3+}\!:\mathrm{Y}_2\mathrm{SiO}_5$ (site 1) & 2.203846 & 4.407692 & 226.9 & 0.128812 & 0.283883 \\
$^{153}\mathrm{Eu}^{3+}\!:\mathrm{Y}_2\mathrm{SiO}_5$ (site 2) & 3.806931 & 7.613861 & 131.3 & 0.079150 & 0.301320 \\
$[{}^{151}\mathrm{Eu}^{3+}(\mathrm{BA})_4](\mathrm{pip})]$ & 2.335000 & 4.670000 & 214.1 & 0.411135 & 0.960000 \\
$[{}^{153}\mathrm{Eu}^{3+}(\mathrm{BA})_4](\mathrm{pip})]$ & 3.473585 & 6.947170 & 143.9 & 0.193994 & 0.673854 \\
$[{}^{151}\mathrm{Eu}^{3+}(\mathrm{TMHD})_3(\mathrm{phen})]$$^{*}$ & 18.555556 & 37.111111 & 26.9 & 0.099132 & 1.839444 \\
\bottomrule
\end{tabular*}
\end{table}

\newpage

\begin{table}[htbp]
\centering
\caption{Independent $\delta$-window fits for $^{151}\mathrm{Eu}^{3+}:\mathrm{Y}_2\mathrm{SiO}_5$ (site 1). $(m_{e,2},m_{e,1})$ \& $(m_{g,2},m_{g,1})$ are multipliers of $\delta$ to match the hyperfine level separations. C$_{\text{norm}}$ represents the mismatch degree between a perfect HAGEM spectrum with a grating distance of $2 \delta$ and the spectrum created by a single SHB with the simulated material. A lower value of C$_{\text{norm}}$ leads to a better HAGEM implementation.}
\label{tab:eu-yso-windows}
\scriptsize
\renewcommand{\arraystretch}{1.10}
\begin{tabular*}{\textwidth}{@{\extracolsep{\fill}}cccccccc}
\toprule
$\delta$ window (MHz) & $\delta_{\rm opt}$ (MHz) & $(m_{e,2},m_{e,1})$ & $(m_{g,2},m_{g,1})$ & $T_{\rm echo}$ (ns) & $C_{\rm norm}$ & $C_{\rm abs}$ (MHz) & Boundary optimum \\
\midrule
$[2, 20]$ & 2.681818 & $(38, 28)$ & $(17, 13)$ & 186.4 & 0.168362 & 0.451515 & False \\
$[1, 2]$ & 1.500000 & $(68, 50)$ & $(31, 23)$ & 333.3 & 0.133333 & 0.200000 & False \\
$[0.5, 1]$ & 0.797297 & $(128, 94)$ & $(58, 43)$ & 627.1 & 0.217272 & 0.173230 & False \\
$[0.2, 0.5]$ & 0.300000 & $(340, 250)$ & $(154, 115)$ & 1666.7 & $<10^{-12}$ & $<10^{-12}$ & False \\
$[0.1, 0.2]$ & 0.100000 & $(1020, 750)$ & $(462, 345)$ & 5000.0 & $<10^{-12}$ & $<10^{-12}$ & True \\
\bottomrule
\end{tabular*}
\end{table}

\newpage

\begin{figure}[!h]
\centerline{\includegraphics[width=1\columnwidth]{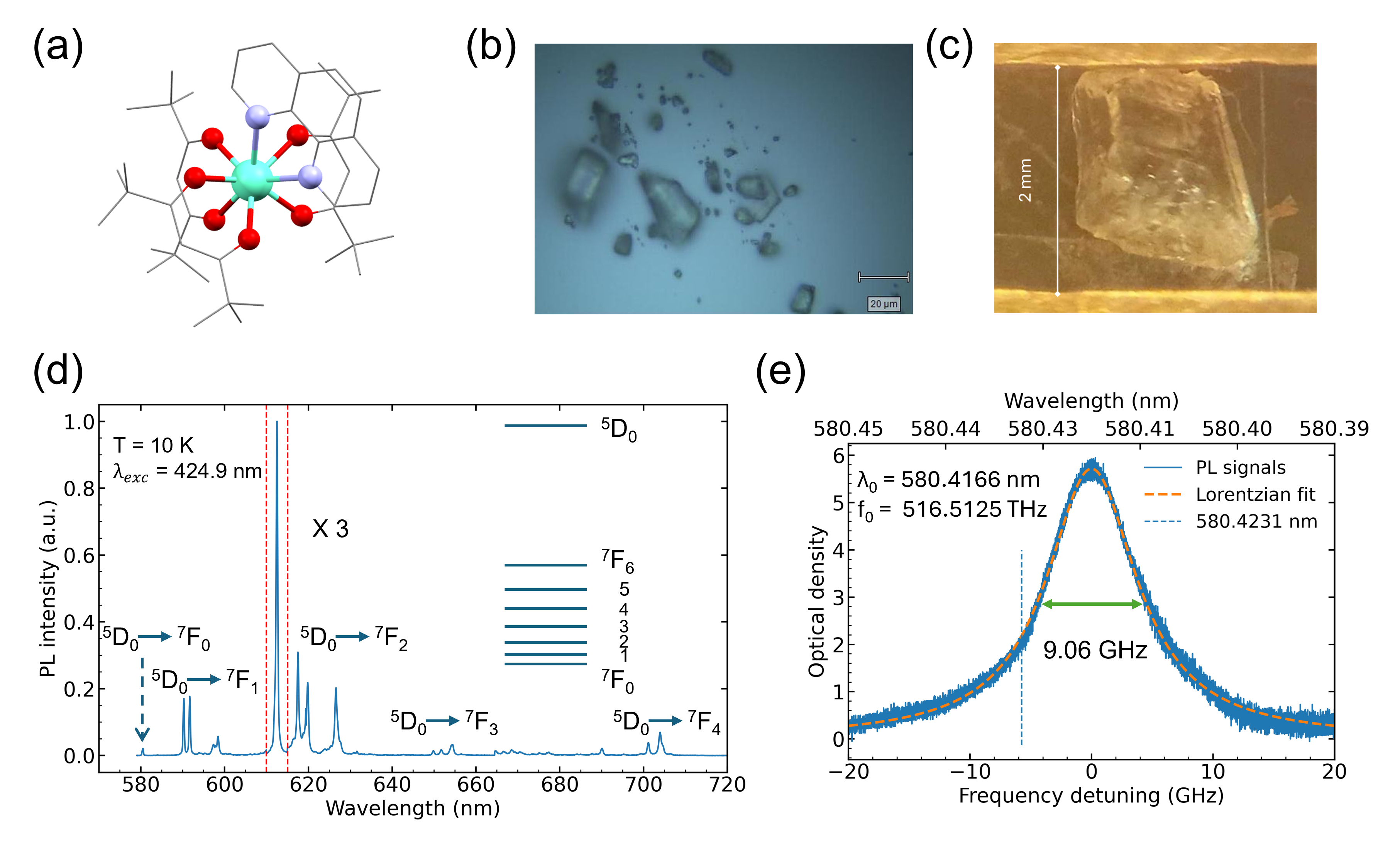}}
	\caption{Molecular material [Eu$^\text{{3+}}$(TMHD)$_\text{3}$(phen)]: (a) Illustration of the chemical composition: oxygen (red), europium (green) and carbon (light purple). (b) Microscope image of the micro-crystalline powder. (c) Microscope image of the millimeter-size single crystal. (d) Photoluminescence spectrum of the single crystal. The actual intensity of the region inside the two red dashed lines is three times the plotted intensity. (e) Inhomogeneous broadening of the single crystal. The vertical line indicates the wavelength at which the quantum storage experiments are performed.}
	\label{Fig_Material}
\end{figure}

\newpage

\begin{figure}[!h]
\centerline{\includegraphics[width=1\columnwidth]{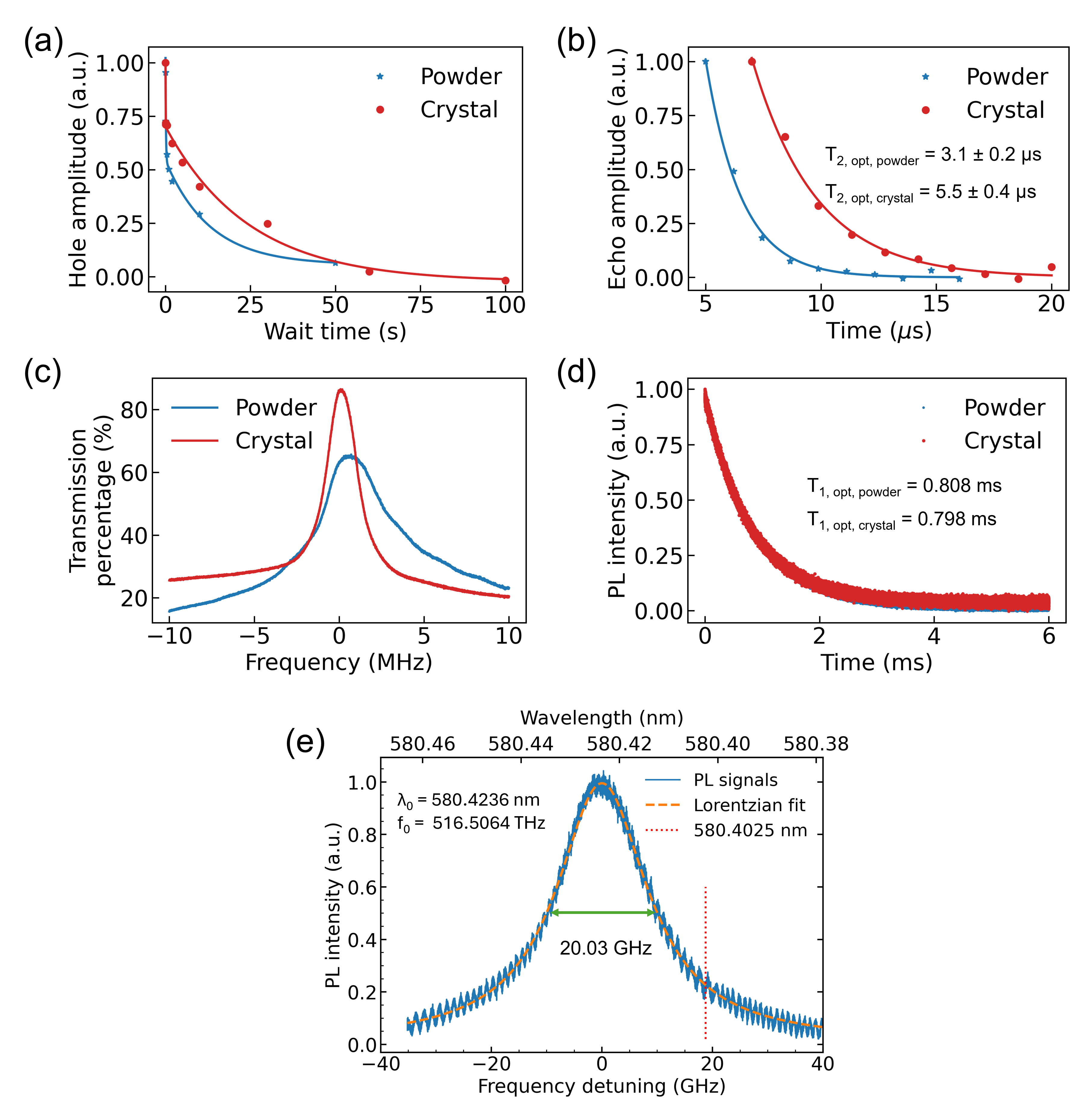}}
	\caption{Optical properties of [Eu$^\text{{3+}}$(TMHD)$_\text{3}$(phen)]: (a) Ground state lifetime measurements by spectral hole burning for powder and crystal. (b) Optical coherence time measurements by two-pulse photon echo for powder and crystal. (c) pumping efficiency of a single spectral hole for powder and crystal. (d) Optical lifetime measurement by optical fluorescence for powder and crystal. (e) Inhomogeneous broadening of the powder. The vertical line indicates the wavelength at which the quantum storage experiments are performed.}
	\label{Fig_OpticalPropertyM1}
\end{figure}

\newpage

\begin{figure}[!h]
\centerline{\includegraphics[width=1\columnwidth]{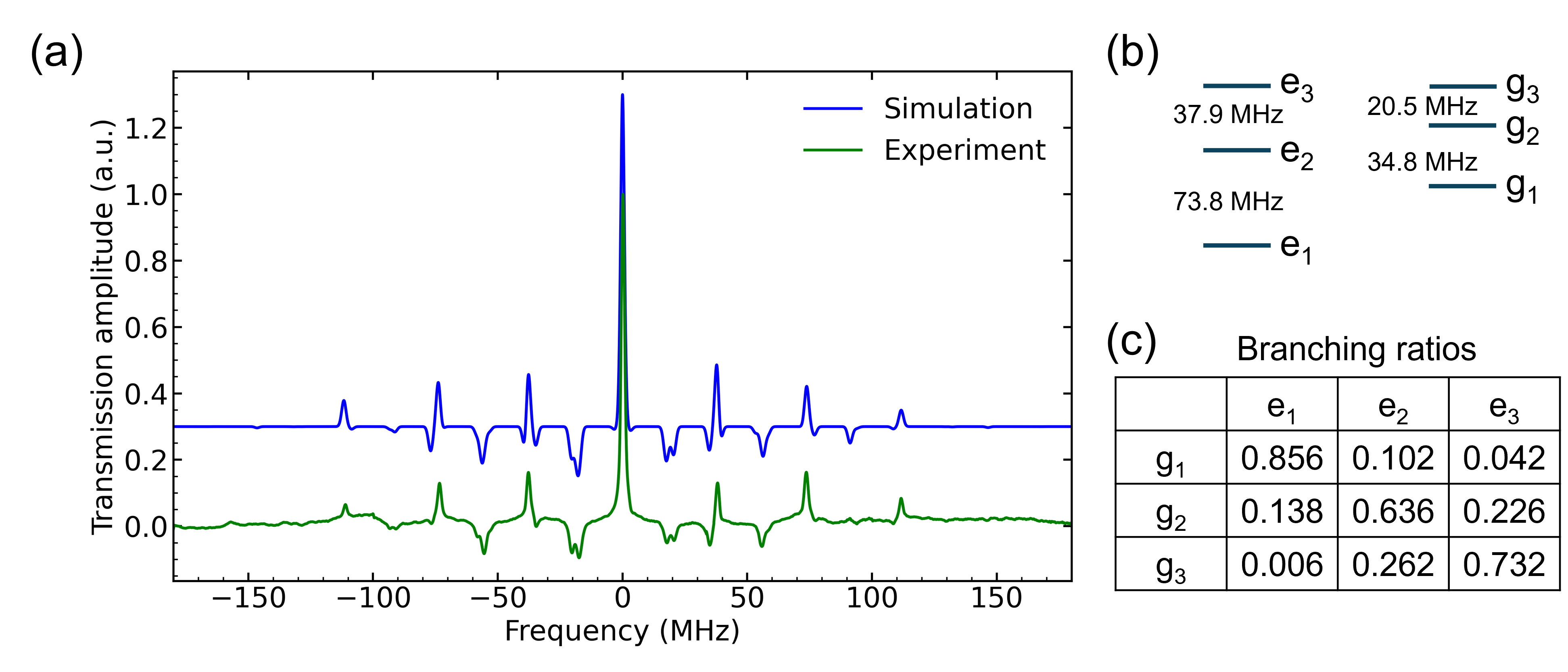}}
	\caption{SHB spectrum simulation for $[\mathrm{Eu}^{3+}(\mathrm{TMHD})_3(\mathrm{phen})]$: (a) A single spectral hole burning: experimental and simulated spectrum of 360~MHz. (b) Hyperfine level separation values. (c) Branching ratio values derived from the spectral simulation. }
	\label{Fig_SHB}
\end{figure}

\newpage

\begin{figure}[!h]
\centering
\includegraphics[width=0.96\textwidth]{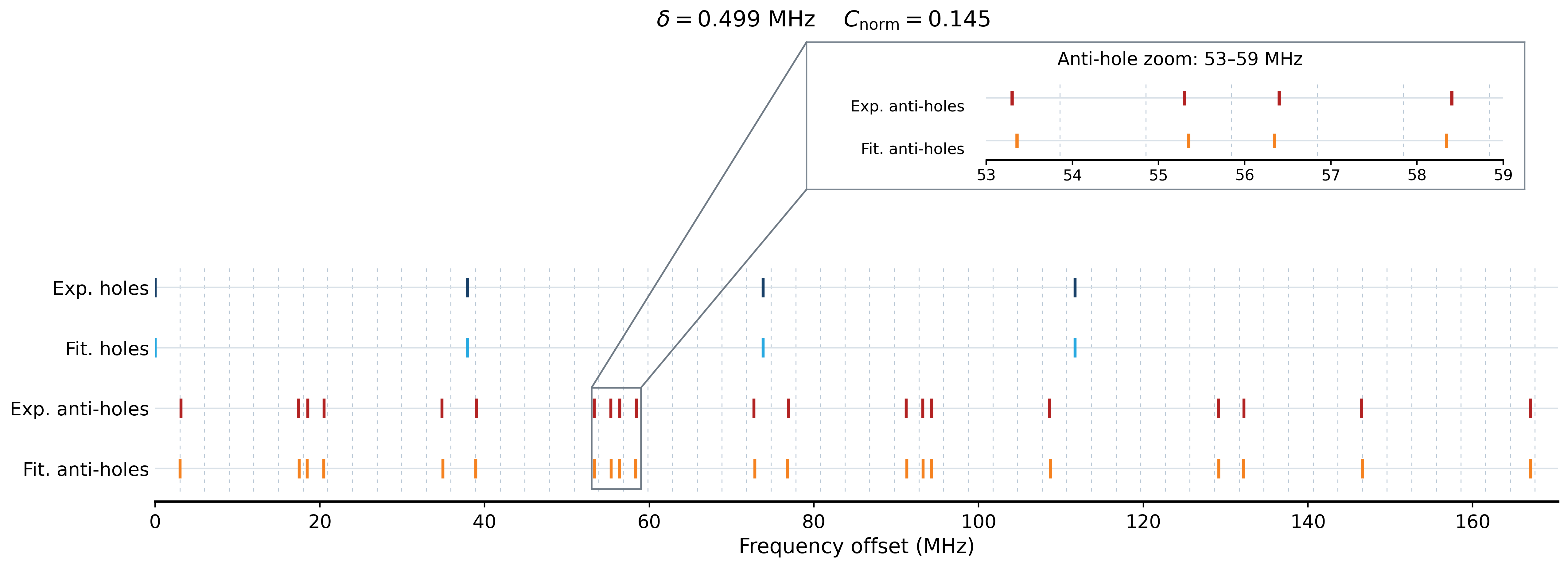}
\caption{Positive-half-spectrum comparison for $[\mathrm{Eu}^{3+}(\mathrm{TMHD})_3(\mathrm{phen})]$ using the $0.4$--$1.5$~MHz search interval for $\delta$. Dashed vertical guides in the main panel show a regularly subsampled $2n\delta$ grid for legibility. The inset figure is the zoom-in $53$--$59$~MHz frequency region showing the experimental and fitted anti-hole positions, with $2\delta = 0.998$~MHz gird dashed lines.}
\label{fig_HAGEMEuMMultimode}

\end{figure}

\newpage

\begin{figure}[!h]
\centerline{\includegraphics[width=1.0\columnwidth]{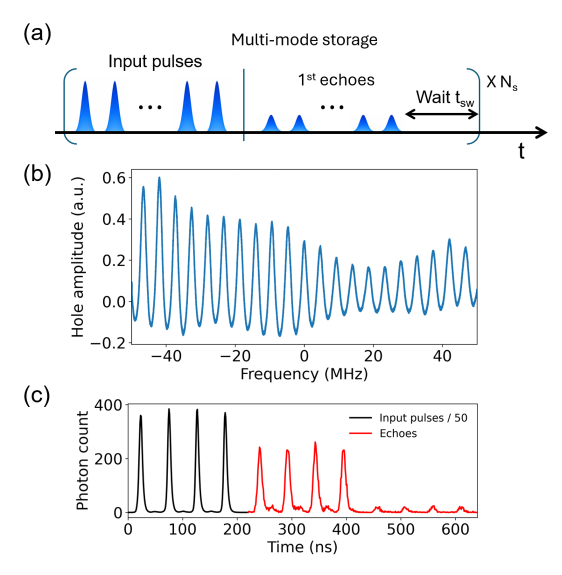}}
	\caption{Multi-mode quantum storage: (a) Experimental sequence of multi-mode quantum storage process. (b) 100~MHz spectrum after spectral preparation with $\Delta_{HAGEM}$ = 4.65~MHz. (c) SPD measurements of 4 temporal-mode quantum storage for crystal.}
	\label{Fig_StorageMultimode}
\end{figure}

\newpage

\begin{table}[t]
    \centering
    \caption{Simulation parameters used for the degenerate 
    $^{167}$Er$^{3+}$:Y$_2$SiO$_5$ two-elementary-link scheme and the 
    non-degenerate Eu$^{3+}$:Y$_2$SiO$_5$ midpoint-heralding scheme.}
    \label{tab:simulation_parameters}
    \begin{tabular}{lccc}
        \toprule
        Parameter & Symbol &
        $^{167}$Er$^{3+}$:Y$_2$SiO$_5$ &
        Eu$^{3+}$:Y$_2$SiO$_5$ \\
        \midrule

        Pair-generation probability
            & \(p\) & 0.050 & 0.050 \\

        Source-attempt rate
            & \(R_{\mathrm{source}}\)
            & \(1.000\times10^{8}\,\mathrm{s}^{-1}\)
            & \(1.000\times10^{9}\,\mathrm{s}^{-1}\) \\

        Storage-mode rate
            & \(R_{\mathrm{storage}}\)
            & \(1.000\times10^{8}\,\mathrm{s}^{-1}\)
            & \(1.000\times10^{9}\,\mathrm{s}^{-1}\) \\

        Maximum memory efficiency
            & \(\eta_{m,\max}\)
            & 0.250
            & 0.300 \\

        Memory lifetime
            & \(\tau_m\)
            & \(4.000\,\mathrm{ms}\)
            & \(2.000\,\mathrm{ms}\) \\

        Fiber attenuation
            & \(\alpha\)
            & \(0.200\,\mathrm{dB/km}\)
            & \(0.200\,\mathrm{dB/km}\) \\

        Fiber group velocity
            & \(v_{\mathrm{fiber}}\)
            & \(2.000\times10^{8}\,\mathrm{m/s}\)
            & \(2.000\times10^{8}\,\mathrm{m/s}\) \\

        Node survival probability
            & \(\eta_{\mathrm{node}}\)
            & 0.900
            & 0.900 \\

        Detector efficiency
            & \(\eta_d\)
            & 0.900
            & 0.900 \\

        Intrinsic BSM success
            & \(P_{\mathrm{BSM}}\)
            & 0.500
            & 0.500 \\

        Swapping-stage duration
            & \(t_{\mathrm{stage}}\)
            & \(1.000\,\mu\mathrm{s}\)
            & \(1.000\,\mu\mathrm{s}\) \\

        Variable-delay modes
            & \(N_v\)
            & 100
            & 100 \\

        Variable-delay efficiency
            & \(\eta_v\)
            & 0.500
            & 0.500 \\

        \bottomrule
    \end{tabular}
\end{table}

\end{bibunit}

\clearpage

\setcounter{section}{0}
\renewcommand{\thesection}{S\arabic{section}}

\setcounter{subsection}{0}
\renewcommand{\thesubsection}{S\arabic{section}.\arabic{subsection}}

\setcounter{figure}{0}
\renewcommand{\thefigure}{S\arabic{figure}}

\setcounter{table}{0}
\renewcommand{\thetable}{S\arabic{table}}

\setcounter{equation}{0}
\renewcommand{\theequation}{S\arabic{equation}}

\begingroup

\renewcommand{\citenumfont}[1]{S#1}
\renewcommand{\bibnumfmt}[1]{[S#1]}

\begin{bibunit}


\date{\today}
\title{}

\author{Yisheng Lei}
\thanks{Corresponding author}
\email{YishengLei2025@u.northwestern.edu}
\affiliation{Institut de Recherche de Chimie Paris, Chimie ParisTech, CNRS, PSL University, 75005 Paris, France}

\author{Senthil Kumar Kuppusamy}
\affiliation{Institute for Quantum Materials and Technologies (IQMT), Karlsruhe Institute of Technology, Karlsruhe, Germany}

\author{Idris Tlemsani}
\affiliation{Institut de Recherche de Chimie Paris, Chimie ParisTech, CNRS, PSL University, 75005 Paris, France}
\affiliation{Institut de Chimie Moléculaire et des Matériaux d’Orsay CNRS, Université Paris-Saclay UMR 8182, 17, avenue des Sciences, 91400, Orsay (France)}

\author{Suma Al-Hunaishi}
\affiliation{Institut de Recherche de Chimie Paris, Chimie ParisTech, CNRS, PSL University, 75005 Paris, France}

\author{Pengrui Jiao}
\affiliation{Institut de Recherche de Chimie Paris, Chimie ParisTech, CNRS, PSL University, 75005 Paris, France}

\author{Olaf Fuhr}
\affiliation{Institute of Nanotechnology (INT), Karlsruhe Institute of Technology, Karlsruhe, Germany}
\affiliation{Karlsruhe Nano Micro Facility (KNMFi), Karlruhe Institute of Technology (KIT), Kaiserstraße 12, 76131 Karlsruhe}


\author{Mario Ruben}
\affiliation{Institute for Quantum Materials and Technologies (IQMT), Karlsruhe Institute of Technology, Karlsruhe, Germany}
\affiliation{Institute of Nanotechnology (INT), Karlsruhe Institute of Technology, Karlsruhe, Germany}
\affiliation{Centre Européen de Sciences Quantiques (CESQ), Institut de Science et d’Ingénierie Supramoléculaires (ISIS), Strasbourg, France}

\author{Philippe Goldner}
\affiliation{Institut de Recherche de Chimie Paris, Chimie ParisTech, CNRS, PSL University, 75005 Paris, France}

\author{Diana Serrano}
\thanks{Corresponding author}
\email{diana.serrano@chimieparistech.psl.eu}
\affiliation{Institut de Recherche de Chimie Paris, Chimie ParisTech, CNRS, PSL University, 75005 Paris, France}

\title{Supplementary Information: Broadband and Large-Multimode Quantum Optical Storage with Rare-Earth Ions in Solids}
\maketitle

\vspace{1em}

%

\makeatletter

\providecommand{\contentsname}{Contents}

\let\supp@origaddcontentsline\addcontentsline

\renewcommand{\addcontentsline}[3]{%
  \def\supp@target{#1}%
  \def\supp@toc{toc}%
  \ifx\supp@target\supp@toc
    \supp@origaddcontentsline{stoc}{#2}{#3}%
  \else
    \supp@origaddcontentsline{#1}{#2}{#3}%
  \fi
}

\section*{\contentsname}
\@starttoc{stoc}

\makeatother

\newpage

\section{Memory protocol}

\subsection{HAGEM implementation simulation method}

The input to each simulation is the adjacent hyperfine level gaps in
the optical ground and excited manifolds. Following the analysis of spectral hole \& anti-hole locations described in the main text, these gaps are used to generate the
positions of all spectral holes and anti-holes produced by optical pumping and
subsequent population redistribution. Side hole offsets are determined by
energy differences within the excited-state manifold, whereas anti-hole
offsets combine ground-state population shifts with the allowed excited-state
offsets. Ion-specific optical transitions and modifications to these rules are
described in the corresponding subsections below.

\begin{equation}
\begin{aligned}
    \Delta_{e} &= m_{e} \cdot \delta, \\
    \Delta_{g} &= m_{g} \cdot \delta.
\end{aligned}
\end{equation}

For simulations of atoms with two hyperfine levels in both the ground and excited states, $m_{e}$, $m_{g}$, and $\delta$ are free parameters with some constraints, for example, $m_{e}$ is an even integer; $m_{g}$ is an odd integer. To find the best spectrum for the implementation of HAGEM, the spectral hole and anti-hole positions produced by these fitted hyperfine level splitting values $\Delta_{e}$ \& $\Delta_{g}$ must match well with those experimental values of the simulated material, since the fitted spectrum is a perfect HAGEM spectrum. 

\begin{equation}
\begin{aligned}
    \Delta_{e,2} &= m_{e,2} \cdot \delta, \\
    \Delta_{e,1} &= m_{e,1} \cdot \delta, \\
    \Delta_{g,2} &= m_{g,2} \cdot \delta, \\
    \Delta_{g,1} &= m_{g,1} \cdot \delta.
\end{aligned}
\end{equation}

For simulations of atoms with three hyperfine levels in both the ground and excited states, $m_{e,2}$, $m_{e,1}$, $m_{g,2}$, $m_{g,1}$, and $\delta$ are free parameters with some constraints, for example, $m_{e,2}$ \& $m_{e,1}$ are even integers; at least one of $m_{g,2}$ \& $m_{g,1}$ is an odd integer, and the other can be an odd or even integer. To find the best spectrum for the implementation of HAGEM, the spectral hole and anti-hole positions produced by these fitted hyperfine level splitting values $\Delta_{e,2}$, $\Delta_{e,1}$, $\Delta_{g,2}$, \& $\Delta_{g,1}$ have to match well with those experimental values of the simulated material, since the fitted spectrum is a perfect HAGEM spectrum. 
Positions that coincide within the numerical merge tolerance are counted as
one spectral position, because overlapping contributions cannot be
distinguished by a position-only comparison. The zero-frequency hole is
retained. When the resulting spectrum is inversion symmetric, only the
positive-frequency half is evaluated; the negative-frequency half is then
fixed by symmetry. The calculation compares discrete spectral positions
directly and does not assume a Gaussian, Lorentzian, or other line-shape model.

For each line family
$X\in\{\mathrm{hole},\mathrm{anti}\}$, let $E_X$ denote the unique reference
positions generated from the measured hyperfine gaps and let $F_X$ denote the
positions generated from a candidate set of grid-constrained gaps. The two
lists are sorted in frequency. If their lengths differ, every position in the
shorter list is matched to a distinct position in the longer list. The
assignment preserves the frequency order and is chosen to minimize the total
absolute displacement. Writing $S_X$ and $L_X$ for the shorter and longer
lists, respectively, the mean displacement of the matched positions is

\begin{equation}
D_X=
\frac{1}{|S_X|}
\sum_{i=1}^{|S_X|}
\left|S_{X,i}-L_{X,\pi_X(i)}\right|,
\label{eq:mean-line-distance}
\end{equation}

where $\pi_X(i)$ identifies the position in the longer list paired with
$S_{X,i}$. Thus, $D_X=0$ only when every matched pair has the same frequency.

The displacement term alone is insufficient when the two spectra contain
different numbers of unique positions: it would compare only the shorter list
and ignore the remaining positions in the longer list. We therefore introduce
the unmatched-line penalty

\begin{equation}
P_X=
2\delta\,
\frac{|L_X|-|S_X|}{|L_X|}.
\label{eq:cardinality-penalty}
\end{equation}

The numerator is the number of positions left unmatched, and the denominator
converts it into the unmatched fraction of the longer list. Multiplication by
$2\delta$, the period of the target HAGEM spectral grid, places this penalty on
the same frequency scale as the position mismatch. Consequently, $P_X=0$ when
the two lists contain the same number of unique positions, and it increases as
a larger fraction of the calculated and reference spectra cannot be paired.

The hole and anti-hole contributions are evaluated separately and then added:

\begin{align}
C_{\rm hole} &= D_{\rm hole}+P_{\rm hole},\\
C_{\rm anti} &= D_{\rm anti}+P_{\rm anti},\\
C_{\rm abs} &= C_{\rm hole}+C_{\rm anti},\\
C_{\rm norm} &= \frac{C_{\rm abs}}{\delta}.
\label{eq:normalized-cost}
\end{align}

Here, $C_{\rm abs}$ is the total spectral-position mismatch in MHz, whereas
$C_{\rm norm}$ expresses that mismatch relative to the candidate spacing
$\delta$ and is therefore dimensionless. A value of $C_{\rm norm}=0$ means
that both the frequencies and the numbers of unique hole and anti-hole
positions agree exactly. The hyperfine-gap errors reported in the tables are
used only as diagnostics and tie breakers; they are not added to the objective
defined above.

For the standard three-level case, candidate excited-state gaps are restricted
to even multiples of $\delta$, whereas candidate ground-state gaps are integer
multiples of $\delta$ with at least one odd multiplier. The search enumerates
the allowed integer multipliers over the specified $\delta$ interval,
constructs the corresponding discrete hole and anti-hole spectra, evaluates
$C_{\rm norm}$, and then refines $\delta$ around the best candidates. The
associated ideal echo time is

\begin{equation}
T_{\rm echo}=\frac{1}{2\delta},
\end{equation}

or $T_{\rm echo}=500/\delta$ ns when $\delta$ is expressed in MHz.
Ion-specific changes to the spectral construction, allowed transitions, and
search intervals are stated below.

Throughout the spectral-position figures, ``Exp.'' denotes the reference line
positions calculated from the experimentally measured hyperfine splittings,
whereas ``Fit.'' denotes the positions calculated from the optimized,
grid-constrained splittings. The displayed values of $\delta$ and
$C_{\rm norm}$ are rounded to three decimal places, whereas the tables retain
the full numerical precision used in the optimization.

\subsection{HAGEM with two hyperfine levels}
We describe the memory protocol with two hyperfine levels. Due to inhomogeneous broadening, the laser pulse interacts with four classes of ions. There are 2 side holes located around f$_\text{0}$ de-tuned by values equal to $\pm \Delta_e$ and six anti-holes located around f$_\text{0}$ de-tuned by values equal to $\pm \Delta_g$, $\pm (\Delta_g - \Delta_e)$ and $\pm (\Delta_g + \Delta_e)$. Here $\Delta_e$ = 20~MHz and $\Delta_g$ = 10~MHz, and the branching ratio values are listed in Fig. \ref{Fig_HAGEM2HFlevels}(a). The single spectral hole burning spectrum is simulated in \ref{Fig_HAGEM2HFlevels}(b) and the HAGEM spectrum is simulated in \ref{Fig_HAGEM2HFlevels}(c). In this case, $\Delta_{HAGEM} = \Delta_g/k$ will always create a perfect HAGEM spectrum, where $k$ is an odd integer.

\begin{figure}[!ht]
\centerline{\includegraphics[width=1\columnwidth]{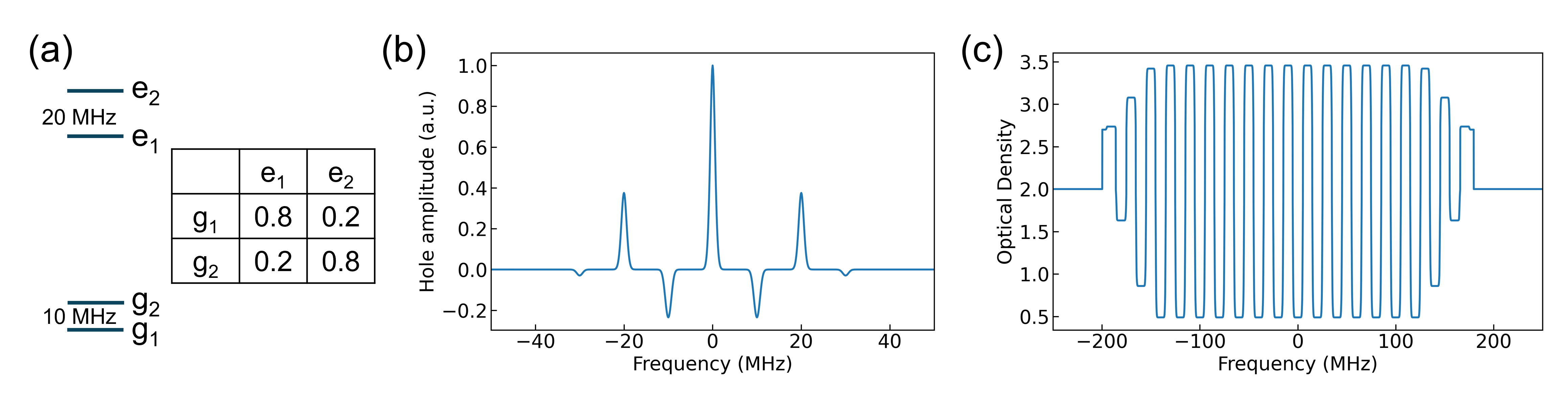}}
	\caption{HAGEM memory protocol for two hyperfine levels: (a) one example of hyperfine level structures and branching ratio values between hyperfine levels. (b) Simulation of a single spectral hole burning. (d) Simulation of HAGEM spectrum.}
	\label{Fig_HAGEM2HFlevels}
\end{figure}

The example above is a special case with correlated hyperfine level separations. The general rules: $\delta$ is a specially selected value (many possible choices); ground state splitting is an odd multiple of $\delta$, and excited state splitting is an even multiple of $\delta$; hyperfine level separations do not need to be correlated and can be any random unrelated values, especially for smaller values of $\delta$. 


\subsection{HAGEM with three hyperfine levels}

\begin{figure}[!ht]
\centerline{\includegraphics[width=1\columnwidth]{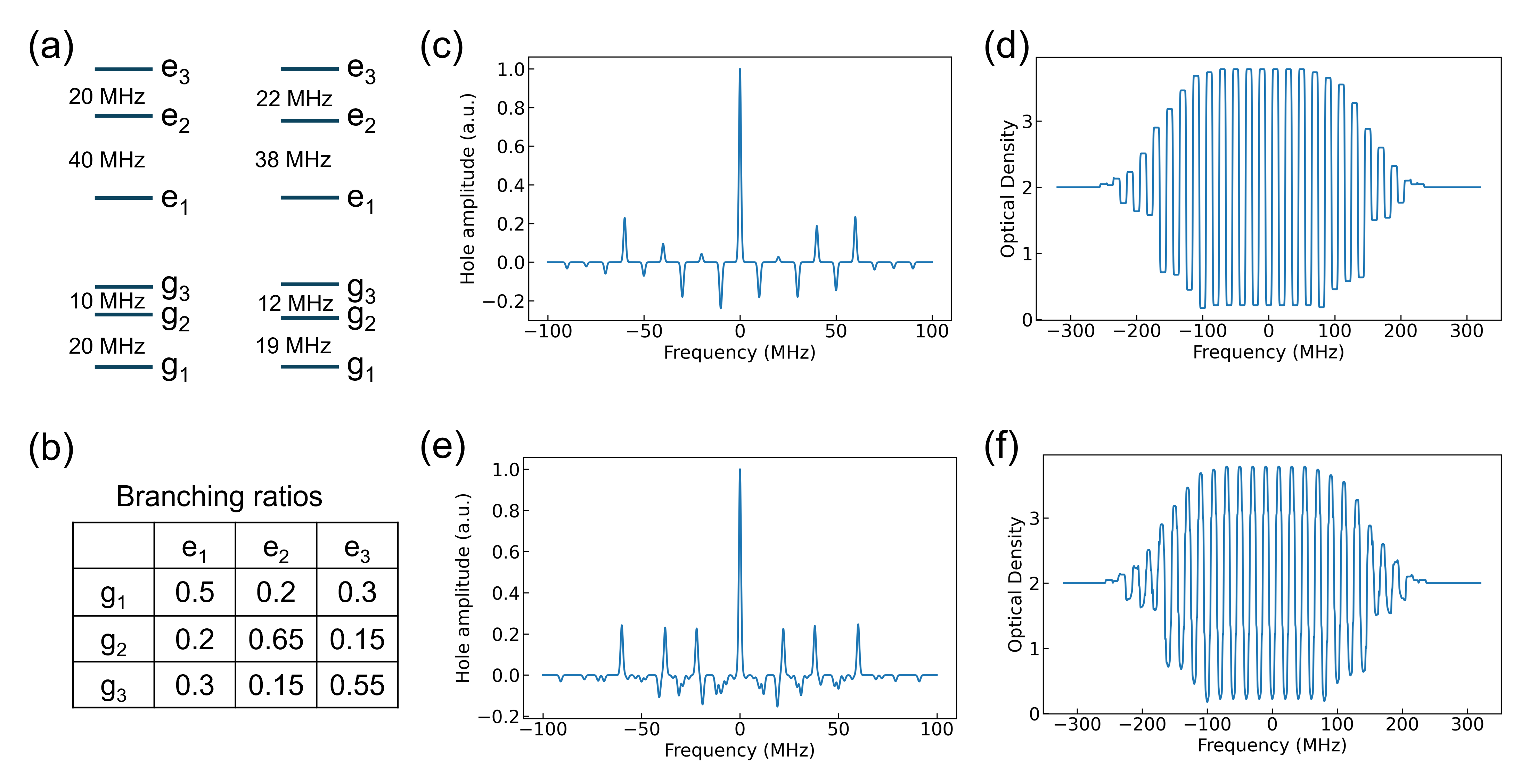}}
	\caption{HAGEM memory protocol principle: (a) Two exemples of HAGEM-compatible three-level hyperfine structures hereinafter referred to as first case (left) and second case (right). The first case presents ideal hyperfine level correlations while the second one shows minor frequency mismatches (b) Branching ratio values for hyperfine levels between the optical transition states used in (c-f) spectral hole burning simulations. (c) Simulation of a single spectral hole burning for the first case. (d) Simulation of HAGEM spectrum for the first case. (e) Simulation of a single spectral hole burning for the second case. (f) Simulation of HAGEM spectrum for the second case.}
	\label{Fig_ExtraMemoryProtocol}
\end{figure}

After a single spectral hole burning (SHB) with laser frequency f$_\text{0}$, there are 6 side holes located around f$_\text{0}$ deviated by values equal to terms in [$\pm\Delta e_{1}$, $\pm\Delta e_{2}$, $\pm\left( \Delta e_{1} + \Delta e_{2} \right)$], and 42 anti-holes located around f$_\text{0}$ deviated by values equal to one term in [$\pm\Delta g_{1}$, $\pm\Delta g_{2}$, $\pm\left( \Delta g_{1} + \Delta g_{2} \right)$] + one term in [$0$, $\pm\Delta e_{1}$, $\pm\Delta e_{2}$, $\pm\left( \Delta e_{1} + \Delta e_{2} \right)$]. In the example of Fig. \ref{Fig_ExtraMemoryProtocol}(a), $\Delta g_{2}$ is assumed to be $\delta$ = 10~MHz, $\Delta g_{1}$ = $2*\delta$ = 20~MHz, $\Delta e_{2}$ = $2*\delta$ = 20~MHz and $\Delta e_{1}$ = $4*\delta$ = 40~MHz. One single spectral hole burning spectrum is simulated with spectral linewidth of 1.5~MHz (Full-Width-Half-Maximum value) and a characteristic pumping duration corresponding to approximately 20 excitation–relaxation cycles: spectral holes are equally distributed in frequency domain and all of them are located at frequencies of even multiples of $\delta$; same as the anti-holes and most of them are located at frequencies of odd multiples of $\delta$, which form a perfect grating structure, as shown in Fig. \ref{Fig_ExtraMemoryProtocol}(b). The side holes near the center holes are much smaller compared to the furthermost side holes, which is due to the overlap by spectral anti-holes. In this case, $\Delta_{HAGEM} = \Delta_{g_2}/k$ will always create a perfect HAGEM spectrum, where $k$ is an odd integer. The second case in Fig. \ref{Fig_ExtraMemoryProtocol}(a) is in a similar correlation (hyperfine level separations are even or odd multiples of a fixed value.) as the first case, but with some minor frequency mismatches. The simulation indicates that a nearly perfect grating structure can still be created, as shown in Fig. \ref{Fig_ExtraMemoryProtocol}(f). 


The examples above are special cases with correlated hyperfine level separations. General rules: $\delta$ is a specially selected value (many possible choices); one ground state energy gap is an odd multiple of $\delta$ and another one can be even or odd multiples of $\delta$; both excited state energy gaps have to be even multiples of $\delta$; hyperfine level separations do not need to be correlated and can be any random unrelated values, especially for smaller values of $\delta$.

\subsection{HAGEM with four hyperfine levels}

\begin{figure}[!ht]
\centerline{\includegraphics[width=1\columnwidth]{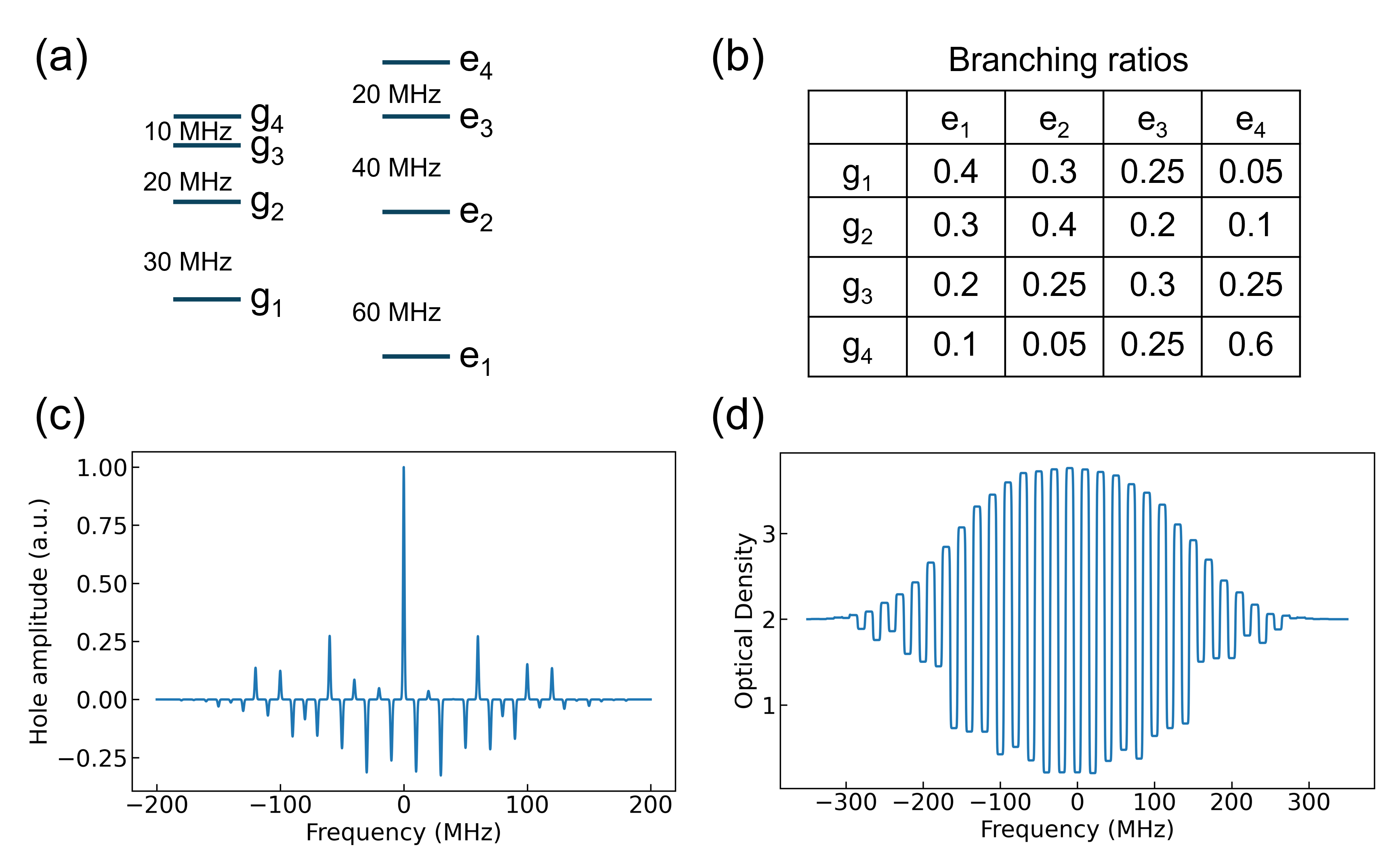}}
	\caption{An example of HAGEM memory protocol for atoms with four hyperfine ground states: (a) One example of four hyperfine level structures. (b) Branching ratio values for hyperfine levels at optical excitied and ground states. (c) Simulation of a single spectral hole burning. (d) Simulation of HAGEM spectrum.}
	\label{Fig_4LevelsHAGEM}
\end{figure}

Following the analysis of HAGEM with three hyperfine levels, we further describe the memory protocol with four hyperfine levels. Due to inhomogeneous broadening, the laser pulse interacts with sixteen classes of ions. There are 12 side holes located around f$_\text{0}$ deviated by values equal to terms in [$\pm\Delta e_{1}$, $\pm\Delta e_{2}$, $\pm\Delta e_{3}$, $\pm\left( \Delta e_{1} + \Delta e_{2} \right)$, $\pm\left( \Delta e_{2} + \Delta e_{3} \right)$, $\pm\left( \Delta e_{1} + \Delta e_{2} + \Delta e_{3} \right)$], and 156 anti-holes located around f$_\text{0}$ deviated by values equal to one term in [$\pm\Delta g_{1}$, $\pm\Delta g_{2}$, $\pm\Delta g_{3}$, $\pm\left( \Delta g_{1} + \Delta g_{2} \right)$, $\pm\left( \Delta g_{2} + \Delta g_{3} \right)$, $\pm\left( \Delta g_{1} + \Delta g_{2} + \Delta g_{3} \right)$] + one term in [$0$, $\pm\Delta e_{1}$, $\pm\Delta e_{2}$, $\pm\Delta e_{3}$, $\pm\left( \Delta e_{1} + \Delta e_{2} \right)$, $\pm\left( \Delta e_{2} + \Delta e_{3} \right)$, $\pm\left( \Delta e_{1} + \Delta e_{2} + \Delta e_{3} \right)$]. Here $\Delta g_{3}$ is assumed to be $\delta$ = 10~MHz, $\Delta g_{2}$ = $2*\delta$ = 20~MHz, $\Delta g_{1}$ = $3*\delta$ = 30~MHz, $\Delta e_{3}$ = $2*\delta$ = 20~MHz, $\Delta e_{2}$ = $4*\delta$ = 40~MHz and $\Delta e_{1}$ = $6*\delta$ = 60~MHz, as shown in Fig. \ref{Fig_4LevelsHAGEM}(a). The branching ratio values used for spectrum simulation are listed in Fig. \ref{Fig_4LevelsHAGEM}(b). The simulation of a single spectral hole burning spectrum is shown in Fig. \ref{Fig_4LevelsHAGEM}(c). The simulation of the HAGEM spectrum is shown in Fig. \ref{Fig_4LevelsHAGEM}(d). The same method can be applied to atoms with more than four hyperfine levels, and the HAGEM memory protocol works with some similar hyperfine level splitting. In this case, $\Delta_{HAGEM} = \Delta_{g_3}/k$ will always create a perfect HAGEM spectrum, where $k$ is an odd integer.

The example above is a special case with correlated hyperfine level separations. General rules: $\delta$ is a specially selected value (many possible choices); at least one ground state energy gap is an odd multiple of $\delta$ and the rest can be even or odd multiples of $\delta$; all excited state energy gaps must be even multiples of $\delta$; hyperfine level separations do not need to be correlated and can be any random unrelated values, especially for smaller values of $\delta$. 
This rule can be extended to RE ions with five or more hyperfine levels. Based on the analysis above, if the atoms have n hyperfine levels in both the ground and excited states, there are in total $n \times (n-1)$ spectral side holes and $\left(n \times (n-1)\right)\times\left(n \times (n-1)+1\right)$ spectral anti-holes. In a more general form, if the atoms have $n_g$ hyperfine levels in the ground states and $n_e$ hyperfine levels in the excited states, there are in total $n_e \times (n_e-1)$ spectral side holes and $\left(n_g \times (n_g-1)\right)\times\left(n_e \times (n_e-1)+1\right)$ spectral anti-holes.

\subsection{Imperfect HAGEM with some hyperfine level structures}

\begin{figure}[!ht]
\centerline{\includegraphics[width=1.0\columnwidth]{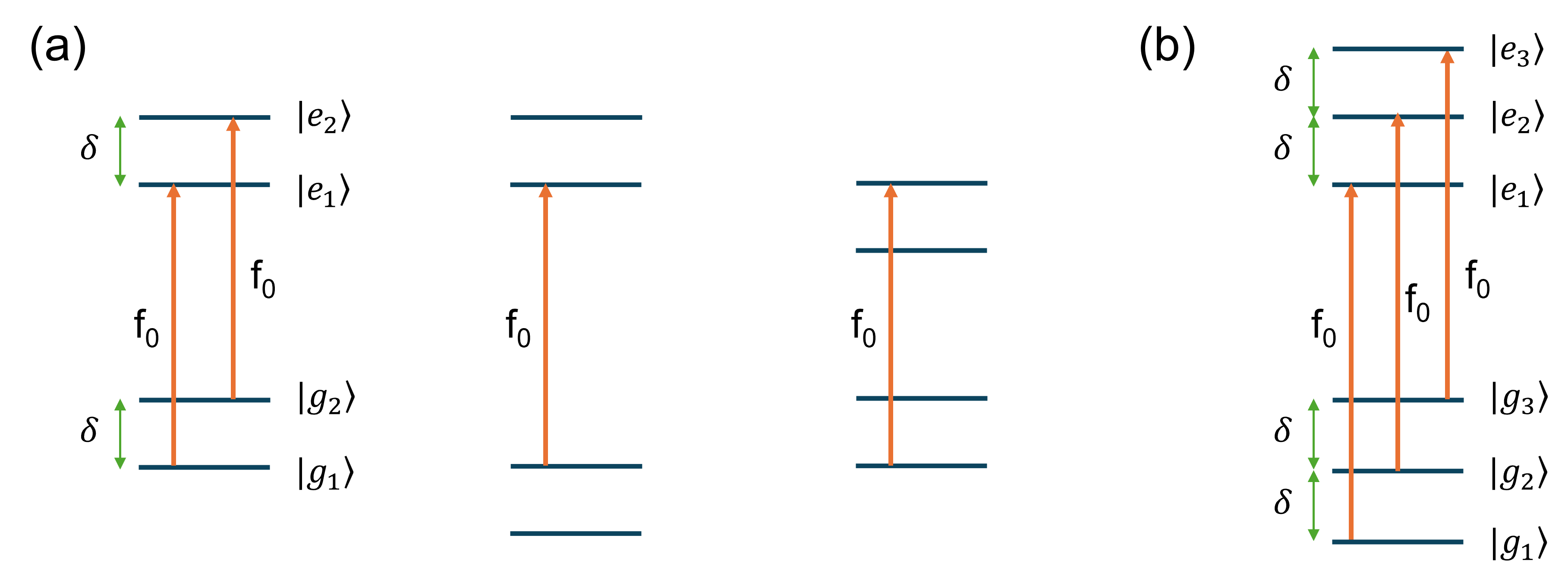}}
	\caption{(a) An example of imperfect HAGEM with two hyperfine levels. (b) An example of imperfect HAGEM with three hyperfine levels.}
	\label{Fig_ImperfectBQS}
\end{figure}

Here we show two examples of hyperfine level structures that HAGEM can be implemented but cannot achieve 100\% pumping efficiency. For two hyperfine levels and the splittings are equal, after spectral hole burning at frequency $f_0$, some of the atoms are pumped away, but there is always a fraction of the atoms is resonant with laser pulse of frequency $f_0$, as shown in Fig. \ref{Fig_ImperfectBQS}(a). HAGEM can be implemented with $\Delta_{HAGEM} = 2\delta$, but the storage efficiency will be limited by background absorption due to inefficient spectral pumping. A similar example with three hyperfine levels is shown in Fig. \ref{Fig_ImperfectBQS}(b), and HAGEM can be implemented with $\Delta_{HAGEM} = 4\delta$, of which the storage efficiency will be limited as well.

\subsection{HAGEM with Kramer ions}

\begin{figure}[!ht]
\centerline{\includegraphics[width=1.0\columnwidth]{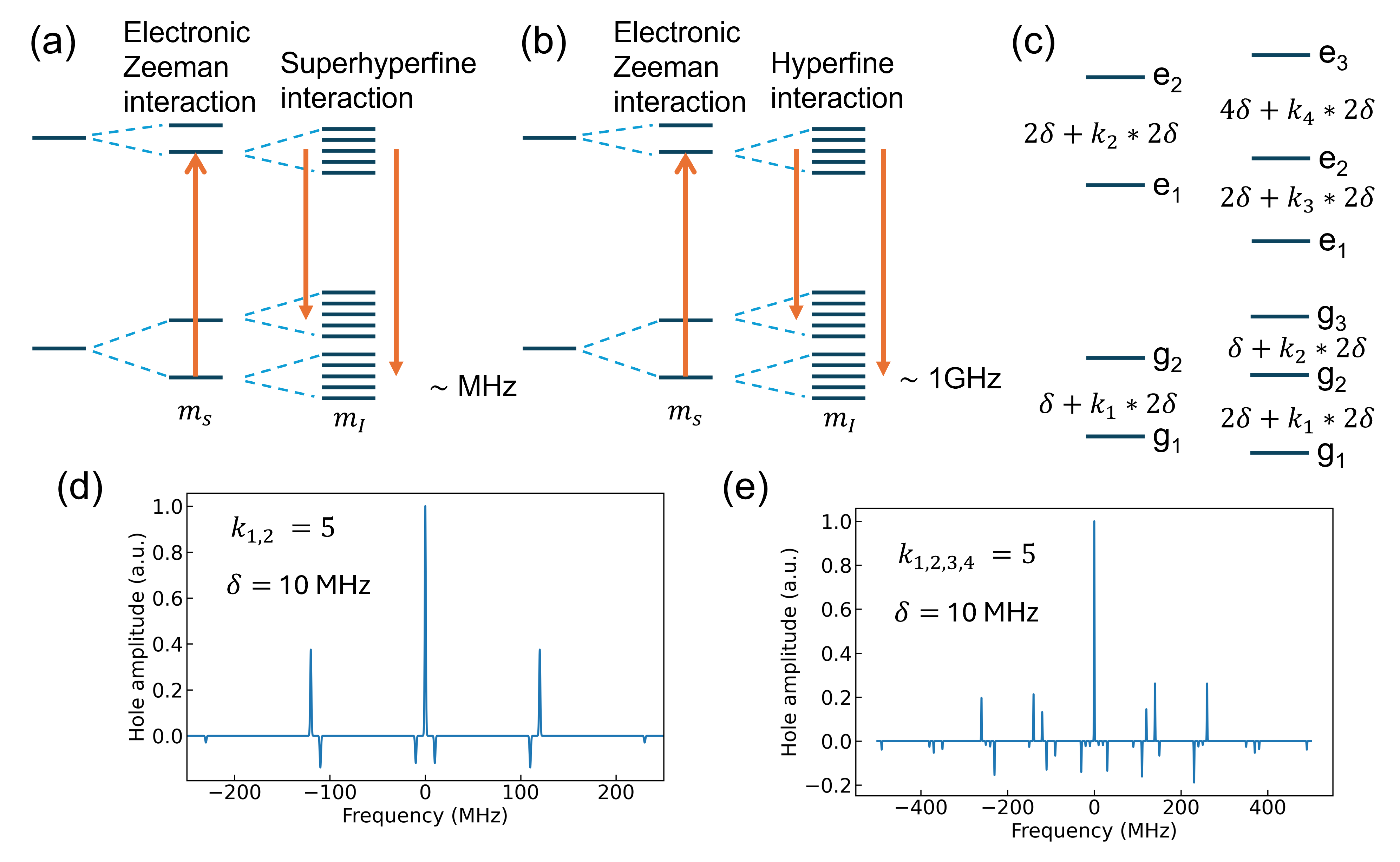}}
	\caption{(a) An example of Kramer ions without nuclear spins and have superhyperfine interactions with nuclear spins from host material. (b) An example of Kramer ions with nuclear spins and have no superhyperfine interactions. (c) Two examples of hyperfine level structures. (d) Simulation of a single spectral hole burning for the first case. (e) Simulation of a single spectral hole burning for the second case.}
	\label{Fig_KramerIon}
\end{figure}

For Kramer ions without nuclear spins, and that have superhyperfine interactions with nuclear spins from host material, the superhyperfine level splittings typically are a few MHz. Due to spin decays into another electron spin level, as shown in Fig. \ref{Fig_KramerIon}(a), the broadband storage scheme is HAGEM mixed with AFC. A $\delta$ value chosen for this superhyperfine level structure needs to match HAGEM memory protocol, and it is combined with AFC since there is a branching ratio into another electron spin level, so some of the atoms are pumped away from the storage frequency window, if the upper electronic spin lifetime is long enough. For Kramer ions with nuclear spins and have no superhyperfine splittings, hyperfine level splittings are typically around GHz ranges,  as shown in Fig. \ref{Fig_KramerIon}(b). Here we consider two special cases with correlated hyperfine level separations, if the hyperfine level splittings match the cases in Fig. \ref{Fig_KramerIon}(c), they can be used for HAGEM memory protocol, as shown in Fig. \ref{Fig_KramerIon}(d)\&(e). Another electron spin hyperfine levels can be emptied during spectral preparation, after HAGEM storage, these levels can be used for spin-wave storage with control pulses, which will enable long storage time and on-demand capability. To implement HAGEM, these hyperfine level separations do not need to be correlated and they can be any random unrelated values. The analysis for some specific types of Kramer ions is discussed in SI S2.2. 

\section{HAGEM material search simulation}

\subsection{Non-Kramer ion materials}

\subsubsection{Eu$^\text{3+}$ materials}

All Eu$^{3+}$ cases have three zero-field hyperfine energies in each manifold
and therefore use the common three-level construction and parity constraints.
We simulate both isotopes at the $C_{4v}$ and $C_{3v}$ sites in
$\mathrm{CaF}_2$\cite{silversmith1986c3v,silversmith1992c4v,smith2022complete},
in $\mathrm{CaWO}_4$\cite{an2025cawo4}, in
$\mathrm{EuCl}_3\!\cdot\!6\mathrm{H}_2\mathrm{O}$
\cite{longdell2006eu,ahlefeldt2013eucl3,smith2022complete}, in the $C_2$ site of
$\mathrm{Y}_2\mathrm{O}_3$\cite{karlsson2017y2o3}, and at both sites of
$\mathrm{Y}_2\mathrm{SiO}_5$\cite{longdell2006eu,lauritzen2012spectroscopic,jin2022site2,yano1991yso}.
The molecular calculations use the hyperfine splittings reported for
$[\mathrm{Eu}^{3+}(\mathrm{BA})_4](\mathrm{pip})$ \cite{serrano2022ultra} and
$[{}^{151}\mathrm{Eu}^{3+}(\mathrm{TMHD})_3(\mathrm{phen})]$.

The default Eu scan is $2\leq\delta\leq20$ MHz.  The lower limit is extended
to $1.5$ MHz for the poorly matched $^{151}\mathrm{Eu}:\mathrm{CaF}_2$
$C_{4v}$ case, due to the smaller hyperfine gap values. For the $[{}^{151}\mathrm{Eu}^{3+}(\mathrm{TMHD})_3(\mathrm{phen})]$ complex, the smallest ground-state multiplier is
also fixed successively to the requested odd values so that each experimentally
relevant subharmonic family is represented.  Finally, the site-1
$^{151}\mathrm{Eu}:\mathrm{Y}_2\mathrm{SiO}_5$ calculation is repeated in five
independent $\delta$ windows down to $0.1$ MHz to test the trade-off between finer
commensurability and longer echo time, and the results are shown in main texts; the same $\delta$ window study is applied to an
exactly commensurate $(40,20,20,10)$ MHz control structure.

The restricted-multiplier calculation in Table~\ref{tab:eu-509-fixed-last}
fixes only the final ground-state multiplier for the molecular platform and
allows the other multipliers to optimize independently. The $\delta$ window study shown in Extended Data Table. 2 of the main text tests whether progressively smaller $\delta$
provides a finer integer grid for the site-1 YSO splittings. Exact numerical
matches at $\delta=0.3$ and $0.1$ MHz follow from the one-decimal precision of
the tabulated experimental gaps and should therefore be tested against the
experimental uncertainties before being interpreted as robust solutions.
Table~\ref{tab:ideal-windows} supplies the requested control for the perfectly
commensurate gaps $(\Delta_{e,2},\Delta_{e,1},\Delta_{g,2},\Delta_{g,1})
=(40,20,20,10)$ MHz. Its zero-cost solutions are degenerate:
$\delta=10/k$ MHz for any positive odd integer $k$ satisfies the excited-state
even-multiplier constraint and leaves an odd ground-state multiplier.

\begin{table*}[!tbp]
\centering
\caption{Restricted-multiplier fits for $[{}^{151}\mathrm{Eu}^{3+}(\mathrm{TMHD})_3(\mathrm{phen})]$. Only the final ground-state multiplier $m_{g,1}$ is fixed; $\delta$ and the remaining three multipliers are optimized independently.}
\label{tab:eu-509-fixed-last}
\scriptsize
\renewcommand{\arraystretch}{1.10}
\begin{tabular*}{\textwidth}{@{\extracolsep{\fill}}cccccccc}
\toprule
Fixed $m_{g,1}$ & $\delta$ (MHz) & $(m_{e,2},m_{e,1})$ & $(m_{g,2},m_{g,1})$ & $T_{\rm echo}$ (ns) & $C_{\rm norm}$ & $C_{\rm abs}$ (MHz) & Max. mismatch (MHz) \\
\midrule
3 & 6.185185 & $(12, 6)$ & $(6, 3)$ & 80.8 & 0.297395 & 1.839444 & 3.100000 \\
5 & 3.723333 & $(20, 10)$ & $(9, 5)$ & 134.3 & 0.446362 & 1.661954 & 3.173333 \\
7 & 3.102778 & $(24, 12)$ & $(11, 7)$ & 161.1 & 0.361926 & 1.122975 & 1.886111 \\
9 & 2.325397 & $(32, 16)$ & $(15, 9)$ & 215.0 & 0.356345 & 0.828642 & 1.122222 \\
11 & 1.925862 & $(38, 20)$ & $(18, 11)$ & 259.6 & 0.447200 & 0.861245 & 1.301724 \\
\bottomrule
\end{tabular*}
\end{table*}

\begin{table*}[!tbp]
\centering
\caption{Cost-function verification for the perfectly commensurate gaps
$(\Delta_{e,2},\Delta_{e,1},\Delta_{g,2},\Delta_{g,1})
=(40,20,20,10)$ MHz. The rows are members of the exact scaling family
$\delta_k=10/k$ MHz for odd $k$; they are evaluated directly and are not
obtained from independent optimizations.}
\label{tab:ideal-windows}
\scriptsize
\renewcommand{\arraystretch}{1.10}
\begin{tabular*}{\textwidth}{@{\extracolsep{\fill}}ccccccc}
\toprule
$k$ &
$\delta_k$ (MHz) &
$(m_{e,2},m_{e,1})$ &
$(m_{g,2},m_{g,1})$ &
$T_{\rm echo}$ (ns) &
$C_{\rm norm}$ &
$C_{\rm abs}$ (MHz) \\
\midrule
1 & 10.000000 & $(4,2)$   & $(2,1)$  & 50.0  & $0$ & $0$ \\
3 & 3.333333  & $(12,6)$  & $(6,3)$  & 150.0 & $0$ & $0$ \\
5 & 2.000000  & $(20,10)$ & $(10,5)$ & 250.0 & $0$ & $0$ \\
7 & 1.428571  & $(28,14)$ & $(14,7)$ & 350.0 & $0$ & $0$ \\
9 & 1.111111  & $(36,18)$ & $(18,9)$ & 450.0 & $0$ & $0$ \\
\bottomrule
\end{tabular*}
\end{table*}

\begin{table*}[!tbp]
\centering
\caption{Detailed Eu-material fits. Rows follow the requested order $\Delta_{e,2},\Delta_{e,1},\Delta_{g,2},\Delta_{g,1}$ where two gaps are present.}
\label{tab:eu-details-1}
\scriptsize
\renewcommand{\arraystretch}{1.08}
\setlength{\tabcolsep}{2.8pt}
\begin{minipage}[t]{0.385\textwidth}
\centering
\textbf{$^{151}\mathrm{Eu}^{3+}\!:\mathrm{CaF}_2$ ($C_{4v}$ site)}\\[-0.15em]
$\delta=1.630625\,\mathrm{MHz}$; $C_{\rm norm}=0.124952$\\[0.25em]
\resizebox{\linewidth}{!}{%
\begin{tabular}{cccccc}
\toprule
Gap & Exp. & $m$ & Fit & $e$ & $|e|/\delta$ \\
 & (MHz) & & (MHz) & (MHz) & \\
\midrule
$\Delta_{e,2}$ & 52.180 & 32 & 52.180 & -0.000 & 0.0000 \\
$\Delta_{e,1}$ & 26.090 & 16 & 26.090 & -0.000 & 0.0000 \\
$\Delta_{g,2}$ & 9.580 & 6 & 9.784 & -0.204 & 0.1250 \\
$\Delta_{g,1}$ & 4.790 & 3 & 4.892 & -0.102 & 0.0625 \\
\bottomrule
\end{tabular}%
}
\end{minipage}
\hfill
\begin{minipage}[t]{0.385\textwidth}
\centering
\textbf{$^{151}\mathrm{Eu}^{3+}\!:\mathrm{CaF}_2$ ($C_{3v}$ site)}\\[-0.15em]
$\delta=5.560000\,\mathrm{MHz}$; $C_{\rm norm}=0.053657$\\[0.25em]
\resizebox{\linewidth}{!}{%
\begin{tabular}{cccccc}
\toprule
Gap & Exp. & $m$ & Fit & $e$ & $|e|/\delta$ \\
 & (MHz) & & (MHz) & (MHz) & \\
\midrule
$\Delta_{e,2}$ & 88.970 & 16 & 88.960 & +0.010 & 0.0018 \\
$\Delta_{e,1}$ & 44.470 & 8 & 44.480 & -0.010 & 0.0018 \\
$\Delta_{g,2}$ & 16.520 & 3 & 16.680 & -0.160 & 0.0288 \\
$\Delta_{g,1}$ & 33.080 & 6 & 33.360 & -0.280 & 0.0504 \\
\bottomrule
\end{tabular}%
}
\end{minipage}
\par\vspace{1.0em}
\begin{minipage}[t]{0.385\textwidth}
\centering
\textbf{$^{153}\mathrm{Eu}^{3+}\!:\mathrm{CaF}_2$ ($C_{4v}$ site)}\\[-0.15em]
$\delta=2.553846\,\mathrm{MHz}$; $C_{\rm norm}=0.337349$\\[0.25em]
\resizebox{\linewidth}{!}{%
\begin{tabular}{cccccc}
\toprule
Gap & Exp. & $m$ & Fit & $e$ & $|e|/\delta$ \\
 & (MHz) & & (MHz) & (MHz) & \\
\midrule
$\Delta_{e,2}$ & 132.800 & 52 & 132.800 & +0.000 & 0.0000 \\
$\Delta_{e,1}$ & 66.400 & 26 & 66.400 & +0.000 & 0.0000 \\
$\Delta_{g,2}$ & 26.400 & 10 & 25.538 & +0.862 & 0.3373 \\
$\Delta_{g,1}$ & 13.200 & 5 & 12.769 & +0.431 & 0.1687 \\
\bottomrule
\end{tabular}%
}
\end{minipage}
\hfill
\begin{minipage}[t]{0.385\textwidth}
\centering
\textbf{$^{153}\mathrm{Eu}^{3+}\!:\mathrm{CaF}_2$ ($C_{3v}$ site)}\\[-0.15em]
$\delta=5.683333\,\mathrm{MHz}$; $C_{\rm norm}=0.033431$\\[0.25em]
\resizebox{\linewidth}{!}{%
\begin{tabular}{cccccc}
\toprule
Gap & Exp. & $m$ & Fit & $e$ & $|e|/\delta$ \\
 & (MHz) & & (MHz) & (MHz) & \\
\midrule
$\Delta_{e,2}$ & 227.300 & 40 & 227.333 & -0.033 & 0.0059 \\
$\Delta_{e,1}$ & 113.700 & 20 & 113.667 & +0.033 & 0.0059 \\
$\Delta_{g,2}$ & 39.850 & 7 & 39.783 & +0.067 & 0.0117 \\
$\Delta_{g,1}$ & 79.760 & 14 & 79.567 & +0.193 & 0.0340 \\
\bottomrule
\end{tabular}%
}
\end{minipage}
\end{table*}

\begin{table*}[!tbp]
\centering
\caption{Detailed Eu-material fits. Rows follow the requested order $\Delta_{e,2},\Delta_{e,1},\Delta_{g,2},\Delta_{g,1}$ where two gaps are present.}
\label{tab:eu-details-2}
\scriptsize
\renewcommand{\arraystretch}{1.08}
\setlength{\tabcolsep}{2.8pt}
\begin{minipage}[t]{0.385\textwidth}
\centering
\textbf{$^{151}\mathrm{Eu}^{3+}\!:\mathrm{CaWO}_4$}\\[-0.15em]
$\delta=2.163333\,\mathrm{MHz}$; $C_{\rm norm}=0.379277$\\[0.25em]
\resizebox{\linewidth}{!}{%
\begin{tabular}{cccccc}
\toprule
Gap & Exp. & $m$ & Fit & $e$ & $|e|/\delta$ \\
 & (MHz) & & (MHz) & (MHz) & \\
\midrule
$\Delta_{e,2}$ & 43.300 & 20 & 43.267 & +0.033 & 0.0154 \\
$\Delta_{e,1}$ & 30.200 & 14 & 30.287 & -0.087 & 0.0401 \\
$\Delta_{g,2}$ & 21.600 & 10 & 21.633 & -0.033 & 0.0154 \\
$\Delta_{g,1}$ & 19.700 & 9 & 19.470 & +0.230 & 0.1063 \\
\bottomrule
\end{tabular}%
}
\end{minipage}
\hfill
\begin{minipage}[t]{0.385\textwidth}
\centering
\textbf{$^{153}\mathrm{Eu}^{3+}\!:\mathrm{CaWO}_4$}\\[-0.15em]
$\delta=5.525000\,\mathrm{MHz}$; $C_{\rm norm}=0.343784$\\[0.25em]
\resizebox{\linewidth}{!}{%
\begin{tabular}{cccccc}
\toprule
Gap & Exp. & $m$ & Fit & $e$ & $|e|/\delta$ \\
 & (MHz) & & (MHz) & (MHz) & \\
\midrule
$\Delta_{e,2}$ & 110.700 & 20 & 110.500 & +0.200 & 0.0362 \\
$\Delta_{e,1}$ & 77.100 & 14 & 77.350 & -0.250 & 0.0452 \\
$\Delta_{g,2}$ & 55.200 & 10 & 55.250 & -0.050 & 0.0090 \\
$\Delta_{g,1}$ & 50.000 & 9 & 49.725 & +0.275 & 0.0498 \\
\bottomrule
\end{tabular}%
}
\end{minipage}
\par\vspace{1.0em}
\begin{minipage}[t]{0.385\textwidth}
\centering
\textbf{$^{151}\mathrm{Eu}^{3+}\!:\mathrm{EuCl}_3\!\cdot\!6\mathrm{H}_2\mathrm{O}$}\\[-0.15em]
$\delta=2.417692\,\mathrm{MHz}$; $C_{\rm norm}=0.425158$\\[0.25em]
\resizebox{\linewidth}{!}{%
\begin{tabular}{cccccc}
\toprule
Gap & Exp. & $m$ & Fit & $e$ & $|e|/\delta$ \\
 & (MHz) & & (MHz) & (MHz) & \\
\midrule
$\Delta_{e,2}$ & 43.090 & 18 & 43.518 & -0.428 & 0.1772 \\
$\Delta_{e,1}$ & 53.600 & 22 & 53.189 & +0.411 & 0.1699 \\
$\Delta_{g,2}$ & 29.030 & 12 & 29.012 & +0.018 & 0.0073 \\
$\Delta_{g,1}$ & 26.990 & 11 & 26.595 & +0.395 & 0.1635 \\
\bottomrule
\end{tabular}%
}
\end{minipage}
\hfill
\begin{minipage}[t]{0.385\textwidth}
\centering
\textbf{$^{153}\mathrm{Eu}^{3+}\!:\mathrm{EuCl}_3\!\cdot\!6\mathrm{H}_2\mathrm{O}$}\\[-0.15em]
$\delta=2.626596\,\mathrm{MHz}$; $C_{\rm norm}=0.358827$\\[0.25em]
\resizebox{\linewidth}{!}{%
\begin{tabular}{cccccc}
\toprule
Gap & Exp. & $m$ & Fit & $e$ & $|e|/\delta$ \\
 & (MHz) & & (MHz) & (MHz) & \\
\midrule
$\Delta_{e,2}$ & 109.900 & 42 & 110.317 & -0.417 & 0.1588 \\
$\Delta_{e,1}$ & 137.000 & 52 & 136.583 & +0.417 & 0.1588 \\
$\Delta_{g,2}$ & 74.580 & 28 & 73.545 & +1.035 & 0.3942 \\
$\Delta_{g,1}$ & 70.070 & 27 & 70.918 & -0.848 & 0.3229 \\
\bottomrule
\end{tabular}%
}
\end{minipage}
\end{table*}

\begin{table*}[!tbp]
\centering
\caption{Detailed Eu-material fits. Rows follow the requested order $\Delta_{e,2},\Delta_{e,1},\Delta_{g,2},\Delta_{g,1}$ where two gaps are present.}
\label{tab:eu-details-3}
\scriptsize
\renewcommand{\arraystretch}{1.08}
\setlength{\tabcolsep}{2.8pt}
\begin{minipage}[t]{0.385\textwidth}
\centering
\textbf{$^{151}\mathrm{Eu}^{3+}\!:\mathrm{Y}_2\mathrm{O}_3$ ($C_2$ site)}\\[-0.15em]
$\delta=2.625000\,\mathrm{MHz}$; $C_{\rm norm}=0.356792$\\[0.25em]
\resizebox{\linewidth}{!}{%
\begin{tabular}{cccccc}
\toprule
Gap & Exp. & $m$ & Fit & $e$ & $|e|/\delta$ \\
 & (MHz) & & (MHz) & (MHz) & \\
\midrule
$\Delta_{e,2}$ & 58.000 & 22 & 57.750 & +0.250 & 0.0952 \\
$\Delta_{e,1}$ & 89.000 & 34 & 89.250 & -0.250 & 0.0952 \\
$\Delta_{g,2}$ & 33.990 & 13 & 34.125 & -0.135 & 0.0514 \\
$\Delta_{g,1}$ & 29.330 & 11 & 28.875 & +0.455 & 0.1733 \\
\bottomrule
\end{tabular}%
}
\end{minipage}
\hfill
\begin{minipage}[t]{0.385\textwidth}
\centering
\textbf{$^{153}\mathrm{Eu}^{3+}\!:\mathrm{Y}_2\mathrm{O}_3$ ($C_2$ site)}\\[-0.15em]
$\delta=4.892637\,\mathrm{MHz}$; $C_{\rm norm}=0.271155$\\[0.25em]
\resizebox{\linewidth}{!}{%
\begin{tabular}{cccccc}
\toprule
Gap & Exp. & $m$ & Fit & $e$ & $|e|/\delta$ \\
 & (MHz) & & (MHz) & (MHz) & \\
\midrule
$\Delta_{e,2}$ & 147.000 & 30 & 146.779 & +0.221 & 0.0451 \\
$\Delta_{e,1}$ & 225.000 & 46 & 225.061 & -0.061 & 0.0125 \\
$\Delta_{g,2}$ & 87.800 & 18 & 88.067 & -0.267 & 0.0547 \\
$\Delta_{g,1}$ & 73.230 & 15 & 73.390 & -0.160 & 0.0326 \\
\bottomrule
\end{tabular}%
}
\end{minipage}
\par\vspace{1.0em}
\begin{minipage}[t]{0.385\textwidth}
\centering
\textbf{$^{151}\mathrm{Eu}^{3+}\!:\mathrm{Y}_2\mathrm{SiO}_5$ (site 1)}\\[-0.15em]
$\delta=2.681818\,\mathrm{MHz}$; $C_{\rm norm}=0.168362$\\[0.25em]
\resizebox{\linewidth}{!}{%
\begin{tabular}{cccccc}
\toprule
Gap & Exp. & $m$ & Fit & $e$ & $|e|/\delta$ \\
 & (MHz) & & (MHz) & (MHz) & \\
\midrule
$\Delta_{e,2}$ & 102.000 & 38 & 101.909 & +0.091 & 0.0339 \\
$\Delta_{e,1}$ & 75.000 & 28 & 75.091 & -0.091 & 0.0339 \\
$\Delta_{g,2}$ & 46.200 & 17 & 45.591 & +0.609 & 0.2271 \\
$\Delta_{g,1}$ & 34.500 & 13 & 34.864 & -0.364 & 0.1356 \\
\bottomrule
\end{tabular}%
}
\end{minipage}
\hfill
\begin{minipage}[t]{0.385\textwidth}
\centering
\textbf{$^{151}\mathrm{Eu}^{3+}\!:\mathrm{Y}_2\mathrm{SiO}_5$ (site 2)}\\[-0.15em]
$\delta=2.227368\,\mathrm{MHz}$; $C_{\rm norm}=0.202426$\\[0.25em]
\resizebox{\linewidth}{!}{%
\begin{tabular}{cccccc}
\toprule
Gap & Exp. & $m$ & Fit & $e$ & $|e|/\delta$ \\
 & (MHz) & & (MHz) & (MHz) & \\
\midrule
$\Delta_{e,2}$ & 62.360 & 28 & 62.366 & -0.006 & 0.0028 \\
$\Delta_{e,1}$ & 106.920 & 48 & 106.914 & +0.006 & 0.0028 \\
$\Delta_{g,2}$ & 29.540 & 13 & 28.956 & +0.584 & 0.2623 \\
$\Delta_{g,1}$ & 57.240 & 26 & 57.912 & -0.672 & 0.3015 \\
\bottomrule
\end{tabular}%
}
\end{minipage}
\end{table*}

\begin{table*}[!tbp]
\centering
\caption{Detailed Eu-material fits. Rows follow the requested order $\Delta_{e,2},\Delta_{e,1},\Delta_{g,2},\Delta_{g,1}$ where two gaps are present.}
\label{tab:eu-details-4}
\scriptsize
\renewcommand{\arraystretch}{1.08}
\setlength{\tabcolsep}{2.8pt}
\begin{minipage}[t]{0.385\textwidth}
\centering
\textbf{$^{153}\mathrm{Eu}^{3+}\!:\mathrm{Y}_2\mathrm{SiO}_5$ (site 1)}\\[-0.15em]
$\delta=2.203846\,\mathrm{MHz}$; $C_{\rm norm}=0.128812$\\[0.25em]
\resizebox{\linewidth}{!}{%
\begin{tabular}{cccccc}
\toprule
Gap & Exp. & $m$ & Fit & $e$ & $|e|/\delta$ \\
 & (MHz) & & (MHz) & (MHz) & \\
\midrule
$\Delta_{e,2}$ & 194.000 & 88 & 193.938 & +0.062 & 0.0279 \\
$\Delta_{e,1}$ & 260.000 & 118 & 260.054 & -0.054 & 0.0244 \\
$\Delta_{g,2}$ & 119.000 & 54 & 119.008 & -0.008 & 0.0035 \\
$\Delta_{g,1}$ & 90.000 & 41 & 90.358 & -0.358 & 0.1623 \\
\bottomrule
\end{tabular}%
}
\end{minipage}
\hfill
\begin{minipage}[t]{0.385\textwidth}
\centering
\textbf{$^{153}\mathrm{Eu}^{3+}\!:\mathrm{Y}_2\mathrm{SiO}_5$ (site 2)}\\[-0.15em]
$\delta=3.806931\,\mathrm{MHz}$; $C_{\rm norm}=0.079150$\\[0.25em]
\resizebox{\linewidth}{!}{%
\begin{tabular}{cccccc}
\toprule
Gap & Exp. & $m$ & Fit & $e$ & $|e|/\delta$ \\
 & (MHz) & & (MHz) & (MHz) & \\
\midrule
$\Delta_{e,2}$ & 160.000 & 42 & 159.891 & +0.109 & 0.0286 \\
$\Delta_{e,1}$ & 274.000 & 72 & 274.099 & -0.099 & 0.0260 \\
$\Delta_{g,2}$ & 148.100 & 39 & 148.470 & -0.370 & 0.0973 \\
$\Delta_{g,1}$ & 76.400 & 20 & 76.139 & +0.261 & 0.0687 \\
\bottomrule
\end{tabular}%
}
\end{minipage}
\par\vspace{1.0em}
\begin{minipage}[t]{0.385\textwidth}
\centering
\textbf{$[{}^{151}\mathrm{Eu}^{3+}(\mathrm{BA})_4](\mathrm{pip})]$}\\[-0.15em]
$\delta=2.335000\,\mathrm{MHz}$; $C_{\rm norm}=0.411135$\\[0.25em]
\resizebox{\linewidth}{!}{%
\begin{tabular}{cccccc}
\toprule
Gap & Exp. & $m$ & Fit & $e$ & $|e|/\delta$ \\
 & (MHz) & & (MHz) & (MHz) & \\
\midrule
$\Delta_{e,2}$ & 46.200 & 20 & 46.700 & -0.500 & 0.2141 \\
$\Delta_{e,1}$ & 37.900 & 16 & 37.360 & +0.540 & 0.2313 \\
$\Delta_{g,2}$ & 21.500 & 9 & 21.015 & +0.485 & 0.2077 \\
$\Delta_{g,1}$ & 34.000 & 15 & 35.025 & -1.025 & 0.4390 \\
\bottomrule
\end{tabular}%
}
\end{minipage}
\hfill
\begin{minipage}[t]{0.385\textwidth}
\centering
\textbf{$[{}^{153}\mathrm{Eu}^{3+}(\mathrm{BA})_4](\mathrm{pip})]$}\\[-0.15em]
$\delta=3.473585\,\mathrm{MHz}$; $C_{\rm norm}=0.193994$\\[0.25em]
\resizebox{\linewidth}{!}{%
\begin{tabular}{cccccc}
\toprule
Gap & Exp. & $m$ & Fit & $e$ & $|e|/\delta$ \\
 & (MHz) & & (MHz) & (MHz) & \\
\midrule
$\Delta_{e,2}$ & 118.200 & 34 & 118.102 & +0.098 & 0.0282 \\
$\Delta_{e,1}$ & 96.800 & 28 & 97.260 & -0.460 & 0.1325 \\
$\Delta_{g,2}$ & 55.700 & 16 & 55.577 & +0.123 & 0.0353 \\
$\Delta_{g,1}$ & 87.300 & 25 & 86.840 & +0.460 & 0.1325 \\
\bottomrule
\end{tabular}%
}
\end{minipage}
\end{table*}

\begin{table*}[!tbp]
\centering
\caption{Detailed Eu-material fits. Rows follow the requested order $\Delta_{e,2},\Delta_{e,1},\Delta_{g,2},\Delta_{g,1}$ where two gaps are present.}
\label{tab:eu-details-5}
\scriptsize
\renewcommand{\arraystretch}{1.08}
\setlength{\tabcolsep}{2.8pt}
\begin{minipage}[t]{0.385\textwidth}
\centering
\textbf{$[{}^{151}\mathrm{Eu}^{3+}(\mathrm{TMHD})_3(\mathrm{phen})]$}\\[-0.15em]
$\delta=18.555556\,\mathrm{MHz}$; $C_{\rm norm}=0.099132$\\[0.25em]
\resizebox{\linewidth}{!}{%
\begin{tabular}{cccccc}
\toprule
Gap & Exp. & $m$ & Fit & $e$ & $|e|/\delta$ \\
 & (MHz) & & (MHz) & (MHz) & \\
\midrule
$\Delta_{e,2}$ & 73.800 & 4 & 74.222 & -0.422 & 0.0228 \\
$\Delta_{e,1}$ & 37.900 & 2 & 37.111 & +0.789 & 0.0425 \\
$\Delta_{g,2}$ & 34.800 & 2 & 37.111 & -2.311 & 0.1246 \\
$\Delta_{g,1}$ & 20.500 & 1 & 18.556 & +1.944 & 0.1048 \\
\bottomrule
\end{tabular}%
}
\end{minipage}
\end{table*}

\clearpage

\subsubsection{Pr$^\text{3+}$ materials}

The same three-level spectral-position construction is applied to
$\mathrm{Pr}^{3+}:\mathrm{La}_2(\mathrm{WO}_4)_3$
\cite{lovric2011prla,lovric2013memory} and
$\mathrm{Pr}^{3+}:\mathrm{Y}_2\mathrm{SiO}_5$\cite{gundogan2015pryso}.
Because their hyperfine gaps are much smaller than the Eu gaps, the $\delta$ searches
use correspondingly smaller intervals: $0.30$--$5.00$ MHz for the tungstate and
$0.10$--$2.00$ MHz for YSO.  No other part of the objective is changed.  The
selected $\delta=0.1$ MHz YSO solution is exactly commensurate at the precision
of the displayed source values, so its apparent zero cost should not be read as
accuracy beyond that precision.

\begin{table*}[!tbp]
\centering
\caption{Optimized Pr$^{3+}$ material fits.}
\label{tab:pr-summary}
\scriptsize
\renewcommand{\arraystretch}{1.10}
\begin{tabular*}{0.98\textwidth}{@{\extracolsep{\fill}}cccccc}
\toprule
Material & $\delta$ (MHz) & $2\delta$ (MHz) & $T_{\rm echo}$ (ns) & $C_{\rm norm}$ & $C_{\rm abs}$ (MHz) \\
\midrule
$\mathrm{Pr}^{3+}\!:\mathrm{La}_2(\mathrm{WO}_4)_3$ & 0.405667 & 0.811333 & 1232.5 & 0.423710 & 0.171885 \\
$\mathrm{Pr}^{3+}\!:\mathrm{Y}_2\mathrm{SiO}_5$ & 0.100000 & 0.200000 & 5000.0 & 0.000000 & 0.000000 \\
\bottomrule
\end{tabular*}
\end{table*}

\begin{table*}[!tbp]
\centering
\caption{Detailed Pr-material fits, using the same row ordering and table typography as the Eu results.}
\label{tab:pr-details}
\scriptsize
\renewcommand{\arraystretch}{1.08}
\setlength{\tabcolsep}{2.8pt}
\begin{minipage}[t]{0.385\textwidth}
\centering
\textbf{$\mathrm{Pr}^{3+}\!:\mathrm{La}_2(\mathrm{WO}_4)_3$}\\[-0.15em]
$\delta=0.405667\,\mathrm{MHz}$; $C_{\rm norm}=0.423710$\\[0.25em]
\resizebox{\linewidth}{!}{%
\begin{tabular}{cccccc}
\toprule
Gap & Exp. & $m$ & Fit & $e$ & $|e|/\delta$ \\
 & (MHz) & & (MHz) & (MHz) & \\
\midrule
$\Delta_{e,2}$ & 4.940 & 12 & 4.868 & +0.072 & 0.1775 \\
$\Delta_{e,1}$ & 7.230 & 18 & 7.302 & -0.072 & 0.1775 \\
$\Delta_{g,2}$ & 24.440 & 60 & 24.340 & +0.100 & 0.2465 \\
$\Delta_{g,1}$ & 14.870 & 37 & 15.010 & -0.140 & 0.3443 \\
\bottomrule
\end{tabular}%
}
\end{minipage}
\hfill
\begin{minipage}[t]{0.385\textwidth}
\centering
\textbf{$\mathrm{Pr}^{3+}\!:\mathrm{Y}_2\mathrm{SiO}_5$}\\[-0.15em]
$\delta=0.100000\,\mathrm{MHz}$; $C_{\rm norm}=0.000000$\\[0.25em]
\resizebox{\linewidth}{!}{%
\begin{tabular}{cccccc}
\toprule
Gap & Exp. & $m$ & Fit & $e$ & $|e|/\delta$ \\
 & (MHz) & & (MHz) & (MHz) & \\
\midrule
$\Delta_{e,2}$ & 4.600 & 46 & 4.600 & -0.000 & 0.0000 \\
$\Delta_{e,1}$ & 4.800 & 48 & 4.800 & -0.000 & 0.0000 \\
$\Delta_{g,2}$ & 17.300 & 173 & 17.300 & +0.000 & 0.0000 \\
$\Delta_{g,1}$ & 10.200 & 102 & 10.200 & -0.000 & 0.0000 \\
\bottomrule
\end{tabular}%
}
\end{minipage}
\end{table*}

\begin{figure}[htbp]
\centering
\includegraphics[width=0.96\textwidth]{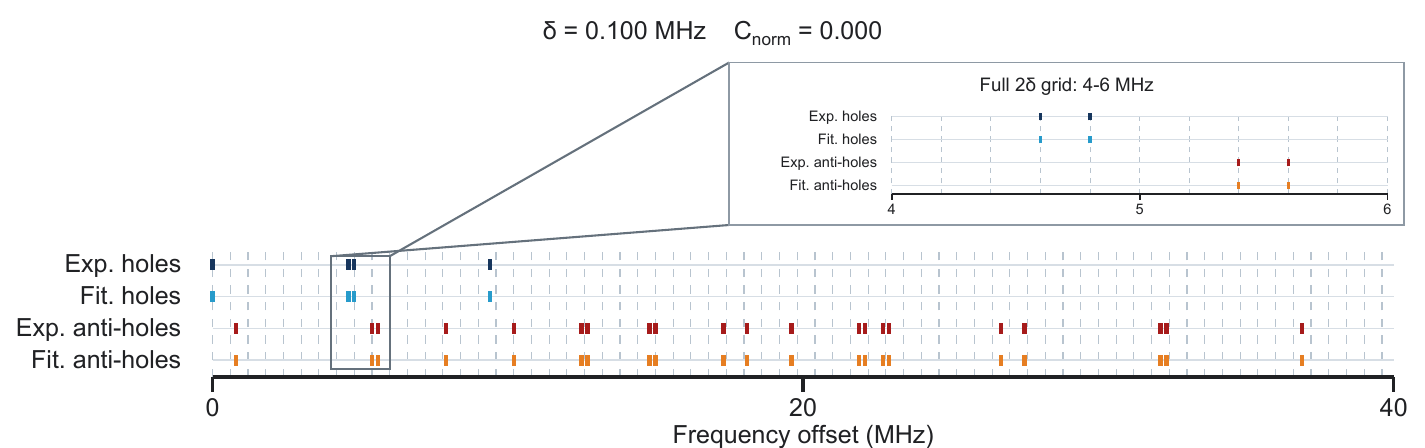}
\caption{Positive-half-spectrum comparison for $\mathrm{Pr}^{3+}:\mathrm{Y}_2\mathrm{SiO}_5$ using the extended $0.1$--$2.0$ MHz $\delta$ search interval. Dashed vertical guides in the main panel show a regularly subsampled \(2n\delta\) grid for legibility. The boxed 4--6 MHz interval is connected to the inset figure showing every grid line, with \(2\delta=0.2\) MHz.}
\label{fig:pr-spectrum}
\label{fig:pr-spectrum}
\end{figure}

\subsubsection{Tm$^{3+}$ materials}

The $^{169}\mathrm{Tm}^{3+}$ isotope has nuclear spin $I=1/2$.
Consequently, under an applied magnetic field, each selected electronic
crystal-field singlet separates into two hyperfine components and is described
by a single splitting. For
$^{169}\mathrm{Tm}^{3+}:\mathrm{Y}_3\mathrm{Al}_5\mathrm{O}_{12}$
(Tm:YAG), we use the calculated splittings of the lowest
$^{3}H_6(0)$ and $^{3}H_4(0)$ crystal-field levels reported by Goldner
\emph{et al.}\cite{goldner2006tmyag}. The first case in their Table~1,
corresponding to $\mathbf{B}_0\parallel[111]$ and sites 1, 3, and 5, gives
\[
\Delta_{g,1}=10.83~\mathrm{MHz},
\qquad
\Delta_{e,1}=1.41~\mathrm{MHz}.
\]

For two hyperfine components in each manifold, the positive-half-spectrum
contains holes at
\[
0,\qquad \Delta_{e,1},
\]
and anti-holes at
\[
\Delta_{g,1}-\Delta_{e,1},\qquad
\Delta_{g,1},\qquad
\Delta_{g,1}+\Delta_{e,1}.
\]
The HAGEM restriction therefore takes the particularly simple form
\[
\Delta_{e,1}=m_e\delta,\qquad
\Delta_{g,1}=m_g\delta,
\]
where $m_e$ is a positive even integer and $m_g$ is a positive odd integer.
Under these restrictions, the side hole is located at an even multiple of
$\delta$, while all three anti-holes are located at odd multiples of
$\delta$.

Because the excited-state splitting is only $1.41$ MHz, the $\delta$ search interval
was set to $0.1$--$2.0$ MHz. The same delta-normalized spectral-position
objective used for the Eu and Pr materials was retained. The optimum is
$\delta=0.140649$ MHz, with $(m_e,m_g)=(10,77)$,
$C_{\rm norm}=0.029086$, and $C_{\rm abs}=0.004091$ MHz. The corresponding
spectral grating distance is $2\delta=0.281299$ MHz and the ideal echo time is
$T_{\rm echo}=3.555~\mu\mathrm{s}$. The residual is smaller than the
$0.01$ MHz precision of the tabulated source splittings; the result should
therefore be regarded as effectively commensurate at the precision of the
reported input data.

\begin{table}[htbp]
\centering
\caption{Optimized two-level HAGEM fit for
$^{169}\mathrm{Tm}^{3+}:\mathrm{Y}_3\mathrm{Al}_5\mathrm{O}_{12}$ with
$\mathbf{B}_0\parallel[111]$ and sites 1, 3, and 5.}
\label{tab:tm-summary}
\scriptsize
\renewcommand{\arraystretch}{1.10}
\begin{tabular*}{\textwidth}{@{\extracolsep{\fill}}lcccccc}
\toprule
Material and field configuration
& $\delta$ (MHz)
& $2\delta$ (MHz)
& $T_{\rm echo}$ (ns)
& $(m_e,m_g)$
& $C_{\rm norm}$
& $C_{\rm abs}$ (MHz) \\
\midrule
$^{169}\mathrm{Tm}^{3+}:\mathrm{Y}_3\mathrm{Al}_5\mathrm{O}_{12}$,
$\mathbf{B}_0\parallel[111]$, sites 1, 3, 5
& 0.140649
& 0.281299
& 3554.9
& $(10,77)$
& 0.029086
& 0.004091 \\
\bottomrule
\end{tabular*}
\end{table}

\begin{table}[htbp]
\centering
\caption{Detailed two-level fit for
$^{169}\mathrm{Tm}^{3+}:\mathrm{Y}_3\mathrm{Al}_5\mathrm{O}_{12}$.
The error is defined as the experimental value minus the fitted value.}
\label{tab:tm-details}
\scriptsize
\renewcommand{\arraystretch}{1.10}
\begin{tabular*}{0.90\textwidth}{@{\extracolsep{\fill}}ccccccc}
\toprule
Gap
& Manifold
& Exp. (MHz)
& $m$
& Parity
& Fit (MHz)
& $e$ (MHz) \\
\midrule
$\Delta_{e,1}$ & $^{3}H_4(0)$ & 1.41000  & 10 & even & 1.40649  & $+0.00351$ \\
$\Delta_{g,1}$ & $^{3}H_6(0)$ & 10.83000 & 77 & odd  & 10.83000 & $-0.00000$ \\
\bottomrule
\end{tabular*}
\end{table}

\begin{figure}[htbp]
\centering
\includegraphics[width=0.96\textwidth]{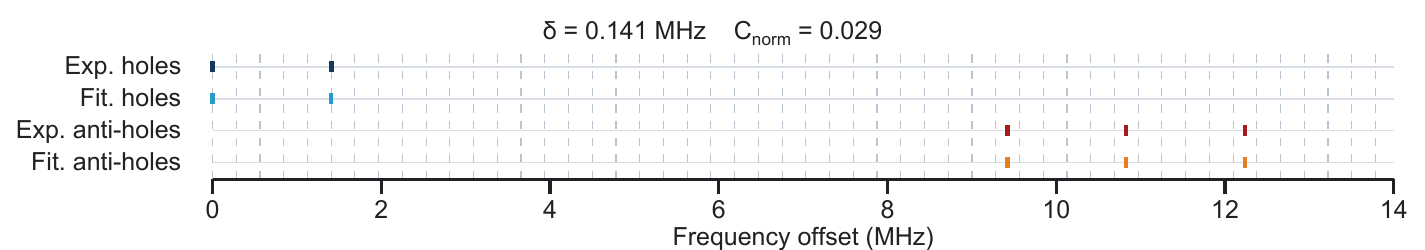}
\caption{Positive-half-spectrum comparison for
$^{169}\mathrm{Tm}^{3+}:\mathrm{Y}_3\mathrm{Al}_5\mathrm{O}_{12}$ with
$\mathbf{B}_0\parallel[111]$ and sites 1, 3, and 5. The experimental and
fitted hole and anti-hole locations are shown separately. Dashed lines mark
the $2n\delta$ reference grid.}
\label{fig:tm-yag-spectrum}
\end{figure}

\clearpage

\subsection{Kramer ion materials}

\subsubsection{Er$^\text{3+}$ materials}

Scenario 1: for $^{167}$Er$^{3+}$: YSO crystal, only the optical transitions $\Delta m_I =0, \pm 1$ are considered and their transition frequencies are separated among three different $\Delta m_I$. If we plan to use $\Delta m_I = 0$ for broadband storage, the side holes are all located outside the transition frequency region of $\Delta m_I = 0$, and the anti-holes within this region are located at [$\pm\left( \Delta e_{1} - \Delta g_{1} \right)$, $\pm\left( \Delta e_{2} - \Delta g_{2} \right)$, $\pm\left( \Delta e_{3} - \Delta g_{3} \right)$, $\pm\left( \Delta e_{4} - \Delta g_{4} \right)$, $\pm\left( \Delta e_{5} - \Delta g_{5} \right)$, $\pm\left( \Delta e_{6} - \Delta g_{6} \right)$, $\pm\left( \Delta e_{7} - \Delta g_{7} \right)$]. To implement HAGEM, at least one anti-hole is located at $k_1*\delta$, where $k_1$ is an odd integer (positive or negative). The rest are located at $k_i*\delta$, where $k_i$ is an odd or even integer (positive or negative). If more anti-holes are located at odd multiples of $\delta$, the HAGEM spectrum is more balanced in frequency domain and requires shorter spectral preparation time.

The resulting positive-half-spectrum comparison is shown in
Fig.~\ref{fig:er-scenario1-spectrum}. At $\delta=1$ MHz, all seven
transition-specific anti-hole multipliers are odd and the experimental and
fitted positions coincide at the precision of the published splittings. The
two transition-specific anti-holes at 93 MHz overlap and are therefore shown
as one spectral location.

\begin{figure}[htbp]
\centering
\includegraphics[width=0.96\textwidth]{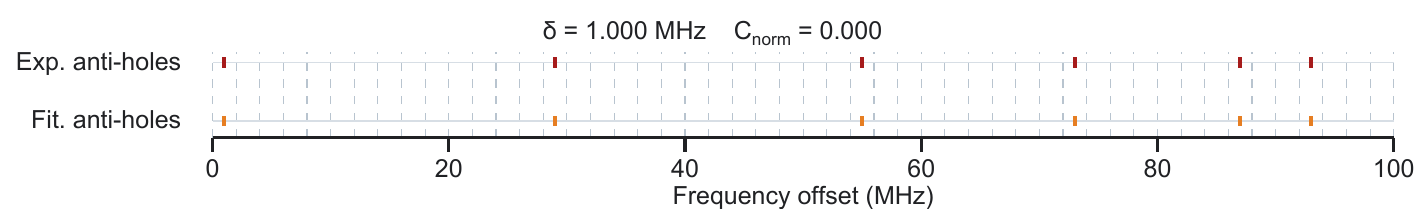}
\caption{Positive-half-spectrum comparison for the first
$^{167}\mathrm{Er}^{3+}:\mathrm{Y}_2\mathrm{SiO}_5$ site-2 scenario.
Only the anti-holes within the central $\Delta m_I=0$ storage-frequency
region are fitted; side holes outside this optical branch are excluded.
For $\delta=1$ MHz, the seven transition-specific multipliers are
$(1,29,55,73,87,93,93)$ and are all odd. The coincident 93 MHz
anti-holes are displayed as one spectral location. Dashed vertical guides mark the $2n\delta$ reference grid.}
\label{fig:er-scenario1-spectrum}
\end{figure}

Because
only the central $\Delta m_I=0$ optical transitions are used, its side holes lie
outside the selected storage-frequency region and are not included in the fit.
The relevant positive anti-hole offsets are therefore
\begin{equation}
a_i=\left|\Delta e_i-\Delta g_i\right|,
\qquad i=1,\ldots,7.
\end{equation}
Coincident positions are merged before comparison. Consequently, the
normalized objective for this scenario is
\begin{equation}
C_{\rm norm}^{(1)}
=
\frac{D_{\rm anti}+P_{\rm anti}}{\delta},
\end{equation}
where $D_{\rm anti}$ is the mean one-to-one displacement between the
experimental and fitted anti-hole positions and $P_{\rm anti}$ is the
corresponding cardinality penalty.

The published hyperfine splittings give the seven transition-specific offsets
\begin{equation}
(1,29,55,73,87,93,93)\ \mathrm{MHz}.
\end{equation}
For $\delta=1$ MHz, the associated integer multipliers are
\begin{equation}
(k_1,k_2,k_3,k_4,k_5,k_6,k_7)
=
(1,29,55,73,87,93,93),
\end{equation}
so all seven multipliers are odd. The two 93 MHz contributions correspond to
one merged spectral location. The experimental and fitted anti-hole positions
therefore coincide exactly at the precision of the tabulated splittings,
giving $C_{\rm norm}^{(1)}=0$ and an ideal echo time of
$T_{\rm echo}=500$ ns.

This zero-cost solution is subharmonically degenerate:
\begin{equation}
\delta=\frac{1}{q}\ \mathrm{MHz},
\qquad q=1,3,5,\ldots,
\end{equation}
also retains integer, odd-valued multipliers for all seven anti-holes. We
report $\delta=1$ MHz because it is the largest member of this exact family
and requires the smallest multipliers. A smaller member may instead be
selected when a longer echo time is preferred. The exact numerical agreement
should not be interpreted more precisely than the approximately 5 MHz
uncertainty of the source level assignments.

\begin{table*}[!tbp]
\centering
\caption{Optimized
$^{167}\mathrm{Er}^{3+}:\mathrm{Y}_2\mathrm{SiO}_5$ site-2 fits.
Scenario 1 fits only the anti-holes within the central $\Delta m_I=0$
storage-frequency region. Scenario 2 fits the combined
$\Delta m_I=0,\pm1$ hole and anti-hole lists.}
\label{tab:er-summary}
\scriptsize
\renewcommand{\arraystretch}{1.15}
\begin{tabular*}{\textwidth}{@{\extracolsep{\fill}}lcccccc}
\toprule
Scenario &
Fitted line families &
$\delta$ (MHz) &
$2\delta$ (MHz) &
$T_{\rm echo}$ (ns) &
$C_{\rm norm}$ &
$C_{\rm abs}$ (MHz) \tabularnewline
\midrule
Central $\Delta m_I=0$ &
Anti-holes only &
1.000000 &
2.000000 &
500.0 &
$<10^{-12}$ &
$<10^{-12}$ \tabularnewline
Combined $\Delta m_I=0,\pm1$ &
Holes and anti-holes &
0.500000 &
1.000000 &
1000.0 &
0.127660 &
0.063830 \tabularnewline
\bottomrule
\end{tabular*}
\end{table*}

\begin{figure}[htbp]
\centering
\includegraphics[width=0.96\textwidth]
{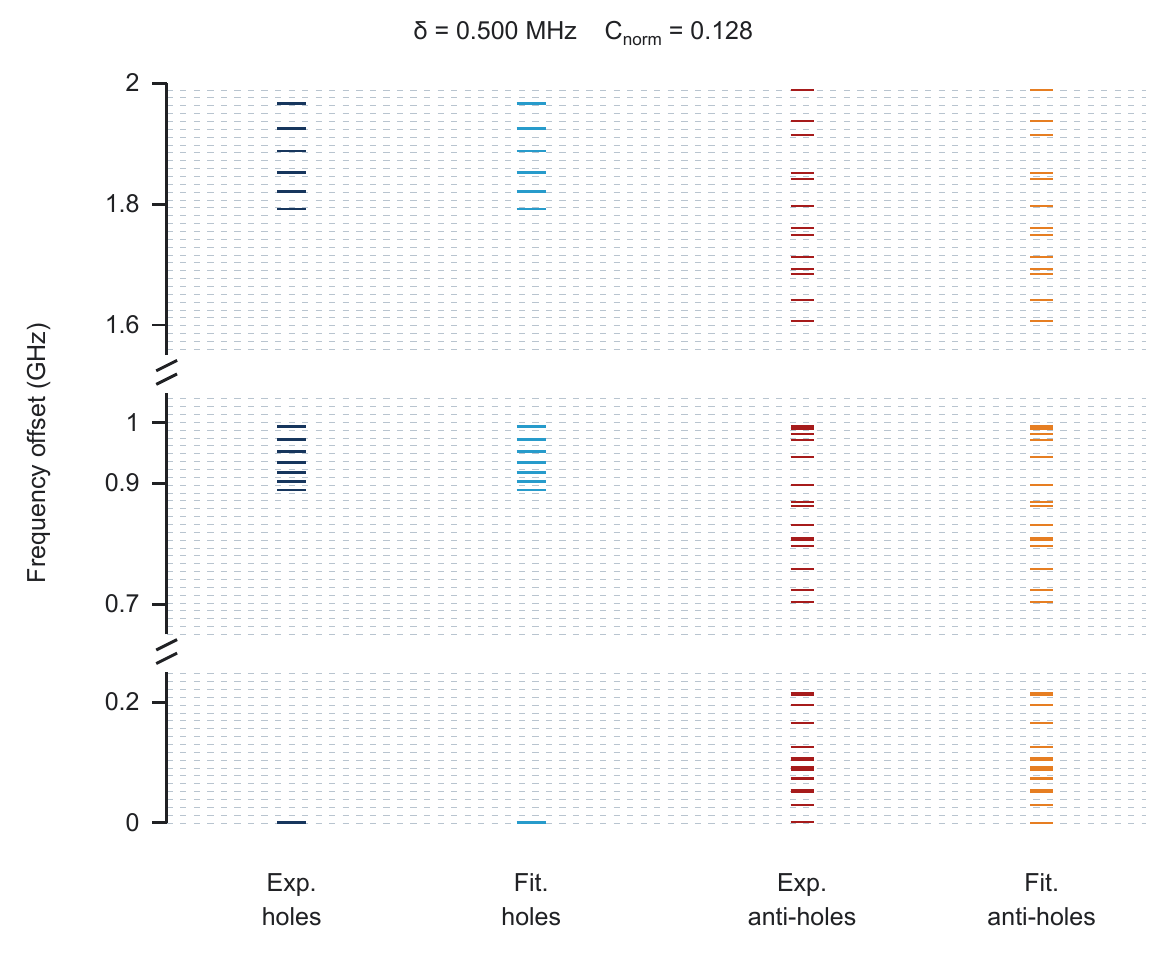}
\caption{Positive-half-spectrum comparison for the implemented second
$^{167}\mathrm{Er}^{3+}:\mathrm{Y}_2\mathrm{SiO}_5$ site-2 scenario,
combining the $\Delta m_I=0,\pm1$ branches. The vertical axis is frequency
offset and the four line families are separated horizontally. The broken vertical axis omits the empty
$0.25$--$0.65$ GHz and $1.05$--$1.55$ GHz intervals. Dashed horizontal guides show a regularly subsampled
$2n\delta$ grid for legibility.}
\label{fig:er-scenario2-spectrum}
\end{figure}

Scenario 2: for $^{167}$Er$^{3+}$: YSO crystal, only the optical transitions $\Delta m_I =0, \pm 1$ are considered. If laser frequency is resonant with optical transitions $\Delta m_I =0$, there are 14 side holes, located at $\pm \Delta e_i$ (i = 1-7); there are 40 anti-holes, located at $\pm \Delta g_i$ \& $\pm (\Delta e_i - \Delta g_i)$ (i = 1-7) and $\pm (\Delta e_i + \Delta e_{i+1} - \Delta g_i)$ \& $\pm (\Delta g_{i+1} - \Delta e_i - \Delta e_{i+1})$ (i = 1-6). If laser frequency is resonant with optical transitions $\Delta m_I =-1$, there are 13 side holes, located at $\Delta e_i$ (i = 1-7) \& $\Delta e_i + \Delta e_{i+1}$ (i = 1-6); there are 37 anti-holes, located at $\Delta g_i$ (i = 1-7), $(\Delta e_i - \Delta g_{i+1})$ (i = 1-6), $(\Delta e_i + \Delta g_i)$ (i = 1-7), $(\Delta g_i + \Delta g_{i+1})$ (i = 1-6), $(\Delta g_i + \Delta g_{i+1} - \Delta e_i)$ (i = 1-6) and $(\Delta g_{i+1} + \Delta g_{i+2} - \Delta e_i - \Delta e_{i+1})$ (i = 1-5). If laser frequency is resonant with optical transitions $\Delta m_I =1$, there are 13 side holes, located at $-\Delta e_i$ (i = 1-7) \& $-(\Delta e_i + \Delta e_{i+1})$ (i = 1-6); there are 37 anti-holes, located at $-\Delta g_i$ (i = 1-7), $(\Delta e_{i+1}- \Delta g_{i})$ (i = 1-6), $-(\Delta e_i + \Delta g_i)$ (i = 1-7), $-(\Delta g_i + \Delta g_{i+1})$ (i = 1-6), $(-\Delta g_i - \Delta g_{i+1} + \Delta e_i)$ (i = 1-6) and $(-\Delta g_{i} - \Delta g_{i+1} + \Delta e_{i+1} + \Delta e_{i+2})$ (i = 1-5). In total, there are 26 side holes and 114 anti-holes. In general, if the excited states and the ground states both have $n$ hyperfine levels, there are $4n-6$ side holes and $18n-30$ anti-holes ($n \geq 4$). To implement HAGEM, at least one ground state hyperfine level gap is odd multiples of $\delta$ \& the rest can be even or odd multiples of $\delta$, and all excited state hyperfine level gaps are even multiples of $\delta$.

The source level structure for
$^{167}\mathrm{Er}^{3+}:\mathrm{Y}_2\mathrm{SiO}_5$ site 2 is taken from the
supplementary spectroscopy of Ran\v{c}i\'c
\emph{et al.}\cite{rancic2018er}. Unlike the three-level cases, the eight
hyperfine energies in each manifold and the selected
$\Delta m_I=0,\pm1$ optical branches require explicit branch-dependent line
rules. The second scenario below has been implemented: all three branches are
combined, coincident lines are merged, and the resulting 26 raw side holes and
114 raw anti-holes are fitted with the common one-to-one objective. Independent $\delta$ windows from $20$--$100$ MHz down to $0.2$--$0.5$ MHz were searched because the
larger-$\delta$ solutions did not give an adequate line-position match. The
best representative reported is $\delta=0.5$ MHz with $C_{\rm norm}=0.127660$.

\subsubsection{Yb$^\text{{3+}}$ materials} 

For $^{171}\mathrm{Yb}^{3+}:\mathrm{CaWO}_4$\cite{tiranov2026yb}, the
zero-field degeneracy gives three unique hyperfine energies per manifold and
the three-level workflow is used.  The $\delta$ search interval is $0.72$--$36$ MHz,
scaled from its 72 MHz smallest gap. For
$^{171}\mathrm{Yb}^{3+}:\mathrm{Y}_2\mathrm{SiO}_5$ site II
\cite{businger2022non}, four hyperfine energies require three adjacent gaps in
each manifold.  Holes are generated from all nonzero pair separations in the
excited manifold (plus zero), while anti-holes combine every nonzero ground
pair separation with every signed excited pair separation. The $\delta$ interval is
$2.88$--$144$ MHz, scaled from its 288 MHz smallest gap.  These wider absolute
ranges are necessary because the Yb splittings are in the GHz regime; the cost
definition and normalization remain unchanged.  The generalized position
construction produces the positive-half-spectrum shown in
Fig.~\ref{fig:yb-spectrum}.

\begin{table}[htbp]
\centering
\caption{Optimized $^{171}$Yb$^{3+}$ material fits.}
\label{tab:yb-summary}
\scriptsize
\renewcommand{\arraystretch}{1.10}
\begin{tabular*}{0.98\textwidth}{@{\extracolsep{\fill}}cccccc}
\toprule
Material & $\delta$ (MHz) & $2\delta$ (MHz) & $T_{\rm echo}$ (ns) & $C_{\rm norm}$ & $C_{\rm abs}$ (MHz) \\
\midrule
$^{171}\mathrm{Yb}^{3+}\!:\mathrm{Y}_2\mathrm{SiO}_5$ (site II) & 16.003778 & 32.007556 & 31.2 & 0.260749 & 4.172976 \\
$^{171}\mathrm{Yb}^{3+}\!:\mathrm{CaWO}_4$ & 12.094828 & 24.189655 & 41.3 & 0.065404 & 0.791051 \\
\bottomrule
\end{tabular*}
\end{table}

\begin{figure}[htbp]
\centering
    \includegraphics[
        width=1.0\textwidth,
        height=0.68\textheight,
        keepaspectratio
    ]{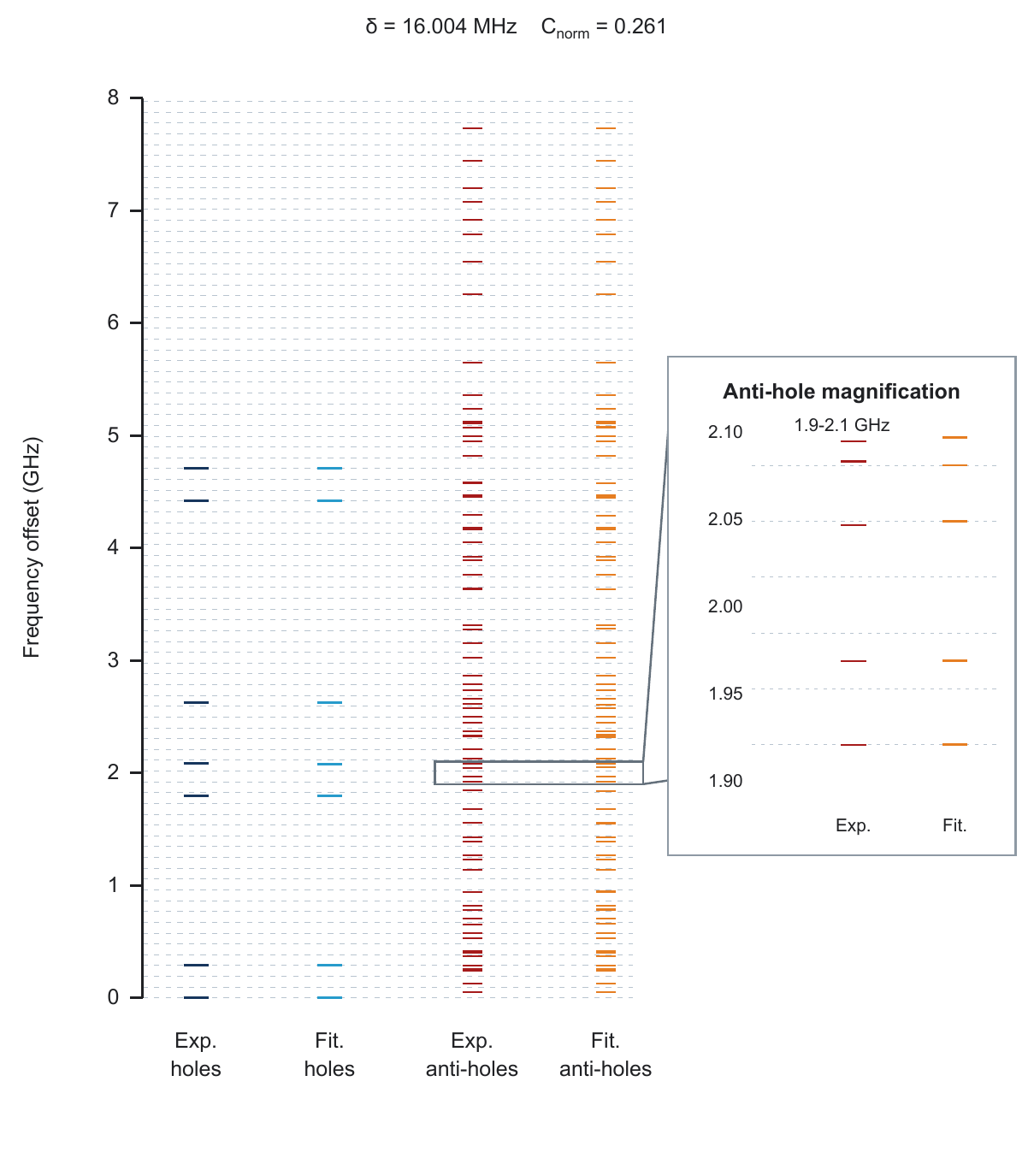}
    \caption{Positive-half-spectrum comparison for
$^{171}\mathrm{Yb}^{3+}:\mathrm{Y}_2\mathrm{SiO}_5$ site II.
The boxed 1.9--2.1 GHz region is connected to a right-hand magnified
inset comparing experimental and fitted anti-holes. Dashed horizontal
guides show a regularly subsampled $2n\delta$ grid for legibility.}
\label{fig:yb-spectrum}
\end{figure}

\begin{table}[htbp]
\centering
\caption{Detailed Yb-material fits. The three-gap site-II rows are ordered $\Delta_{e,3}$ to $\Delta_{e,1}$ and $\Delta_{g,3}$ to $\Delta_{g,1}$.}
\label{tab:yb-details}
\scriptsize
\renewcommand{\arraystretch}{1.08}
\setlength{\tabcolsep}{2.8pt}
\begin{minipage}[t]{0.485\textwidth}
\centering
\textbf{$^{171}\mathrm{Yb}^{3+}\!:\mathrm{Y}_2\mathrm{SiO}_5$ (site II)}\\[-0.15em]
$\delta=16.003778\,\mathrm{MHz}$; $C_{\rm norm}=0.260749$\\[0.25em]
\resizebox{\linewidth}{!}{%
\begin{tabular}{cccccc}
\toprule
Gap & Exp. & $m$ & Fit & $e$ & $|e|/\delta$ \\
 & (MHz) & & (MHz) & (MHz) & \\
\midrule
$\Delta_{e,3}$ & 2623.000 & 164 & 2624.620 & -1.620 & 0.1012 \\
$\Delta_{e,2}$ & 1794.000 & 112 & 1792.423 & +1.577 & 0.0985 \\
$\Delta_{e,1}$ & 288.000 & 18 & 288.068 & -0.068 & 0.0042 \\
$\Delta_{g,3}$ & 528.800 & 33 & 528.125 & +0.675 & 0.0422 \\
$\Delta_{g,2}$ & 1842.000 & 115 & 1840.434 & +1.566 & 0.0978 \\
$\Delta_{g,1}$ & 654.700 & 41 & 656.155 & -1.455 & 0.0909 \\
\bottomrule
\end{tabular}%
}
\end{minipage}
\hfill
\begin{minipage}[t]{0.485\textwidth}
\centering
\textbf{$^{171}\mathrm{Yb}^{3+}\!:\mathrm{CaWO}_4$}\\[-0.15em]
$\delta=12.094828\,\mathrm{MHz}$; $C_{\rm norm}=0.065404$\\[0.25em]
\resizebox{\linewidth}{!}{%
\begin{tabular}{cccccc}
\toprule
Gap & Exp. & $m$ & Fit & $e$ & $|e|/\delta$ \\
 & (MHz) & & (MHz) & (MHz) & \\
\midrule
$\Delta_{e,2}$ & 72.000 & 6 & 72.569 & -0.569 & 0.0470 \\
$\Delta_{e,1}$ & 2734.000 & 226 & 2733.431 & +0.569 & 0.0470 \\
$\Delta_{g,2}$ & 1935.400 & 160 & 1935.172 & +0.228 & 0.0188 \\
$\Delta_{g,1}$ & 1148.400 & 95 & 1149.009 & -0.609 & 0.0503 \\
\bottomrule
\end{tabular}%
}
\end{minipage}
\end{table}

\subsection{Summary}

All the materials simulated above can be found with some $\delta$ values to implement HAGEM. Especially with small $\delta$ values, nearly perfect HAGEM matching conditions can be reached, which shows a wide usage of HAGEM with most RE ion materials. 

There are some special cases, in which nearly perfect HAGEM spectrum can never be created. For example, all the ground state hyperfine level splittings are integer multiples of some of the excited state hyperfine level splittings, of which a special case: all the hyperfine level separations are identical; some of the excited state hyperfine levels are odd integer multiples of all the ground state hyperfine level splittings. These cases can rarely happen.

\section{Experimental setups \& analysis}

\subsection{Spectral hole burning}
Here I show a simple example to explain persistent spectral hole burning. If we want to initialize the atoms into one hyperfine state, we send laser pulses with frequencies f$_\text{1}$ and f$_\text{2}$ resonant with two optical transition, the atoms will go to the excited states and they will decay into all three ground states, if the pumping process keeps for some time, the atoms will be initialized into one ground state, as illustrated in Fig. \ref{Fig_PersistentSHB}. Spectral hole burning is persistent burning with a single laser frequency. 

\begin{figure}[htbp]
\centerline{\includegraphics[width=0.8\columnwidth]{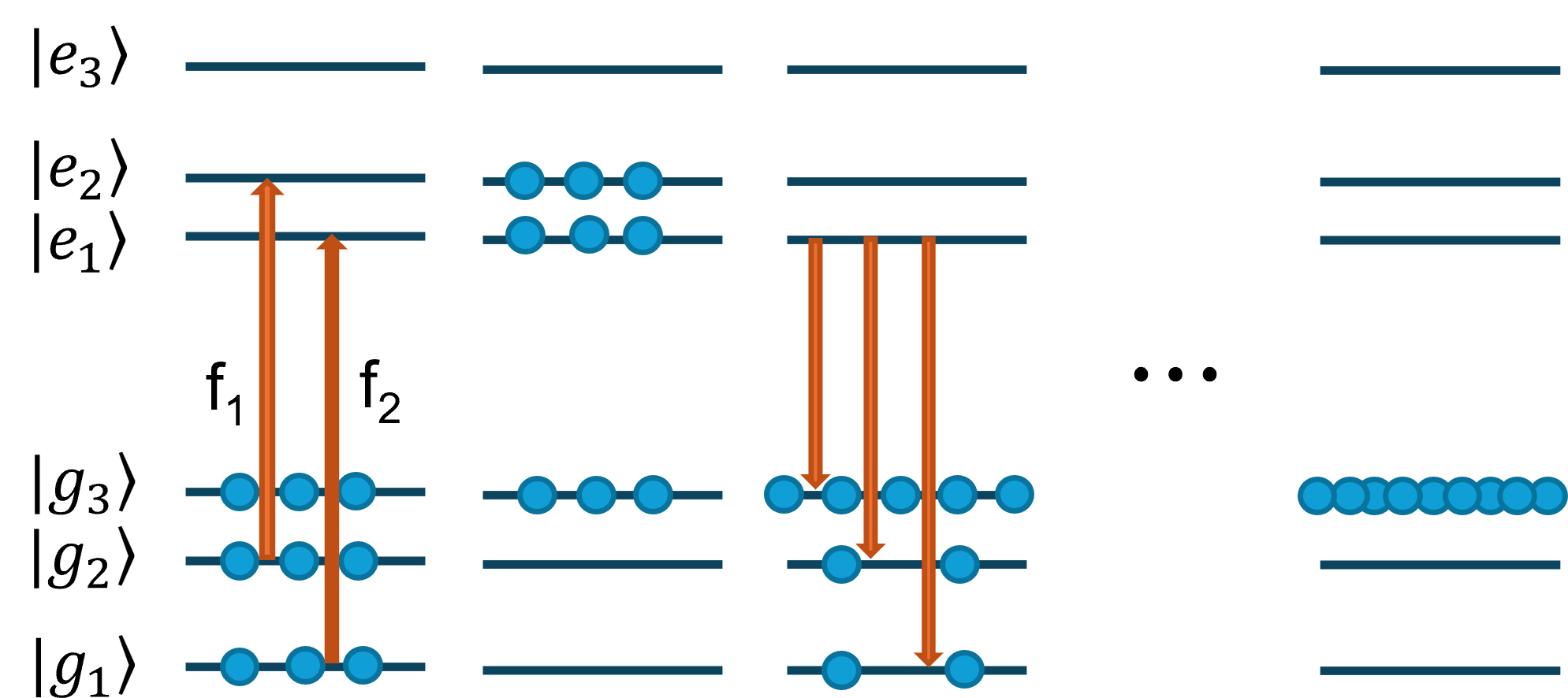}}
	\caption{Persistent spectral hole burning with two frequencies f$_\text{1}$ and f$_\text{2}$.}
	\label{Fig_PersistentSHB}
\end{figure}

\subsection{Determination of hyperfine state splittings}

\begin{figure}[htbp]
\centerline{\includegraphics[width=0.8\columnwidth]{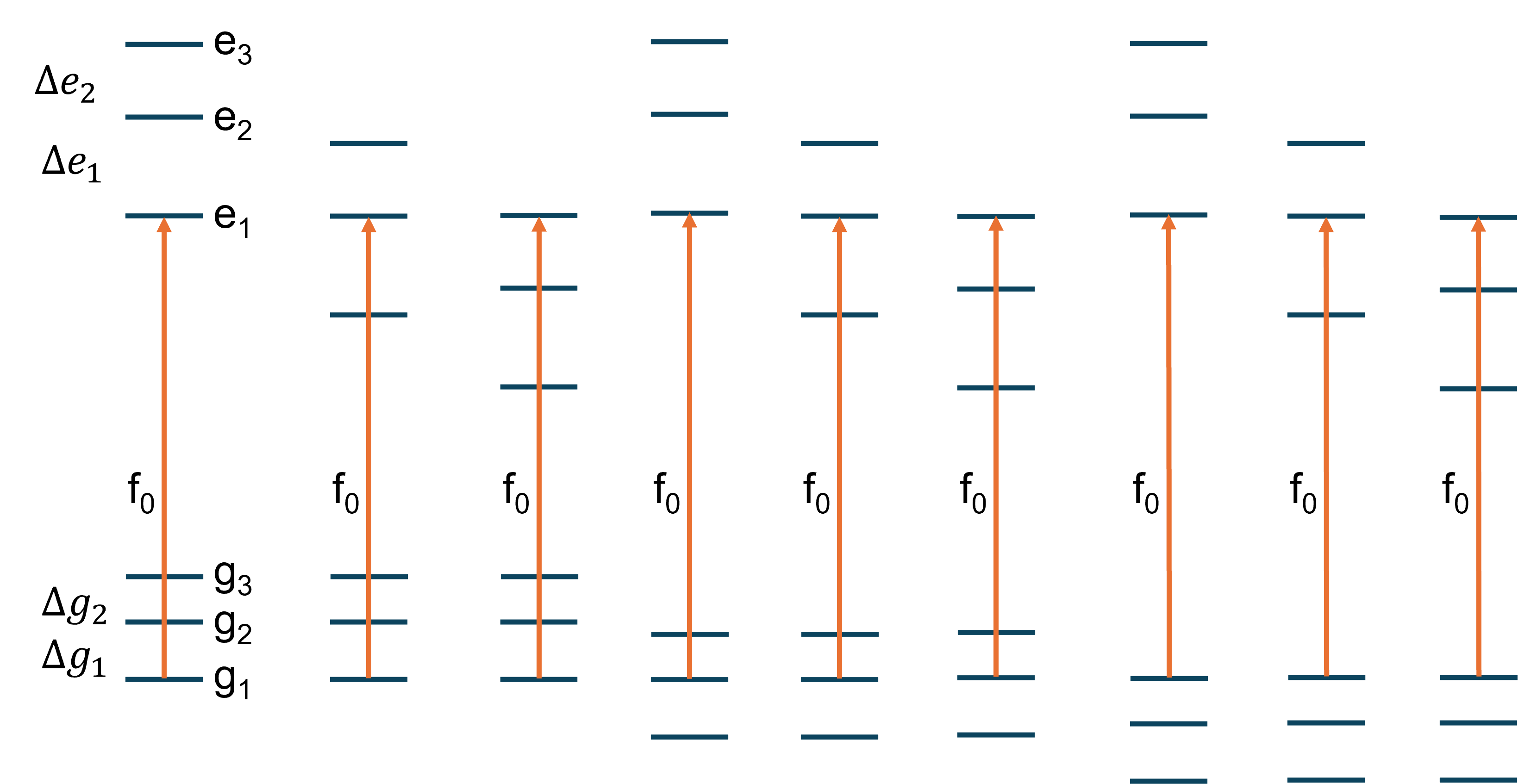}}
	\caption{Nine classes of ions due to inhomogeneous broadening in solids.}
	\label{Fig_NineClass}
\end{figure}

When RE ions are placed inside a solid-state material, at each Eu$^{3+}$ site, the electromagnetic environments are slightly different, which will shift the energy levels differently, as a result the optical transitions have different frequencies. This is called inhomogeneous broadening. Due to this effect, when a laser pulse with frequency f$_\text{0}$ interacts with the atomic medium, there are optical transitions corresponding to nine classes of ions, as shown in Fig. \ref{Fig_NineClass}.

Let's analyze spectral hole burning with one class of ions as shown in Fig. \ref{Fig_OneClassPump}. The laser with frequency f$_\text{0}$ is resonant with the optical transition between g$_\text{1}$ and e$_\text{1}$. The laser pulse will excite the atoms into the excited state e$_\text{1}$, then the atoms will decay into all three ground states. If the process continues for some time, the atoms will be shifted from g$_\text{1}$ to g$_\text{2}$ and g$_\text{2}$. There will be less atoms in g$_\text{1}$, so there will be less absorption between g$_\text{1}$ and e$_\text{1}$, between g$_\text{1}$ and e$_\text{2}$, \& between g$_\text{1}$ and e$_\text{3}$. In result, there will be three spectral holes at frequencies of f$_\text{0}$, f$_\text{0}$ + $\Delta e_{1}$ and f$_\text{0}$ + $\Delta e_{1}$ + $\Delta e_{2}$. At the same time, there will be more atoms in g$_\text{2}$ and g$_\text{3}$, so there will be more absorption from the two ground states to the three excited states, which will lead to six spectral anti-holes at frequencies of f$_\text{0}$ - $\Delta g_{1}$, f$_\text{0}$ - $\Delta g_{1}$ + $\Delta e_{1}$, f$_\text{0}$ - $\Delta g_{1}$ + $\Delta e_{1}$ + $\Delta e_{2}$, f$_\text{0}$ - $\Delta g_{1}$ - $\Delta g_{2}$, f$_\text{0}$ - $\Delta g_{1}$ - $\Delta g_{2}$ + $\Delta e_{1}$, and f$_\text{0}$ - $\Delta g_{1}$ - $\Delta g_{2}$ + $\Delta e_{1}$ + $\Delta e_{2}$.

\begin{figure}[!h]
\centerline{\includegraphics[width=0.8\columnwidth]{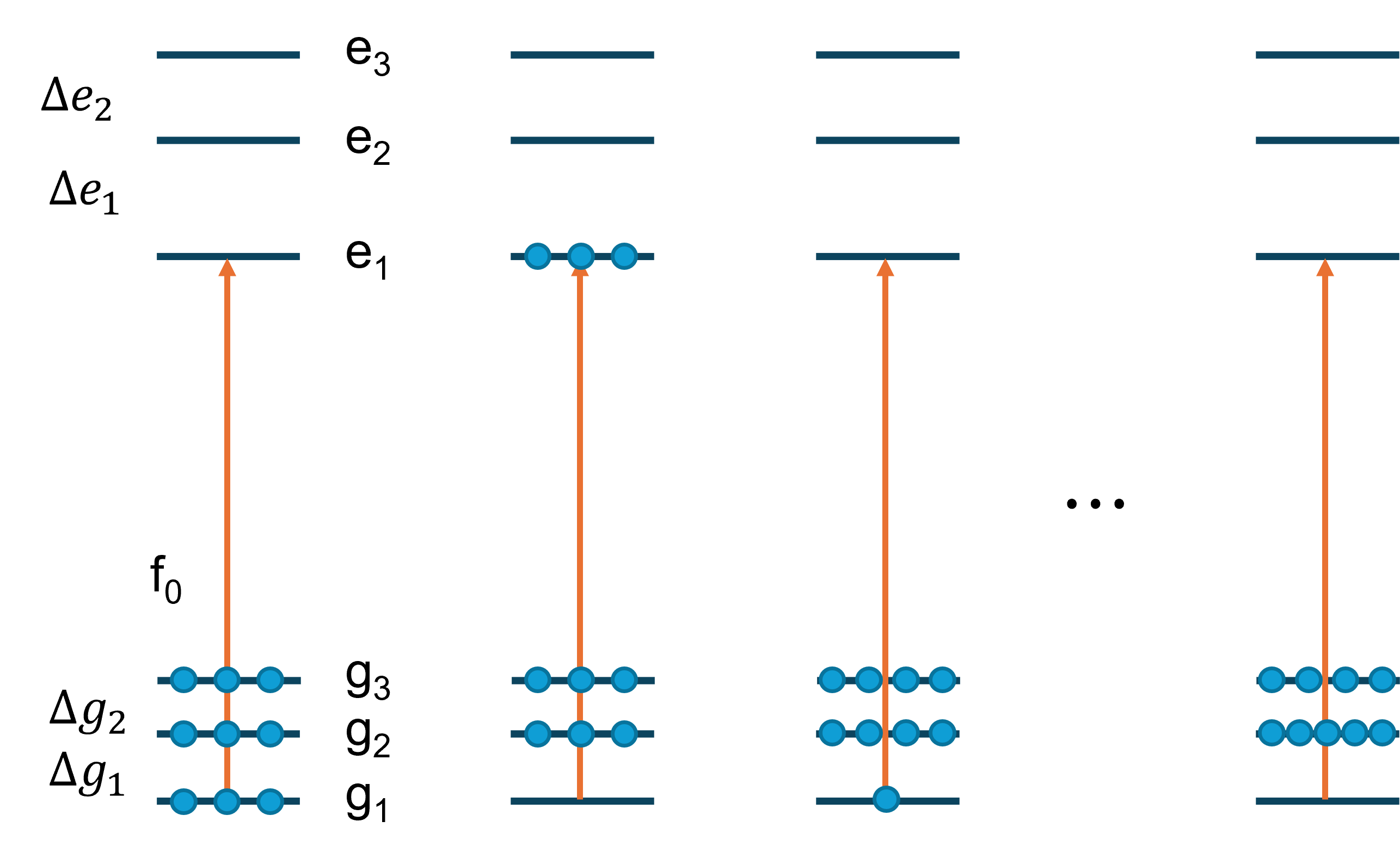}}
	\caption{Spectral hole burning with one class of ions and analysis of the side holes and anti-holes.}
	\label{Fig_OneClassPump}
\end{figure}

Analyzing all nine classes of ions, there are 6 side holes located around f$_\text{0}$ deviated by values equal to terms in [$\pm\Delta e_{1}$, $\pm\Delta e_{2}$, $\pm\left( \Delta e_{1} + \Delta e_{2} \right)$], and 42 anti-holes located around f$_\text{0}$ deviated by values equal to one term in [$\pm\Delta g_{1}$, $\pm\Delta g_{2}$, $\pm\left( \Delta g_{1} + \Delta g_{2} \right)$] + one term in [$0$, $\pm\Delta e_{1}$, $\pm\Delta e_{2}$, $\pm\left( \Delta e_{1} + \Delta e_{2} \right)$].

\begin{figure}[!h]
\centerline{\includegraphics[width=0.8\columnwidth]{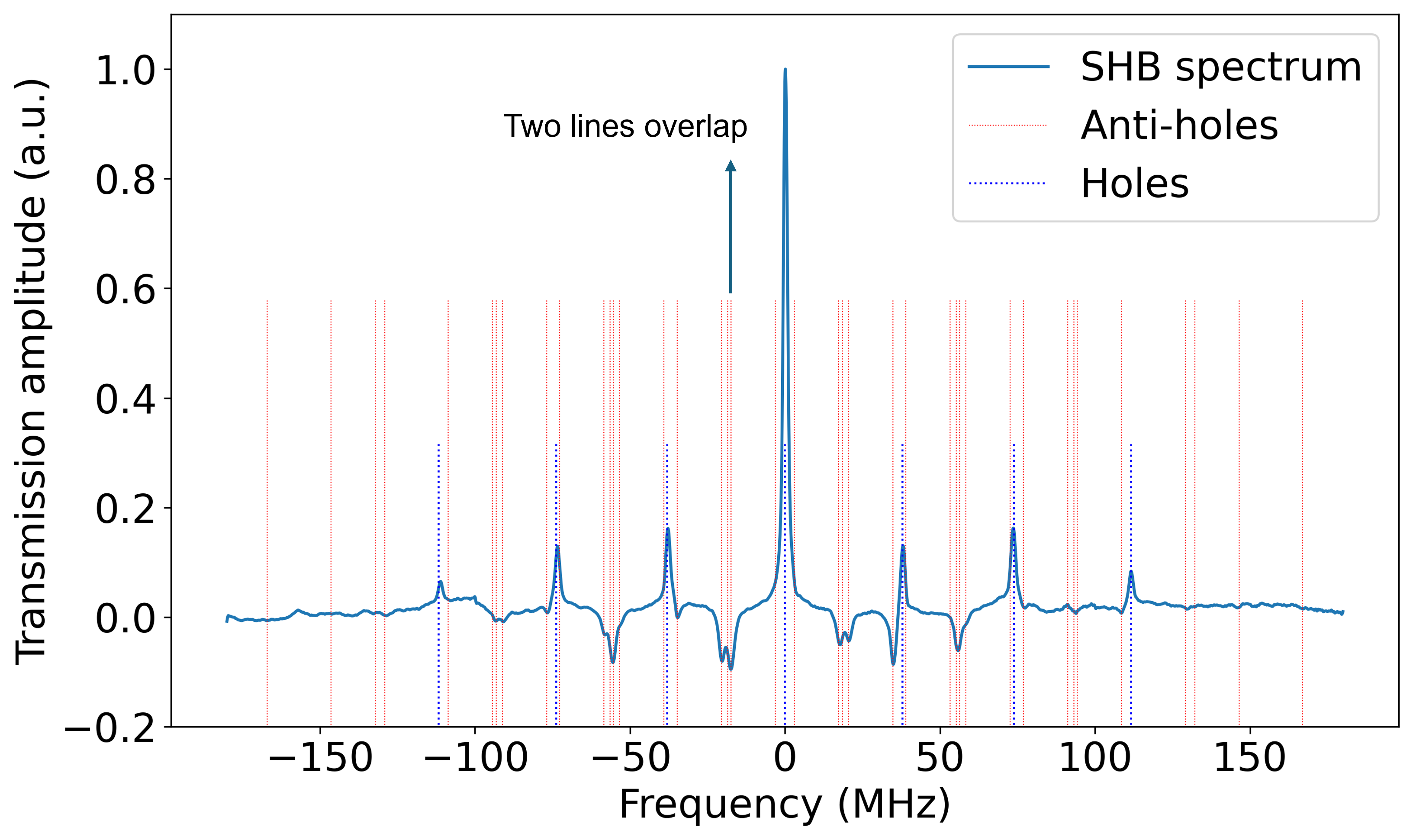}}
	\caption{Hyperfine level splitting: the blue curve is the experimental SHB spectrum; the dashed blue lines are the simulated locations of spectral holes; the dashed red lines are the simulated locations of anti-holes.}
	\label{Fig_HyperfineSplitting}
\end{figure}

As explained above, the locations of side holes and anti-holes can be easily calculated. By simulating different values of $\Delta e_{1}$, $\Delta e_{2}$, $\Delta g_{1}$ and $\Delta g_{2}$, we try to find the best match between the simulated spectral hole locations and the experimental values obtained from the SHB spectrum. As shown in Fig. \ref{Fig_HyperfineSplitting}, the spectral holes are located in blue dashed lines, and the anti-holes are in orange dashed lines.

\subsection{Simulation of SHB spectrum}

\begin{figure}[!h]
\centerline{\includegraphics[width=1\columnwidth]{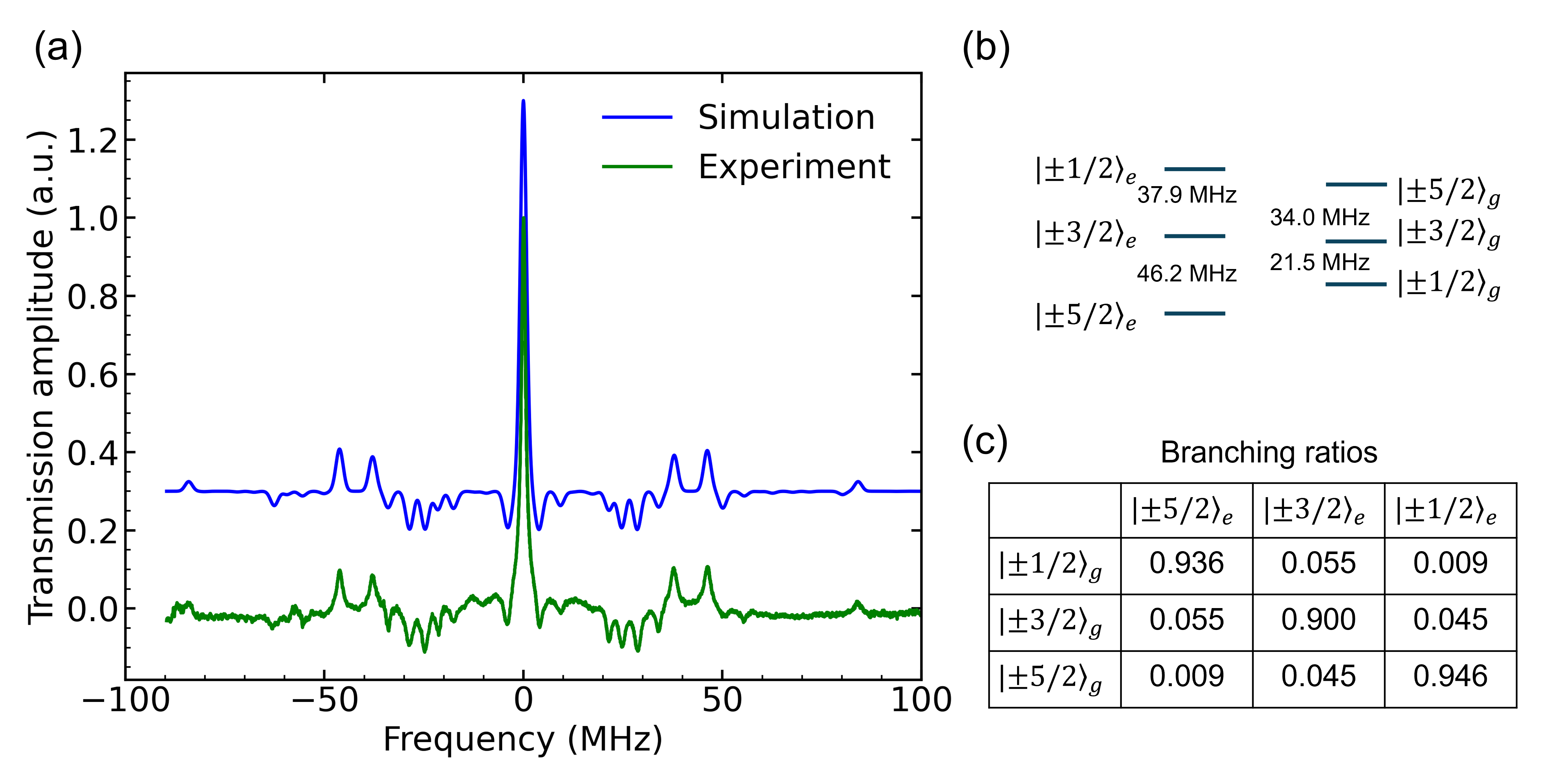}}
	\caption{SHB spectrum simulation for [Eu$^\text{{3+}}$(BA)$_\text{4}$(pip)]: (a) A single spectral hole burning: experimental and simulated spectrum of 200~MHz. (b) Hyperfine level separation values. (c) Branching ratios from the spectral simulation.}
	\label{Fig2_SHB520}
\end{figure}

Eu$^{3+}$ ion has a nuclear spin of 5/2. Without external magnetic field, the energy states of $^{5}$D$_{0}$ and $^{5}$F$_{0}$ have three degenerate levels. Considering optical transitions between $^{5}$D$_{0}$ and $^{5}$F$_{0}$, a long laser pulse is sent through the atomic ensemble. Due to inhomogeneous broadening, the laser pulse interacts with nine classes of ions. The class of ions with the transition from the i level to the j level is resonant with the applied laser, and its branching ratio is BR$_{ij}$ (here we can call the three ground levels as i, k $\&$ h, and the three excited state as j, m $\&$ n.). After a pumping period of t, some amount of the atoms at i level will be shifted into other ground levels. Each class of ions initially has N$_{0}$ atoms. $dN_{i}(t)/dt \propto -N_{i}(t)*BR_{ij}*2/3$, solving the equation, $N(t)=N_{0}e^{-2/3BR_{ij}t}$. There are many energy states of J and crystal field splittings, which makes the atoms equally decay from the excited state to the three ground states, so it means that 2/3 of the atoms decay into the other two ground levels. $\Delta N_{i}(t)=N_{0}(e^{-2/3BR_{ij}t}-1)$, and the other two ground levels $\Delta N_{k}(t) = \Delta N_{h}(t) = N_{0}/2(1-e^{-2/3BR_{ij}t})$. The center hole is at frequency f$_{0}$ with effective number of atoms $Neff_{ij}(t)=BR_{ij}N_{0}(e^{-2/3BR_{ij}t}-1)$, the two side holes corresponding to the transitions from level i to levels m and n are at frequencies f$_{0}$+E$_{jm}$ and f$_{0}$+E$_{jn}$, where E$_{jm}$ and E$_{jn}$ are the energy gaps between level j to levels m and n  respectively, and their effective numbers of atoms are $Neff_{im}(t)=BR_{im}N_{0}(e^{-2/3BR_{ij}t}-1)$ and $Neff_{in}(t)=BR_{in}N_{0}(e^{-2/3BR_{ij}t}-1)$ respectively. There are three antiholes corresponding to the transitions from level k to levels j, m $\&$ n at frequencies f$_{0}$+E$_{ik}$, f$_{0}$+E$_{ik}$+E$_{jm}$ and f$_{0}$+E$_{ik}$+E$_{jn}$, and their effective number of atoms are $Neff_{kj}(t)=BR_{kj}N_{0}/2(1-e^{-2/3BR_{ij}t})$, $Neff_{km}(t)=BR_{km}N_{0}/2(1-e^{-2/3BR_{ij}t})$ and $Neff_{kn}(t)=BR_{kn}N_{0}/2(1-e^{-2/3BR_{ij}t})$. In addition, there are three antiholes corresponding to the transitions from level h to levels j, m $\&$ n at frequencies f$_{0}$+E$_{ih}$, f$_{0}$+E$_{ih}$+E$_{jm}$ and f$_{0}$+E$_{ih}$+E$_{jn}$, and their effective number of atoms are $Neff_{hj}(t)=BR_{hj}N_{0}/2(1-e^{-2/3BR_{ij}t})$, $Neff_{hm}(t)=BR_{hm}N_{0}/2(1-e^{-2/3BR_{ij}t})$ and $Neff_{hn}(t)=BR_{hn}N_{0}/2(1-e^{-2/3BR_{ij}t})$. Due to spectral diffusion, $Neff_{ij}(f)$ with pumping time t is modeled as a simple exponential function $Neff_{ij}(f)=Neff_{ij}(t)/\sigma*e^{-(f-f_{h})^2/\sigma^2}$, where $f_{h}$ is the frequency of the corresponding hole or antihole and $\sigma$ is the spectral linewidth (for simplicity, we assume the same spectral linewidth for all the holes and antiholes.). The same analysis can be applied to the other eight classes of ions. $\Delta Neff(f)$ will be the sum of all holes and antiholes of the nine classes of ions. 

Before pumping, a scanning spectrum of the atomic ensemble is I$_{0}$(f), and after pumping, the spectrum is I$_{p}$(f). By definition, I$_{p}$(f) / I $_{i}$ (f) = e$^{-\Delta OD}$, where $\Delta OD = \Delta Neff(f)*\alpha$ and OD is the optical density and $\alpha$ is an absorption strength constant. Using a simple least squares fitting, we simulate the values of the branching ratio. The best fitting results suggest that t = 15 and $\sigma$ = 2.2~MHz. The measured spectrum typically contains experimental noises and the step size of the simulation may not be small enough due to time constraints, which may cause the results to be not optimized, but it should be quite close to the actual values. After that, some of the values can be slightly adjusted, and performing a few rounds of trials and errors, in the end, the optimized results should be obtained. This is a modified simulation method of the original proposal \cite{lauritzen2012spectroscopic}. SHB spectrum for [Eu$^\text{{3+}}$(BA)$_\text{4}$(pip)] \cite{serrano2022ultra} is simulated and shown in Fig. \ref{Fig2_SHB520}

\subsection{Quantum storage}
The experimental setup is shown in Fig. \ref{Fig_Setup}. Due to the limited extinction ratio of a typical AOM, to perform single-photon-level quantum storage, the laser beam has to be split into two paths: one path for spectral preparation, and another path for sending storage pulses. During storage processes, the pumping path has to be closed to block the leakage from the laser, so an electronically controlled shutter is placed in the pumping path. The single photon detector has a dead time of 45~ns and the storage laser pulse is 10~ns, so the average photon numbers of the input pulses have to be well below one to avoid the detector saturation.  

\begin{figure}[!h]
\centerline{\includegraphics[width=01\columnwidth]{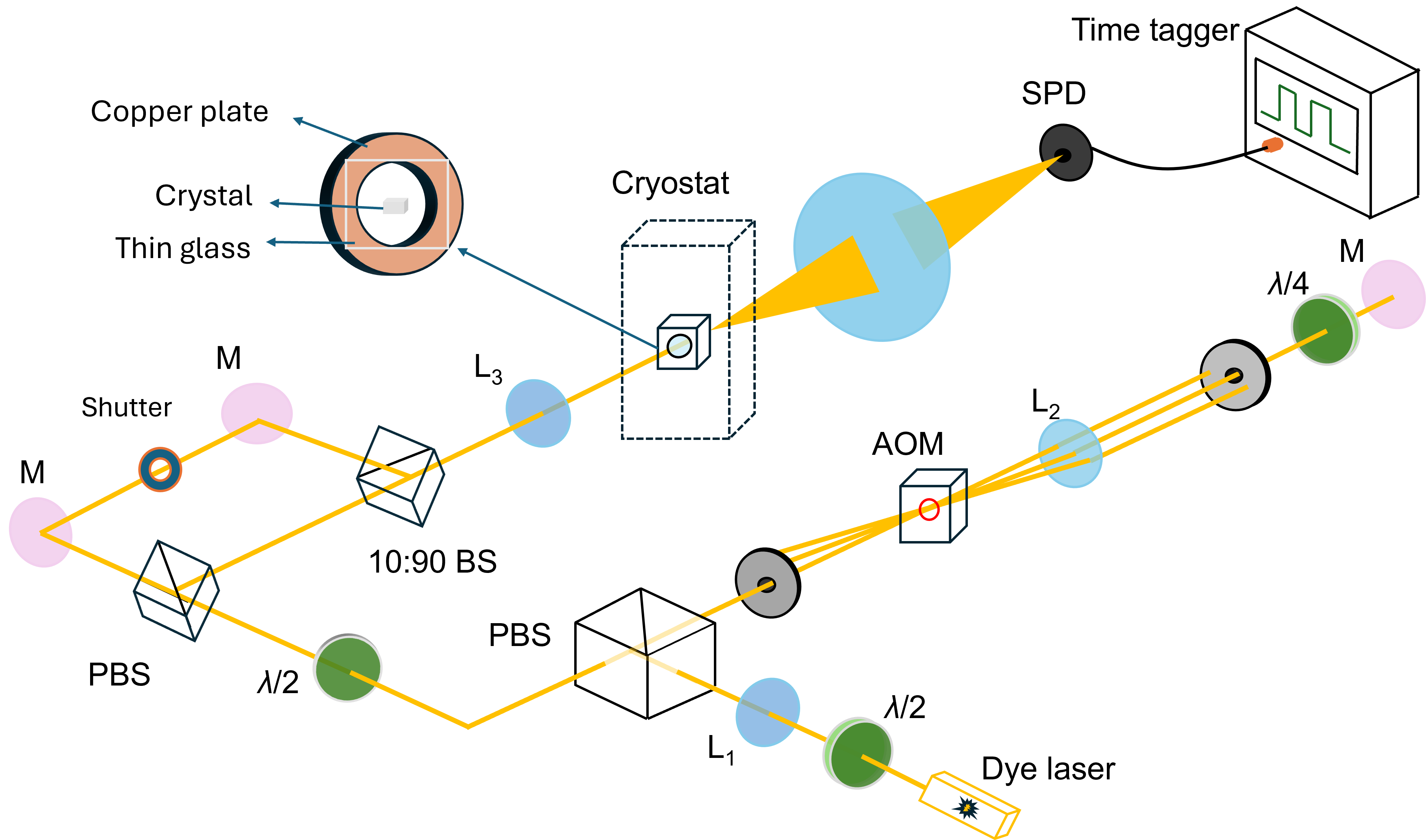}}
	\caption{Experimental setup: mirror (M), lens (L), polarization beam splitter (PBS), beam splitter (BS), single photon detector (SPD), half-wave plate ($\lambda$/2), quarter-wave plate ($\lambda$/4) and acoustic optical modulator (AOM).}
	\label{Fig_Setup}
\end{figure}

\subsection{Analysis of HAGEM spectral preparation}

For HAGEM, the anti-holes are right next to the holes (where pumping frequencies are set), so as long as there is a laser power difference between the anti-holes and the holes, some of the atoms are eventually shifted to the anti-holes after many cycles of pumping until reaching a balance. With a higher ratio of laser power at the spectral holes and anti-holes, HAGEM spectrum with less background absorption can be spectrally tailored. Here we give a simple illustration: if the optical coherence times of the atomic ensemble are 1~ms, 1/2~ms \& 1/3~ms, the corresponding spectral linewidth are $1/\pi$, $2/\pi$ \& $3/\pi$ in kHz, as shown in Fig. \ref{Fig_spectrallinewidth}. If the optical coherence time of the atomic ensemble is 1/3~ms (the corresponding spectral linewidth is $3/\pi$ in kHz.), the laser power at the spectral holes is still larger than the total laser power at the anti-holes with a few neighboring pumping pulses, so a proper HAGEM spectrum with $\Delta_{HAGEM}$ = 1~kHz can still be created (corresponding to a storage time of 1~ms.), but with stronger background absorption and less squarish grating shapes, which will reduce the storage efficiency. In addition, the storage efficiency will be further reduced by atomic dephasing, since the storage time is three times longer than the atomic coherence time. 

\begin{figure}[!h]
\centerline{\includegraphics[width=0.8\columnwidth]{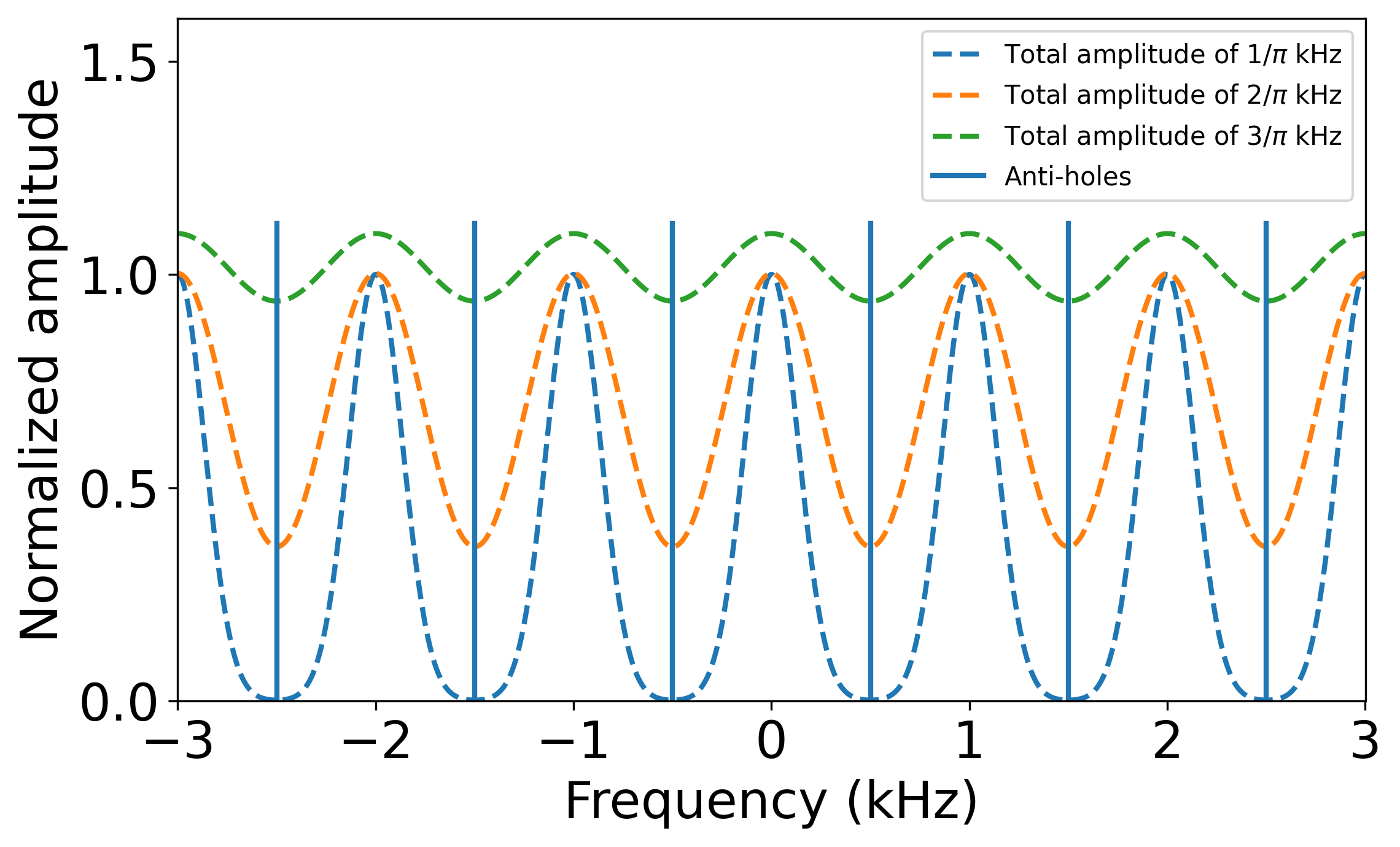}}
	\caption{Analysis of HAGEM spectral preparation: the dashed curves represent the total normalized spectral amplitudes for atomic ensembles with linewidth of $1/\pi$, $2/\pi$, and $3/\pi$~kHz, corresponding to optical coherence times of $1$, $1/2$, and $1/3$~ms, respectively. The six vertical lines mark the anti-hole positions with a spacing of $\Delta_{\mathrm{HAGEM}} = 1~\mathrm{kHz}$.}
	\label{Fig_spectrallinewidth}
\end{figure}

\section{Quantum repeater performance}

\begin{figure}[!h]
\centerline{\includegraphics[width=1\columnwidth]{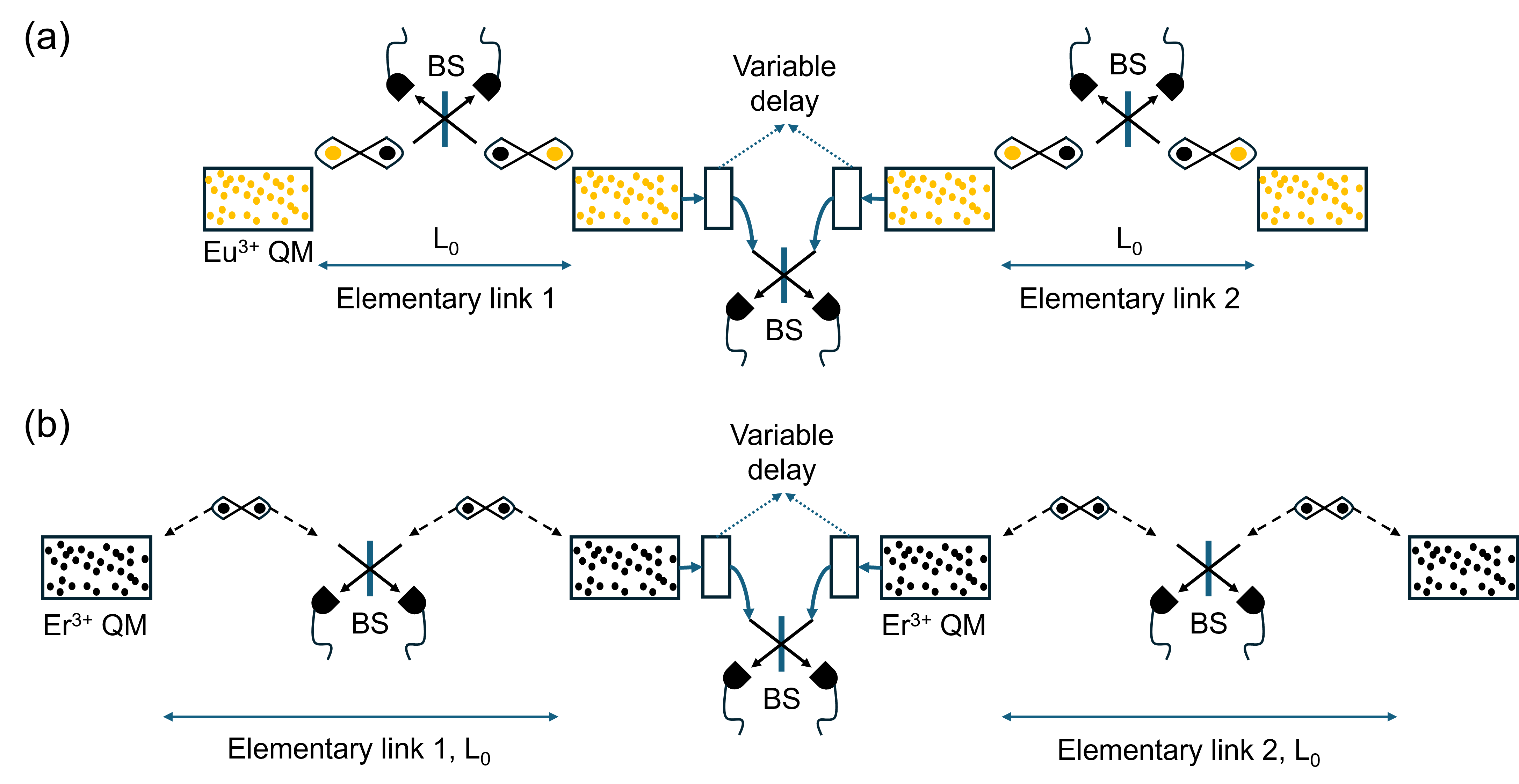}}
	\caption{Two quantum repeater configurations.}
	\label{Fig_QRArchtecture}
\end{figure}

The degenerate architecture is modeled with \(\mathrm{Er}^{\text{3+}}:\mathrm{YSO}\) quantum memories, whereas the non-degenerate architecture is modeled with \(\mathrm{Eu}^{\text{3+}}:\mathrm{YSO}\) quantum memories, as shown in Fig. \ref{Fig_QRArchtecture}.

We compare two probabilistic quantum-repeater models implemented in the accompanying
simulation notebooks. The first model uses a weak degenerate vacuum--pair source and
an explicit two-elementary-link architecture. The second model uses weak
non-degenerate pair sources with midpoint heralding and restricts the repeater search
to nesting levels \(n\geq 1\), so that at least two elementary links are always
present. In both simulations, the fiber transmission over a propagation distance
\(x\) is
\begin{equation}
T(x)=10^{-\alpha x/10},
\end{equation}
where \(\alpha\) is the fiber attenuation coefficient in dB/km. For total
end-to-end distance \(D\) and repeater level \(n\),
\begin{equation}
N_{\rm link}=2^n,
\qquad
L_0=\frac{D}{2^n},
\end{equation}
with \(L_0\) the elementary-link length.

The full simulation cycle includes the classical propagation delay over the total
distance and the sequential swapping-stage time,
\begin{equation}
t_{\rm cycle}
=
(n+1)\,t_{\rm stage}
+
\frac{1000D}{v_{\rm fiber}},
\end{equation}
where \(v_{\rm fiber}\) is the fiber group velocity ,\(t_{\rm stage}\) is the
duration assigned to one swapping stage and 1000 is the correction factor from kilometer to meter. The total storage-and-retrieval efficiency
is modeled as
\begin{equation}
\eta_m(D,n)
=
\eta_{m,\max}
\exp\!\left(
-\frac{4t_{\rm cycle}}{\tau_m}
\right),
\end{equation}
where \(\eta_{m,\max}\) is the zero-delay memory efficiency and \(\tau_m\) is the
memory lifetime. The occupied temporal-mode rate is
\begin{equation}
R_{\rm mode}
=
\min\!\left(
R_{\rm source},
R_{\rm storage}
\right).
\end{equation}

\subsection{Non-degenerate \(^{151}\mathrm{Eu}^{3+}:\mathrm{Y_2SiO_5}\) midpoint-heralding model}
\label{subsec:nondegenerate_model}

To simulate the quantum repeater performance with Eu$^{\text{3+}}$: YSO quantum memory devices, the quantum repeater architecture listed in Fig. \ref{Fig_QRArchtecture}(a) is used. For each elementary link, a weak
non-degenerate source generates a memory-compatible photon and a telecom photon.
The memory-compatible photon is stored locally, while the telecom photon propagates
toward the midpoint heralding station. In the single-photon heralding model considered
here, only one telecom photon must successfully reach the midpoint and be detected.

The source factor used for one elementary-link event is
\begin{equation}
P_{\rm src}=2p(1-p).
\end{equation}
The elementary-link success probability is therefore
\begin{equation}
\boxed{
P_{\rm elem}
=
2p(1-p)\,
T(L_0/2)\,
\eta_{\rm node}\,
\eta_d\,
P_{\rm BSM}
}.
\end{equation}
Here, \(\eta_m\) is the total storage-and-retrieval efficiency of the local
quantum memory, \(T(L_0/2)\) is the fiber transmission for the telecom photon
from the source to the midpoint, \(\eta_{\rm node}\) is the local optical/node
survival probability, \(\eta_d\) is the detector efficiency, and
\(P_{\rm BSM}\) is the intrinsic success probability of the heralding
Bell-state measurement. Because only one telecom photon is required to reach
the midpoint, the transmission/detection factor appears only once and is not
squared.

The higher-level entanglement-swapping probability is modeled as
\begin{equation}
\boxed{
P_{\rm swap}
=
P_{\rm BSM}\,
\eta_{\rm node}\,
\eta_d\,
\eta_{m, R}\,
N_{v}\,
\eta_v
}
\end{equation}
where $\eta_{m, R}$ = $\eta_{m,\max}
\exp\!\left(
-\frac{2t_{\rm cycle}}{\tau_m}
\right)$ is the memory efficiency at the repeater station, $N_v$ is the number of modes for the variable delay, and $\eta_v$ is the average efficiency of the variable delay.
For repeater level \(n\), the number of elementary links is
\begin{equation}
N_{\rm link}=2^n,
\end{equation}
and the number of physical swapping operations is
\begin{equation}
N_{\rm swap}=2^n-1.
\end{equation}
The end-to-end success probability is therefore
\begin{equation}
\boxed{
P_{\rm end}^{(\mathrm{nondeg})}
=
P_{\rm elem}^{\,2^n}
P_{\rm swap}^{\,2^n-1}\,
\eta_d\,
\eta_{\rm node}\,
\eta_m
}.
\end{equation}

The number of occupied temporal modes per cycle is
\begin{equation}
M
=
R_{\rm mode}\,
t_{\rm cycle},
\end{equation}
and the corresponding repeater distribution rate is
\begin{equation}
R_{\rm rep}^{(\mathrm{nondeg})}
=
\frac{M P_{\rm end}^{(\mathrm{nondeg})}}
{t_{\rm cycle}}
=
R_{\rm mode}
P_{\rm end}^{(\mathrm{nondeg})}.
\end{equation}

The direct-transmission baseline places one non-degenerate source at the
transmitting end. The telecom photon propagates over the full distance \(D\), and
the direct success probability is
\begin{equation}
P_{\rm dir}^{(\mathrm{nondeg})}
=
p\,T(D)\,
\eta_{\rm node}^{\,2}\,
\eta_d^{\,2}.
\end{equation}
The corresponding direct-transmission rate is
\begin{equation}
R_{\rm dir}^{(\mathrm{nondeg})}
=
R_{\rm source}
P_{\rm dir}^{(\mathrm{nondeg})}.
\end{equation}
No quantum-memory efficiency factor is included in this direct reference.

Table~\ref{tab:nondeg_params} summarizes the parameters used in the
non-degenerate \(\mathrm{Eu}^{\text{3+}}:\mathrm{YSO}\) simulation. The optimization was
performed over repeater levels \(n=1,\ldots,8\). For all distances considered,
the maximum repeater rate was obtained for \(n=1\), corresponding to two
elementary links and one physical entanglement-swapping operation.

\begin{table}[t]
\centering
\caption{Parameters used in the non-degenerate
\(\mathrm{Eu}^{\text{3+}}:\mathrm{YSO}\) midpoint-heralding simulation.}
\label{tab:nondeg_params}
\small
\begin{tabular}{ccc}
\toprule
\textbf{Parameter} & \textbf{Symbol} & \textbf{Value} \\
\midrule
Pair-generation probability & \(p\) & 0.05 \\
Source-attempt rate & \(R_{\rm source}\) & \(1.0\times10^{9}\,\mathrm{s^{-1}}\) \\
Storage-mode rate & \(R_{\rm storage}\) & \(1.0\times10^{9}\,\mathrm{s^{-1}}\) \\
Maximum memory efficiency & \(\eta_{m,\max}\) & 0.30 \\
Memory lifetime & \(\tau_m\) & 2 ms \\
Fiber attenuation & \(\alpha\) & 0.20 dB/km \\
Fiber group velocity & \(v_{\rm fiber}\) & \(2.0\times10^{8}\,\mathrm{m/s}\) \\
Node survival probability & \(\eta_{\rm node}\) & 0.90 \\
Detector efficiency & \(\eta_d\) & 0.90 \\
Intrinsic BSM success & \(P_{\rm BSM}\) & 0.50 \\
Swapping-stage duration & \(t_{\rm stage}\) & \(1\,\mu\mathrm{s}\) \\
Variable-delay modes & \(N_v\) & 100 \\
Variable-delay efficiency & \(\eta_v\) & 0.50 \\
\bottomrule
\end{tabular}
\end{table}

The labels in Table~\ref{tab:nondeg_results} are defined as follows:
\(D\) is the total end-to-end communication distance;
\(\eta_m\) is the memory efficiency at the selected repeater level;
\(P_{\rm elem}\) is the success probability of one elementary-link heralding event;
\(P_{\rm swap}\) is the success probability of one physical entanglement-swapping
operation;
\(P_{\rm end}\) is the complete end-to-end success probability per occupied
temporal mode;
\(R_{\rm rep}\) is the optimized repeater distribution rate; and
\(R_{\rm dir}\) is the corresponding direct-transmission rate;
\(R_{\rm dir}\) is the direct midpoint-source transmission rate; \(N_{\rm v}\) is the variable delay's mode number; and \(\eta_{\rm v}\) is the variable delay's average storage efficiency. For all rows shown below, the selected level is \(n=1\), so
\(N_{\rm link}=2\) and \(N_{\rm swap}=1\).

\begin{table}[t]
\centering
\caption{Numerical results for the optimized non-degenerate
\(\mathrm{Eu}:\mathrm{YSO}\) model.}
\label{tab:nondeg_results}
\scriptsize
\begin{tabular}{ccccccc}
\toprule
\(D\) (km) &
\(\eta_m\) &
\(P_{\rm elem}\) &
\(P_{\rm swap}\) &
\(P_{\rm end}\) &
\(R_{\rm rep}\) (Hz) &
\(R_{\rm dir}\) (Hz) \\
\midrule

80
& 0.134
& \(1.532\times10^{-2}\)
& 4.064
& \(1.037\times10^{-4}\)
& \(1.037\times10^{5}\)
& \(8.238\times10^{5}\) \\

100
& 0.110
& \(1.217\times10^{-2}\)
& 3.677
& \(4.847\times10^{-5}\)
& \(4.847\times10^{4}\)
& \(3.281\times10^{5}\) \\

120
& 0.090
& \(9.664\times10^{-3}\)
& 3.327
& \(2.266\times10^{-5}\)
& \(2.266\times10^{4}\)
& \(1.306\times10^{5}\) \\

140
& 0.074
& \(7.677\times10^{-3}\)
& 3.011
& \(1.059\times10^{-5}\)
& \(1.059\times10^{4}\)
& \(5.199\times10^{4}\) \\

160
& 0.060
& \(6.098\times10^{-3}\)
& 2.724
& \(4.950\times10^{-6}\)
& \(4.950\times10^{3}\)
& \(2.070\times10^{4}\) \\

180
& 0.049
& \(4.844\times10^{-3}\)
& 2.465
& \(2.314\times10^{-6}\)
& \(2.314\times10^{3}\)
& \(8.238\times10^{3}\) \\

200
& 0.040
& \(3.848\times10^{-3}\)
& 2.230
& \(1.081\times10^{-6}\)
& \(1.081\times10^{3}\)
& \(3.281\times10^{3}\) \\

\bottomrule
\end{tabular}
\end{table}

\subsection{Degenerate \(^{167}\mathrm{Er}^{3+}:\mathrm{Y_2SiO_5}\) midpoint-heralding model}
\label{subsec:degenerate_model}

To simulate the quantum repeater performance with Er$^{\text{3+}}$: YSO quantum memory devices, the quantum repeater architecture listed in Fig. \ref{Fig_QRArchtecture}(b) is used.
The degenerate source is described by the weak vacuum--pair state
\begin{equation}
|\psi\rangle
=
\sqrt{1-p}\,|00\rangle
+
\sqrt{p}\,|11\rangle,
\end{equation}
where \(p\) is the photon-pair generation probability per source attempt.

For the two-elementary-link network considered here, the useful Fock sector
contains exactly one generated photon pair. The pair can be generated in the
left photon source while the right remains in vacuum, or vice versa.
The total probability of occupying this useful sector is therefore
\begin{equation}
\boxed{
P_{\rm src}
=
2p(1-p)
}.
\end{equation}

The two-link calculation corresponds to repeater level \(n=1\), for which
\begin{equation}
L_0=\frac{D}{2},
\end{equation}
where \(D\) is the total end-to-end distance. The degenerate source is located
at the midpoint of an elementary link, so the two photons belonging to the
generated pair each propagate over a distance \(L_0/4\).

The elementary-link success probability is therefore
\begin{equation}
\boxed{
P_{\rm elem}
=
2p(1-p)\,
(T(L_0/4))^{\,2}\,
\eta_{\rm node}\,
\eta_d\,
P_{\rm BSM}
}.
\end{equation}
Here, \(\eta_m\) is the total storage-and-retrieval efficiency of the local
quantum memory, \(T(L_0/4)\) is the fiber transmission for the telecom photon
from the source to the heralding midpoint, \(\eta_{\rm node}\) is the local optical/node
survival probability (accounting for fiber coupling loss, device scattering loss, etc.), \(\eta_d\) is the detector efficiency, and
\(P_{\rm BSM}\) is the intrinsic success probability of the heralding
Bell-state measurement.

The higher-level entanglement-swapping probability is modeled as
\begin{equation}
\boxed{
P_{\rm swap}
=
P_{\rm BSM}\,
\eta_{\rm node}\,
\eta_d\,
\eta_{m, R}\,
N_{v}\,
\eta_v
}
\end{equation}
where $\eta_{m, R}$ = $\eta_{m,\max}
\exp\!\left(
-\frac{2t_{\rm cycle}}{\tau_m}
\right)$ is the memory efficiency at the repeater station, $N_v$ is the number of modes for the variable delay and $\eta_v$ is the average efficiency of the variable delay.
For repeater level \(n\), the number of elementary links is
\begin{equation}
N_{\rm link}=2^n,
\end{equation}
and the number of physical swapping operations is
\begin{equation}
N_{\rm swap}=2^n-1.
\end{equation}

The end-to-end success probability is therefore
\begin{equation}
\boxed{
P_{\rm end}^{(\mathrm{nondeg})}
=
P_{\rm elem}^{\,2^n}
P_{\rm swap}^{\,2^n-1}\,
\eta_d\,
\eta_{\rm node}\,
\eta_m
}.
\end{equation}

The number of occupied temporal modes per cycle is
\begin{equation}
M
=
R_{\rm mode}\,
t_{\rm cycle},
\end{equation}
and the corresponding repeater distribution rate is
\begin{equation}
R_{\rm rep}^{(\mathrm{nondeg})}
=
\frac{M P_{\rm end}^{(\mathrm{nondeg})}}
{t_{\rm cycle}}
=
R_{\rm mode}
P_{\rm end}^{(\mathrm{nondeg})}.
\end{equation}

The direct-transmission baseline places one non-degenerate source at the midpoint. The telecom photon propagates over the distance \(D/2\), and
the direct success probability is
\begin{equation}
P_{\rm dir}^{(\mathrm{nondeg})}
=
p\,T(D/2)^2\,
\eta_{\rm node}^{\,2}\,
\eta_d^{\,2}.
\end{equation}
The corresponding direct-transmission rate is
\begin{equation}
R_{\rm dir}^{(\mathrm{nondeg})}
=
R_{\rm source}
P_{\rm dir}^{(\mathrm{nondeg})}.
\end{equation}
No quantum-memory efficiency factor is included in this direct reference.

Table~\ref{tab:deg_params} summarizes the parameters used in the
\(^{167}\mathrm{Er}:\mathrm{YSO}\) simulation. Here,
\(R_{\rm source}\) is the source-attempt rate,
\(R_{\rm storage}\) is the temporal storage-mode capacity,
\(\eta_{m,\max}\) is the zero-delay total storage-and-retrieval efficiency,
\(\tau_m\) is the memory lifetime,
\(\alpha\) is the fiber attenuation coefficient,
\(v_{\rm fiber}\) is the fiber group velocity,
\(\eta_{\rm node}\) is the local optical/node survival probability,
\(\eta_d\) is the detector efficiency,
\(P_{\rm BSM}\) is the intrinsic Bell-state-measurement success probability,
and \(t_{\rm stage}\) is the duration of one swapping stage.

\begin{table}[t]
\centering
\caption{Parameters used in the degenerate
\(^{167}\mathrm{Er}:\mathrm{YSO}\) two-elementary-link simulation.}
\label{tab:deg_params}
\small
\begin{tabular}{ccc}
\toprule
\textbf{Parameter} & \textbf{Symbol} & \textbf{Value} \\
\midrule
Pair-generation probability & \(p\) & 0.05 \\
Source-attempt rate & \(R_{\rm source}\) & \(1.0\times10^{8}\,\mathrm{s^{-1}}\) \\
Storage-mode rate & \(R_{\rm storage}\) & \(1.0\times10^{8}\,\mathrm{s^{-1}}\) \\
Maximum memory efficiency & \(\eta_{m,\max}\) & 0.25 \\
Memory lifetime & \(\tau_m\) & 4 ms \\
Fiber attenuation & \(\alpha\) & 0.20 dB/km \\
Fiber group velocity & \(v_{\rm fiber}\) & \(2.0\times10^{8}\,\mathrm{m/s}\) \\
Node survival probability & \(\eta_{\rm node}\) & 0.90 \\
Detector efficiency & \(\eta_d\) & 0.90 \\
Intrinsic BSM success & \(P_{\rm BSM}\) & 0.50 \\
Swapping-stage duration & \(t_{\rm stage}\) & \(1\,\mu\mathrm{s}\) \\
Variable-delay modes & \(N_v\) & 100 \\
Variable-delay efficiency & \(\eta_v\) & 0.50 \\
\bottomrule
\end{tabular}
\end{table}

The quantities reported in Table~\ref{tab:deg_results} are defined as follows.
\(D\) is the total end-to-end fiber distance;
\(\eta_m\) is the cycle-time-dependent memory storage-and-retrieval efficiency;
\(P_{\rm elem}\) is the success probability of one elementary-link heralding event;
\(P_{\rm swap}=P_{\rm BSM}\eta_{\rm node}\eta_d\) is the probability of the
single-photon swapping/heralding event;
\(P_{\rm end}\) is the complete end-to-end success probability per occupied
temporal mode;
\(R_{\rm rep}\) is the corresponding repeater distribution rate;
\(R_{\rm dir}\) is the direct midpoint-source transmission rate; \(N_{\rm v}\) is the variable delay's mode number; and \(\eta_{\rm v}\) is the variable delay's average storage efficiency.

\begin{table}[t]
\centering
\caption{Numerical results for the degenerate
\(^{167}\mathrm{Er}:\mathrm{YSO}\) two-elementary-link model,
evaluated using the simulation code as written.}

\label{tab:deg_results}
\scriptsize
\begin{tabular}{ccccccc}
\toprule
\(D\) (km) &
\(\eta_m\) &
\(P_{\rm elem}\) &
\(P_{\rm swap}\) &
\(P_{\rm end}\) &
\(R_{\rm rep}\) (Hz) &
\(R_{\rm dir}\) (Hz) \\
\midrule

80
& 0.204
& \(1.532\times10^{-2}\)
& 4.141
& \(1.609\times10^{-4}\)
& \(1.609\times10^{4}\)
& \(8.238\times10^{4}\) \\

100
& 0.195
& \(1.217\times10^{-2}\)
& 3.939
& \(9.186\times10^{-5}\)
& \(9.186\times10^{3}\)
& \(3.281\times10^{4}\) \\

120
& 0.185
& \(9.664\times10^{-3}\)
& 3.747
& \(5.244\times10^{-5}\)
& \(5.244\times10^{3}\)
& \(1.306\times10^{4}\) \\

140
& 0.176
& \(7.677\times10^{-3}\)
& 3.564
& \(2.994\times10^{-5}\)
& \(2.994\times10^{3}\)
& \(5.199\times10^{3}\) \\

160
& 0.167
& \(6.098\times10^{-3}\)
& 3.390
& \(1.709\times10^{-5}\)
& \(1.709\times10^{3}\)
& \(2.070\times10^{3}\) \\

180
& 0.159
& \(4.844\times10^{-3}\)
& 3.225
& \(9.759\times10^{-6}\)
& \(9.759\times10^{2}\)
& \(8.238\times10^{2}\) \\

200
& 0.151
& \(3.848\times10^{-3}\)
& 3.067
& \(5.572\times10^{-6}\)
& \(5.572\times10^{2}\)
& \(3.281\times10^{2}\) \\

\bottomrule
\end{tabular}
\end{table}

\subsection{Multiplexing}

\begin{figure}[!h]
\centerline{\includegraphics[width=0.95\columnwidth]{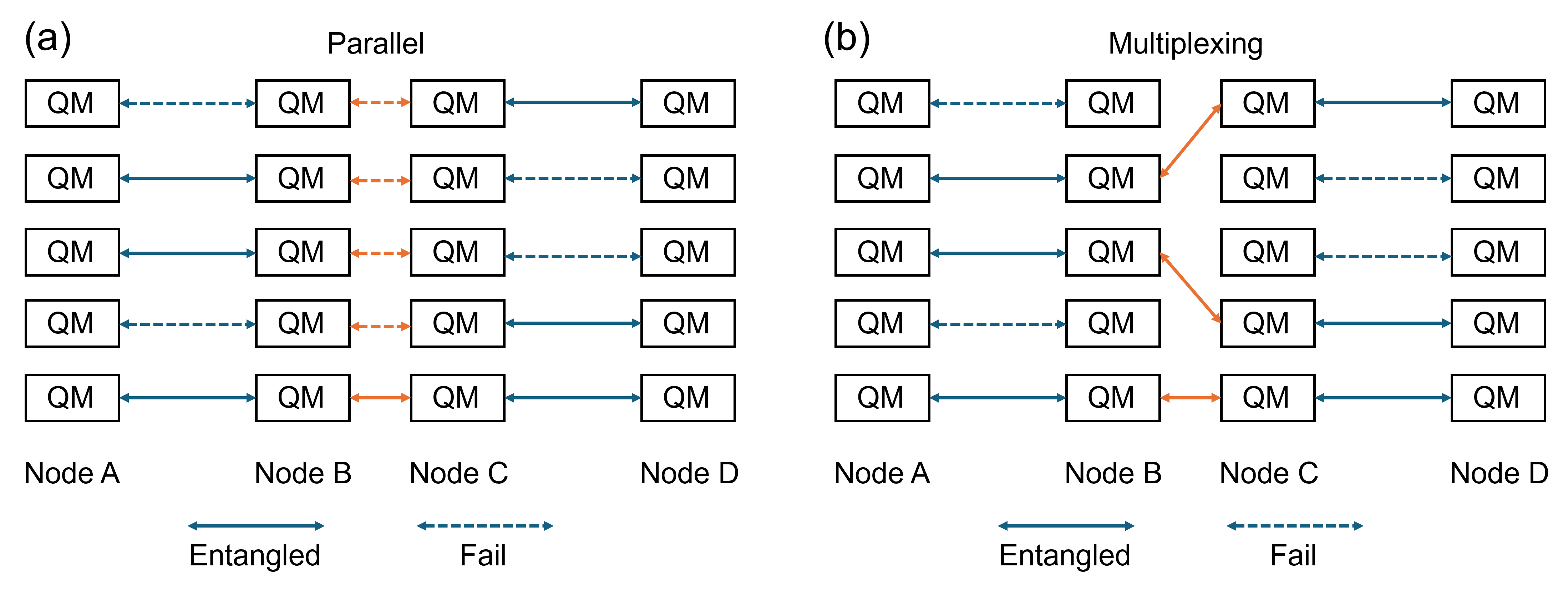}}
	\caption{(a) Parallel operation. (b) Multiplexing operation.}
	\label{Fig_multiplexing}
\end{figure}

Parallel operation can only increase the entanglement distribution rate linearly, while multiplexing operation can enhance the entanglement distribution rate exponentially. To achieve a long-distance entanglement distribution, multiple stages of entanglement swapping are required, so multiplexing operation is crucial to enhance the success probabilities of entanglement swapping, leading to a higher entanglement distribution rate. As discussed in the Main Text, this can be done by improving the performance of spin-wave storage, variable fiber delay lines, or variable free-space cavity delays.

============================================================
\makeatletter
\let\addcontentsline\supp@origaddcontentsline
\makeatother

\renewcommand{\refname}{Supplementary References}
\putbib

\end{bibunit}

\endgroup

\end{document}